\documentclass[useAMS,usenatbib]{mn2e}  
\usepackage{amssymb, amsmath, graphics}
\usepackage{graphicx}  
\usepackage[usenames]{color}  
\usepackage{aas_macros}

\title[Kepler non-Blazhko RR~Lyr Stars]
{Fourier analysis of non-Blazhko ab-type RR~Lyrae stars observed with the {\it Kepler} space telescope}

\author[Nemec, Smolec, Benk\H{o} {\it et al.}]
{J.~M.~Nemec$^{1,2}$\thanks{E-mail:
nemec@camosun.bc.ca},
R. Smolec$^{3}$,
J.~M.~Benk\H{o}$^{4}$,
P. Moskalik$^{5}$,
K. Kolenberg$^{3,6}$, 
\and 
R. Szab\'o$^{4}$,   
D. W. Kurtz$^{7}$, 
S. Bryson$^{8}$,
E. Guggenberger$^{3}$,
M. Chadid$^{9}$,
\and
Y.-B. Jeon$^{10}$,
A. Kunder$^{12}$,
A. C. Layden$^{13}$,
K. Kinemuchi$^{8}$,
L. L. Kiss$^{4}$,
\and
E. Poretti$^{14}$,
J. Christensen-Dalsgaard$^{11}$, 
H. Kjeldsen$^{11}$,
D. Caldwell$^{15}$,
V. Ripepi$^{16}$,
\and
A. Derekas$^{4}$,
J. Nuspl$^{4}$,
F. Mullally$^{15}$,
S.~E.~Thompson$^{15}$, 
W. J. Borucki$^{8}$ 
\\
\\
$^{1} $Department of Physics \& Astronomy, Camosun College, Victoria, British Columbia, V8P~5J2, Canada\\
$^{2} $International Statistics \& Research Corporation, PO Box 39, Brentwood Bay, British Columbia, V8M~1R3, Canada\\
$^{3} $Institute for Astronomy, University of Vienna, T\"urkenschanzstrasse 17, A-1180 Vienna, Austria\\
$^{4} $Konkoly Observatory of the Hungarian Academy of Sciences, Konkoly Thege Mikl\'os \'ut 15-17, H-1121 Budapest, Hungary\\
$^{5} $Copernicus Astronomical Center, ul. Bartycka 18, 00-716 Warsaw, Poland\\
$^{6} $Harvard College Observatory, 60 Garden Street, Cambridge, MA 02138, USA\\
$^{7} $Jeremiah Horrocks Institute of Astrophysics, University of Central Lancashire, Preston PR1 2HE, UK\\
$^{8} $NASA Ames Research Center, MS 244-30, Moffett Field, CA 94035, USA\\
$^{9} $Observatoire de la C\^ote d�Azur, Universit\'e Nice Sophia-Antipolis, UMR 6525, Parc Valrose, 06108 Nice Cedex 02, France\\
$^{10}$Korea Astronomy and Space Science Institute, Daejeon, 305-348, Korea\\
$^{11}$Department of Physics and Astronomy, Aarhus University, DK-8000 Aarhus C, Denmark\\
$^{12}$Cerro Tololo Inter-American Observatory, Cassila 603, La Serena, Chile\\
$^{13}$Physics \& Astronomy Department, Bowling Green State University, Bowling Green, Ohio OH 43403\\ 
$^{14}$Osservatorio Astronomico di Brera, Via E. Bianchi 46, 23807 Merate, Italy\\
$^{15}$SETI Institute/NASA Ames Research Center, MS 244-30, Moffett Field, CA 94025, USA\\ 
$^{16}$INAF-Osservatorio Astronomico di Capodimonte, Via Moiariello 16, I-80131, Napoli, Italy\\
}

\date{Accepted 2011 June 24.  
      Received 2011 June 23;
      in original form 2011 March 28}
  
\begin{document}  
  
  
\pagerange{\pageref{firstpage}--\pageref{lastpage}} \pubyear{2011}  
\maketitle

\label{firstpage}  
  
\begin{abstract}
Nineteen of the $\sim$40 RR~Lyrae stars in the {\it Kepler} field have been 
identified as candidate non-Blazhko (or unmodulated) stars. 
In this paper we present the results of Fourier decomposition of the 
time-series photometry of these stars acquired during the first 417 days of 
operation (Q0-Q5) of the {\it Kepler} telescope.   
Fourier parameters based on $\sim$18400 long-cadence observations per star
(and $\sim$150000 short-cadence observations for FN~Lyr and for AW~Dra)
are derived. None of the stars shows the recently discovered `period-doubling'
effect seen in Blazhko variables; however, KIC~7021124 has been found to pulsate 
simultaneously in the fundamental and second overtone modes with a period 
ratio $P_2/P_0 \sim 0.59305$ and is similar to the double-mode star V350~Lyr.
Period change rates are derived from 
O$-$C diagrams spanning, in some cases, over 100 years; these are compared
with high-precision
periods derived from the {\it Kepler} data alone. 
Extant Fourier correlations by Kov\'acs, Jurcsik {\it et al.}
(with minor transformations from the $V$ to the {\it Kp} passband) have been used 
to derive underlying physical characteristics for all the stars. 
This procedure seems to be validated through 
comparisons of the {\it Kepler} variables with galactic and LMC RR~Lyrae stars. 
The most metal-poor star in the sample is NR~Lyr, with [Fe/H]$=-2.3$ dex; and the 
four most metal-rich stars have [Fe/H] ranging from $-0.6$ to $+0.1$ dex.    
Pulsational luminosities and masses
are found to be systematically smaller than $L$ and ${\cal M}$ values derived from  
stellar evolution models, and are favoured over the evolutionary values when periods are
computed with the Warsaw linear hydrodynamics code.  Finally, the Fourier parameters are compared 
with theoretical values derived using the Warsaw non-linear convective pulsation code.   
\end{abstract} 
 

\section{INTRODUCTION}

The {\it Kepler} Mission is designed to detect ({\it via} transits) Earth-like planets around solar-type 
stars (Koch {\it et al.} 2010). To achieve this goal the 1.4-m {\it Kepler} telescope has been monitoring almost
continuously the light variations of over 150\,000 stars in a field located at
(RA,DEC)$_{\rm J2000}$ = (19:22:40, +44:30) or ($l$,$b$)=(76.3, +13.5).  The mandate of the
Kepler Asteroseismic Science Consortium (KASC) is to use these data to better understand the 
astrophysics of the stars seen in the {\it Kepler} field, with Working Group 13 being responsible 
for the study of the RR~Lyr variable stars.

Kolenberg {\it et al.} (2010) discussed the initial selection of 28 target RR~Lyr stars 
in the {\it Kepler} field and presented preliminary results based on the earliest {\it Kepler}
observations for RR~Lyr and V783~Cyg, both of which 
exhibit the Blazhko (1907) effect, and for the `non-Blazhko' star NR~Lyr.  An important discovery
was the phenomenon of `period doubling', a cycle-to-cycle amplitude variation that occurs at certain Blazhko phases.
The effect has been explained by Szab\'o {\it et al.} (2010) as a 9:2 resonance between the fundamental mode
and the 9th-order radial overtone.  More recently Koll\'ath, Moln\'ar \& Szab\'o (2011) showed that
the resonance destabilises the fundamental mode limit cycle leading to period doubling.
Among RR~Lyr stars period doubling has been observed to date only in the Blazhko variables.      
Benk\H{o} {\it et al.} (2010, hereafter B10) studied a sample of 29 modulated and unmodulated RR~Lyr stars 
using {\it Kepler} Q0-Q2 long-cadence data, and identified 14 of the 29 stars (48\%) as 
exhibiting the Blazhko effect.  Most recently, a more detailed investigation of RR~Lyrae itself 
has been made by Kolenberg {\it et al.} (2011), and Smolec {\it et al.} (2011) have used 
non-linear hydrodynamic models to model Blazhko RR~Lyr stars. 

In the present paper we
look more closely at the unmodulated ab-type RR~Lyr stars -- that is, those 
stars pulsating in the fundamental mode and believed to exhibit no amplitude or phase modulations and 
to have the most stable light curves.  According to the notation established by Alcock {\it et al.} (2000) our sample stars are of type RR0.
The analysis is based primarily on the {\it Kepler} long-cadence (30-min)
photometry acquired in quarters Q0-Q5 (over 18000 measurements per star made over 417~d), and 
the short-cadence (1-min) photometry for FN~Lyr and AW~Dra acquired in Q0
(over 14000 high-precision brightness measurements per star made over 9.7~d) and 
Q5 ($\sim$135,000 observations per star made over $\sim$90 days)\footnote[1]{Public 
data are available from the internet website `http://archive.stsci.edu/kepler/'.  
It is the policy of the {\it Kepler}
Mission to make public all the data mentioned in published research papers.  Thus the Q0-Q5 
long and short cadence data analysed here will become publicly available upon publication of the present paper.}.

We first discuss the sample selection ($\S2.1$) and
procedure for transforming the raw fluxes delivered from the {\it Kepler} Science Office 
to magnitudes on the {\it Kp} system ($\S$2.2).  Before analyzing the {\it Kepler} data a search  ($\S3.1$) was conducted
of available historical data for the stars ({\it e.g.}, photometry, periods, light curves, and times of maximum light).
Accurate periods were derived using all the available {\it Kp} photometry ($\S3.2$).   
Then, from the plotted light curves, total amplitudes and risetimes were
measured ($\S3.3$).   Period change rates (d$P$/d$t$) and revised periods were calculated from O$-$C diagrams
constructed with the historical data combined with the {\it Kepler} data ($\S3.4$).
Fourier analysis of light curves ($\S4.1$) was carried out for both the {\it Kp} data
($\S4.2$) and for the extant $V$ data ($\S4.4$).   
The long time base and intensive sampling of the {\it Kepler} data provided an opportunity
to examine the light curves for cycle-to-cycle variations.  To test for such
variations the data were divided into single-cycle blocks, and for each
block a separate Fourier analysis was conducted ($\S4.3$). 
For the {\it Kepler} stars observed correlations among the Fourier parameters (all cycles combined) were examined ($\S5$)
for both the {\it Kp} and $V$ passbands ($\S5.1$).  The resulting $V$-{\it Kp} offsets ($\S5.2$) allowed comparisons to be 
made between the {\it Kepler} stars and RR~Lyr stars in galactic and Magellanic Cloud 
globular clusters ($\S5.3$) and field stars in the inner regions of
the Large Magellanic Cloud ($\S5.4$).  Physical characteristics 
were also derived from suitably-modified extant $V$-band correlations ($\S6$). 
These include iron (or metal) abundance [Fe/H], dereddened colour $(B-V)_0$, 
effective temperature
$T_{\rm eff}$,  distance $d$, absolute magnitude $M_V$, pulsational and evolutionary luminosities, $L$(puls) and 
$L$(evol), pulsational and evolutionary masses, ${\cal M}$(puls) and ${\cal M}$(evol), 
and location in the instability strip.  Finally, new 
non-linear hydrodynamical models (Warsaw code) are presented and compared with the observational data ($\S7$), and  
our results are summarized ($\S8$).  In Appendix A the newly discovered double-mode
star KIC~7021124 is discussed, and in Appendix B the cyclic behaviour of nine non-Blazhko stars in the ($V-I, V$)-diagram 
(based on ASAS-North data) is discussed.

\section{DATA PROCESSING}

\begin{table*}
\caption[]{Basic data for the 19 {\it Kepler} non-Blazhko ab-type RR~Lyr stars}
\label{Table1}
\begin{flushleft}
\begin{tabular}{clcllcccc}
\hline
\noalign{\smallskip}
KIC     & \multicolumn{1}{c}{ Star}  &$\langle Kp \rangle$&\multicolumn{1}{c}{ Period [d]}     &\multicolumn{1}{c}{Freq. [d$^{-1}$]}  &  $t_0$ &  RT({\it Kp})  & {\it Kp}-range & $A_{\rm tot}$({\it Kp})   \\
        & \multicolumn{1}{c}{ name}  &  [mag]             &\multicolumn{1}{c}{  (this paper) } &\multicolumn{1}{c}{  (this paper)   } &  [BJD] &  [phase]   &  [mag]     &   [mag]      \\
 (1)    & \multicolumn{1}{c}{ (2) }  &   (3)              &\multicolumn{1}{c}{    (4)    }     &\multicolumn{1}{c}{     (5)       }   &   (6)  &  (7)       &     (8)    &   (9)    \\           
\noalign{\smallskip}
\hline
\noalign{\smallskip}
 3733346&  NR~Lyr         & 12.684    & 0.6820264(2)  &1.4662189(4) & 54964.7381 &   0.144&  12.215-12.982& 0.767  \\
 3866709&  V715~Cyg       & 16.265    & 0.47070494(4) &2.1244731(2) & 54964.6037 &   0.136&  15.612-16.600& 0.988  \\
 5299596&  V782~Cyg       & 15.392    & 0.5236377(1)  &1.9097173(4) & 54964.5059 &   0.200&  15.108-15.631& 0.523  \\
 6070714&  V784~Cyg       & 15.370    & 0.5340941(1)  &1.8723292(4) & 54964.8067 &   0.195&  15.036-15.670& 0.634  \\
 6100702&  KIC~6100702    & 13.458    & 0.4881457(2)  &2.0485687(8) & 54953.8399 &   0.200&  13.140-13.715& 0.575  \\
 6763132&  NQ~Lyr         & 13.075    & 0.5877887(1)  &1.7012916(3) & 54954.0702 &   0.144&  12.576-13.387& 0.811  \\
 6936115&  FN~Lyr         & 12.876    & 0.527398471(4)&1.89609954(1)& 54953.2690 &   0.118&  12.146-13.227& 1.081  \\
 7021124&  KIC~7021124    & 13.550    & 0.6224925(7)  &1.606445(2)  & 54965.6471 &   0.139&  13.018-13.849& 0.831  \\
 7030715&  KIC~7030715    & 13.452    & 0.6836137(2)  &1.4628145(4) & 54953.8434 &   0.177&  13.092-13.739& 0.647  \\
 7176080&  V349~Lyr       & 17.433    & 0.5070740(2)  &1.9720987(8) & 54964.9555 &   0.130&  16.780-17.768& 0.988  \\
 7742534&  V368~Lyr       & 16.002    & 0.4564851(1)  &2.1906520(5) & 54964.7828 &   0.134&  15.237-16.370& 1.133  \\
 7988343&  V1510~Cyg      & 14.494    & 0.5811436(1)  &1.7207451(3) & 54964.6695 &   0.138&  13.868-14.848& 0.980  \\
 8344381&  V346~Lyr       & 16.421    & 0.5768288(1)  &1.7336166(3) & 54964.9211 &   0.141&  15.809-16.773& 0.964  \\
 9508655&  V350~Lyr       & 15.696    & 0.5942369(1)  &1.6828305(3) & 54964.7795 &   0.143&  15.086-16.059& 0.973  \\
 9591503&  V894~Cyg       & 13.293    & 0.5713866(2)  &1.7501285(6) & 54953.5627 &   0.144&  12.579-13.684& 1.105  \\
 9947026&  V2470~Cyg      & 13.300    & 0.5485894(1)  &1.8228569(3) & 54953.7808 &   0.193&  12.971-13.570& 0.599  \\
10136240&  V1107~Cyg      & 15.648    & 0.5657781(1)  &1.7674774(3) & 54964.7532 &   0.151&  15.151-15.969& 0.818  \\
10789273&  V838~Cyg       & 13.770    & 0.4802799(1)  &2.0821192(4) & 54964.5731 &   0.134&  13.040-14.138& 1.098  \\
11802860&  AW~Dra         & 13.053    & 0.6872160(2)  &1.4551466(6) & 54954.2160 &   0.165&  12.538-13.430& 0.892  \\
\hline
\end{tabular}
\end{flushleft}
\end{table*}

\subsection{Selection of non-Blazhko stars}

Approximately 40 ~RR~Lyr stars are presently known in the {\it Kepler} field, up from the 
29 stars discussed by Kolenberg {\it et al.} (2010) and B10.    
Four are multiperiodic c-type RR~Lyr stars and are the subject of another paper
(Moskalik {\it et al.} 2011, in preparation), and $\sim$20 are Blazhko stars,
many of which have quite complex modulated light curves (see B10). 
We are interested here in the unmodulated non-Blazhko ab-type RR~Lyr stars with the most stationary light
curves.   When the Fourier parameters and physical characteristics of these stars are known
this information will provide a useful reference for future comparisons with the more complex Blazhko stars
(see Jurcsik {\it et al.} 2009).

  Table~1 of B10 gives coordinates for 16 of the 19 stars considered here,  
and pulsation periods to five decimal places.
Three stars are not in the B10:  FN~Lyr (KIC~6936115)
and AW~Dra (KIC~11802860), both of which were observed at short-cadence (SC) in Q0 with 
long-cadence Q1 data and short-cadence Q5 data released later;  and KIC~7021124, which was discovered
subsequently.  The RA and DEC coordinates (J2000) of these three stars are as follows:
FN~Lyr (19:10:22.25,  +42:27:31.6),
AW~Dra  (19:00:48.00, +50:05:31.3)  and
KIC~7021124 (19:10:26.69, +42:33:37.0).

Three of the stars included in the present study are special cases:
(1) {\bf V349~Lyr} was identified by B10 as a
Blazhko star with a small amplitude modulation and Blazhko-period longer than 127 days.  It is one of the faintest
RR~Lyr stars in the {\it Kepler} field, at $\langle Kp \rangle = 17.433$~mag.  The
detrended light curve in fig.~1 of B10 shows the amplitude decreasing slowly throughout the Q1-Q2 epochs.
Classification of this star depends on the  
detrending procedure that is used to transform
the raw fluxes to magnitudes (see next section and Fig.~1).
It has been included here as a borderline non-Blazhko star.
(2) {\bf V350~Lyr} was discovered by B10 to be pulsating simultaneously
in the fundamental and second 
overtone modes ({\it i.e.}, an RR02 star), with a secondary period $P_2=0.35209$~d and a period ratio $P_2/P_0 = 0.592$.  
So far there has been no evidence for the Blazhko effect.  
With the additional $\sim$270 days provided by the
Q3-Q5 observations this star's behaviour has been 
confirmed and strengthened.  One question of interest concerns the
stability of the $f_2$ frequency -- B10 determined that the
$f_2$ amplitude was varying between 0.00 and 0.05 mag.  This result is confirmed by the newer data.  
Recent work to better understand this phenomenon has been presented by Koll\'ath {\it et al.} (2011). 
(3) {\bf KIC~7021124} was found during the course of the present study to be 
pulsating simultaneously in the fundamental and second-overtone modes, and to be
similar to V350~Lyr.  The details of its pulsational behaviour are given in Appendix~A.  

Basic data for the 19 unmodulated {\it Kepler} RR~Lyrae stars is given in {\bf Table~1}.   Column~3 contains
the mean {\it Kp} magnitude given in the {\it Kepler} Input Catalog (KIC).   
Columns~4-5 contain our best estimate of the pulsation period and frequency derived
using the {\it Kp} data (see Table~3 for details). 
Column~6 contains accurate values for the times of maximum light, 
and column~7 lists precise risetimes (RTs) derived from the
{\it Kp}-photometry.   Columns~8-9 contain the {\it Kp} magnitudes at which
the light curve reaches maximum and minimum light, the difference 
being the total amplitude in the {\it Kp} passband.

\subsection{From raw fluxes to {\it Kp} magnitudes}

The preprocessing phases of the long-cadence (LC) {\it Kepler} data have been described by Jenkins {\it et al.} (2010a,b) and
B10.  Since then three more `seasons' (or $\sim$270 days) of LC data (Q3-Q5) have become 
available for 18 of the 19 stars in our sample.  Also available now are Q0 ($\sim$10 days) and Q5 ($\sim$90 days) SC 
photometry (see Gilliland {\it et al.} 2010) for FN~Lyr and AW~Dra.   The procedures used here for transforming the raw fluxes to {\it Kp} magnitudes
are described next.

\begin{figure*}
\begin{center}$
\begin{array}{cc}
\includegraphics[width=8.0cm]{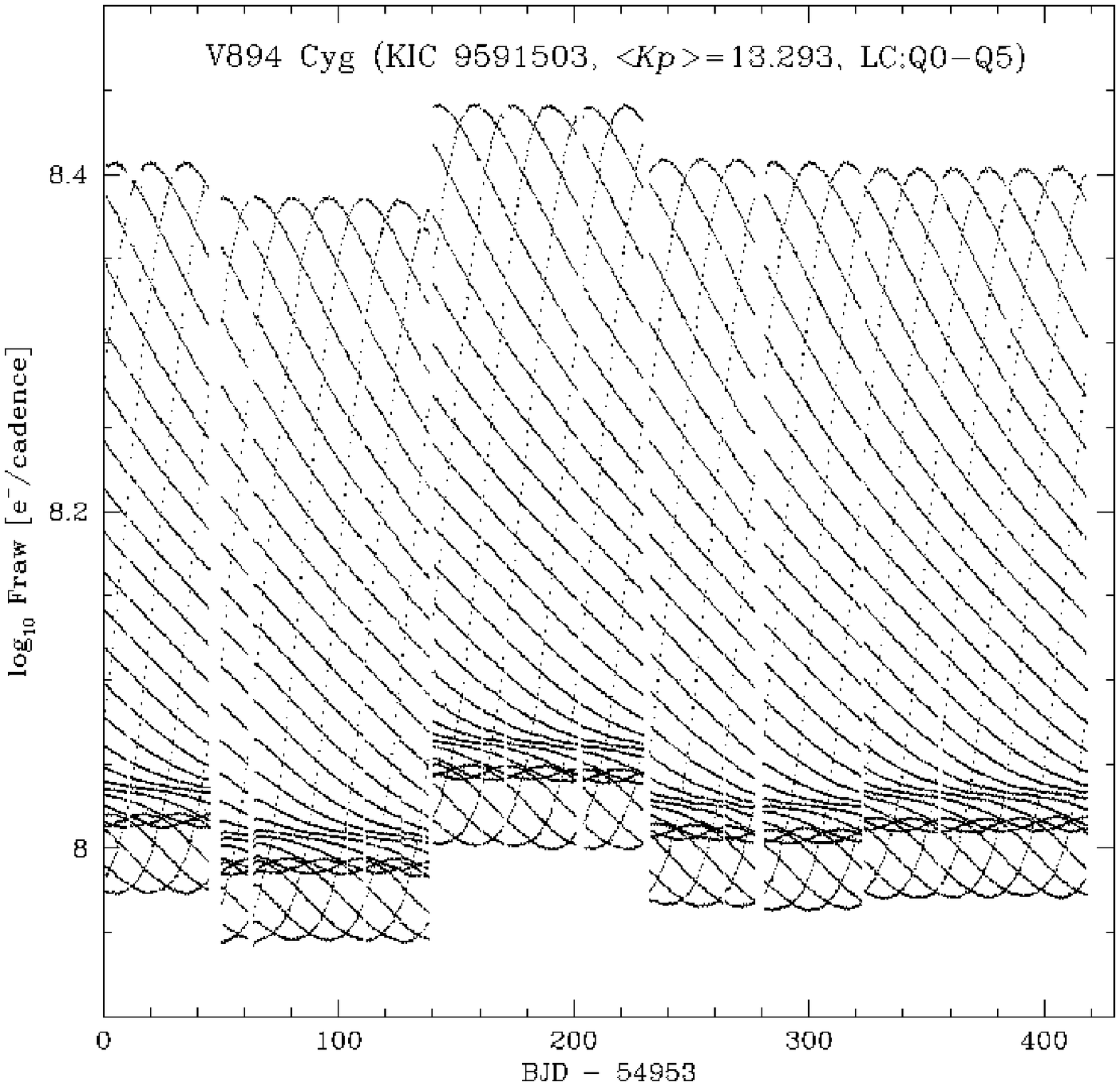}&
\includegraphics[width=8.0cm]{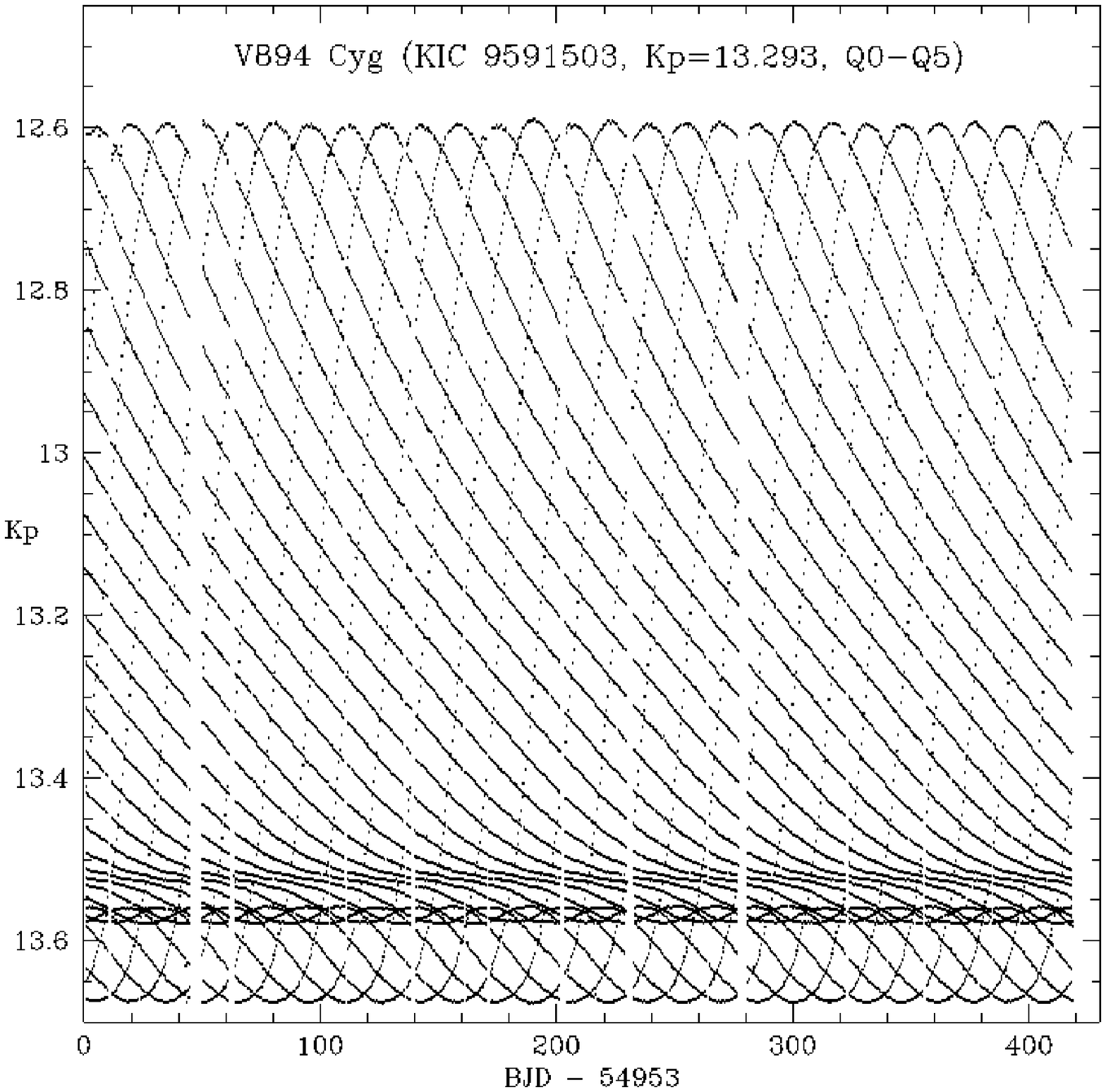} \\
\includegraphics[width=8.0cm]{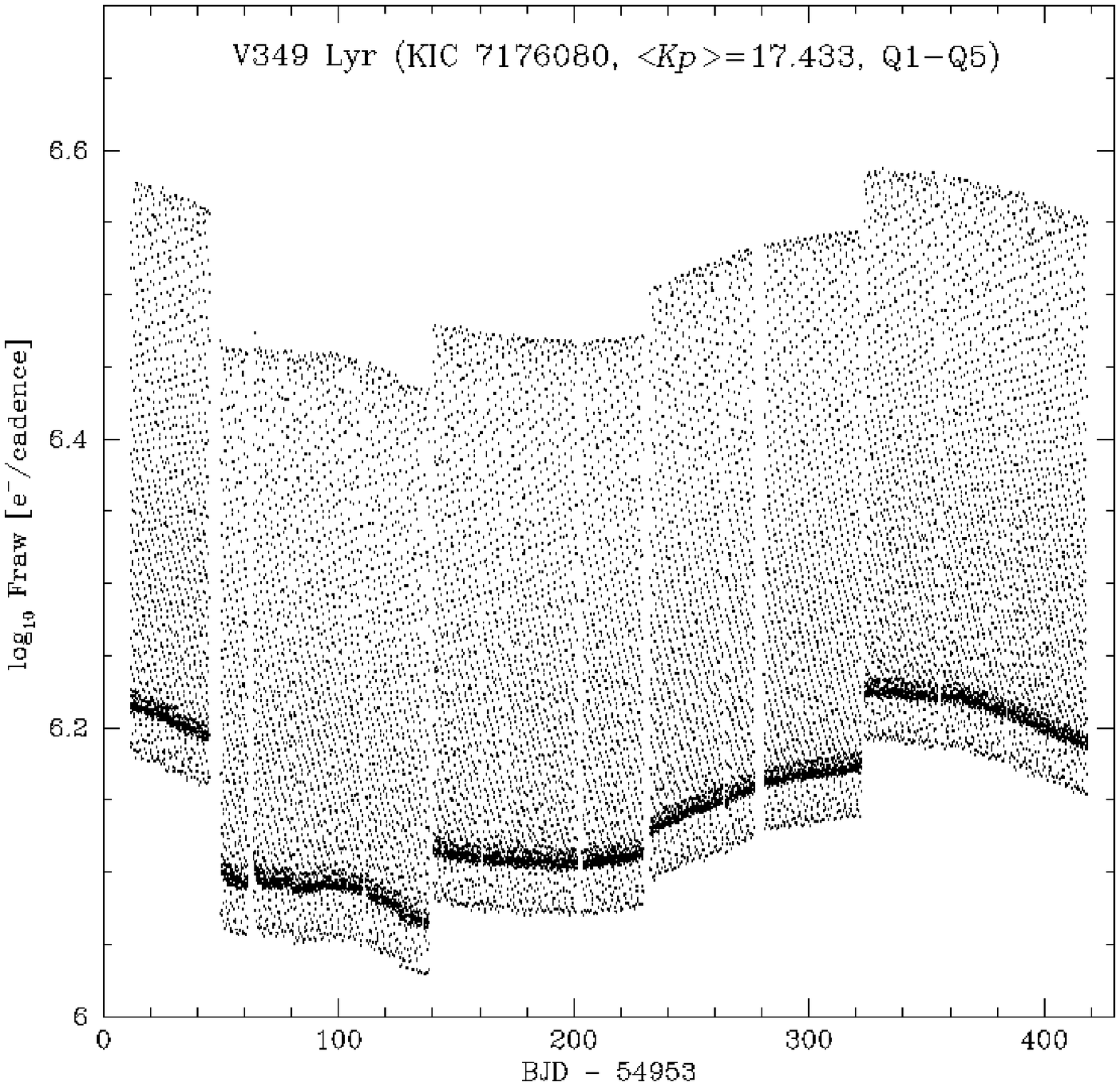}&
\includegraphics[width=8.0cm]{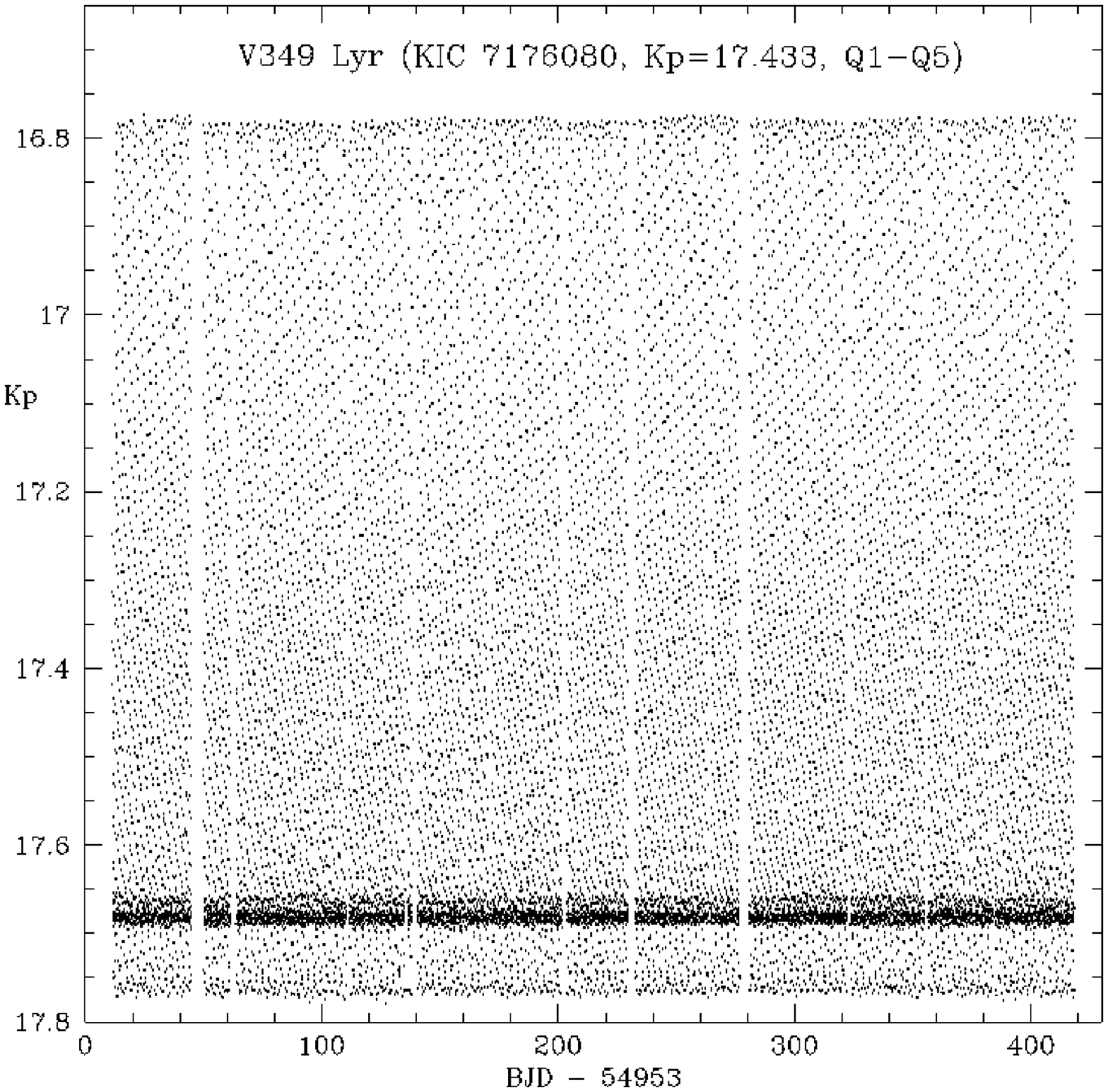}
\end{array}$
\end{center}
\caption{ Raw (left) and processed (right) photometry for V894~Cyg (top) and V349~Lyr (bottom).
V894~Cyg was observed 18911 times at long cadence in quarters Q0-Q5 and is of intermediate brightness with raw flux density
$\sim  1.6 \times 10^8$ e$^{-}$/cadence.  V349~Lyr is the faintest RR~Lyr star in the {\it Kepler} field and was
observed 18387 times at LC:Q1-Q5 with raw flux density $\sim 2 \times 10^6$ e$^{-}$/cadence.   }
\label{AWDra_raw_RTLC}
\end{figure*}

\subsubsection{Long-cadence data}

The present analysis of the stars in Table~1 is based on the raw fluxes.   
Sixteen of the stars were observed at LC (30 min) in `seasons' Q1 to Q5, and
of these, five were also observed in Q0.  KIC~6100702 was not observed in Q5, 
and KIC~7021124 was observed only in Q1. 
The Q2-Q4 LC observations of FN~Lyr were acquired as part of the {\it Kepler} Guest 
Observer Program and Drs.~K. Mighell and S. Howell kindly shared their 
observations with us. 
The original data consisted of raw flux counts as a function of time,
with times given as barycentric Julian date (BJD).
For convenience all the times were translated by subtracting 54953~d so that 
the first Q0 observations occur at time $t=0.53$~d.   

{\bf Figure~1} helps to illustrate the procedures used to convert long cadence fluxes to {\it Kp} magnitudes.
The left panels show the observed raw fluxes (log10 scale) of V894~Cyg (top) and V349~Lyr (bottom) plotted 
against time.  Both stars were observed every 30 minutes in `quarters' Q1 to Q5
and V894~Lyr was also observed in Q0. 
One sees zero point differences (sensitivity variations) resulting from quarterly `rolls' of 
the {\it Kepler} telescope causing the light from a given star
to fall on different CCD chips and pixels each quarter, and trends (both linear and non-linear, and 
increasing and decreasing) within a given quarter.
The scalloped pattern seen for V894~Cyg (with amplitude larger at maximum 
than at minimum light) is due to the 30-min interval
between the LC observations going in and out of phase with the light cycle determined by
the pulsation period of 0.57~d. 
The time ranges for each LC `season' were as follows:
BJD~54953.53 to 54963.24, corresponding to time 0.53 to 10.24~d (Q0),
BJD~54964.51 to 54997.98, or 11.51 to 44.98~d (Q1),
BJD~55002.94 to 55091.47, or 49.94 to 138.47~d (Q2),
BJD~5093.52 to 55182.50, or 140.22 to 229.50~d (Q3),   
BJD~55185.38 to 55275.20,  or 232.38 to 322.20~d (Q4), and
BJD~55275.99 to 55370.66, or 322.99 to 417.66~d (Q5). 
 
The right panels of Fig.~1 show the corresponding detrended and normalized {\it Kp} apparent magnitudes.
The procedure for transforming the raw 
fluxes into apparent magnitudes was as follows: 
(1) the data were separated into blocks, usually by `quarter' but various
blocks were tried (see the left panels of Fig.~1);
(2) the raw fluxes were converted to apparent magnitudes using `mag = $33.5 - 2.5 \times$log10(Fraw)', where
the magnitude zero point is as yet only approximate but gives mean values roughly in accord with the
KIC mean magnitudes;  
(3) within a block the magnitudes were detrended by fitting and subtracting either a linear or polynomial fit; 
(4) the detrended data for each block were shifted to the {\it Kp} mean magnitude level given in the KIC;  
(5) within each block the magnitudes were stretched or compressed by multiplying by an appropriate scale factor
so that the minimum and maximum magnitudes were in agreement with an adopted reference quarter (usually either Q4 or Q5);
(6) obvious outliers ($\sigma > 5$) were removed -- before this was done the total number of 
LC observations per `quarter' was as follows:  476 (Q0), 1639 (Q1), 4354 (Q2), 4370 (Q3), 4397 (Q4) 
and 4633 (Q5);
(7) the resulting light curves were fitted according to their Fourier descriptions and
secondary trending (usually at the 1-2 sigma level) was removed.

This method of normalization works well for unmodulated stars but might 
explain the observed apparent absence of amplitude modulation for V349~Lyr.  Owing to its relative 
faintness (V349~Lyr is the faintest star in our sample), and, if it is a Blazhko star as suggested
by B10, its long 
Blazhko period and low Blazhko amplitude, we may have to 
wait another year or longer to decide if the variations are intrinsic or due to instrumental effects.

\subsubsection{Short-cadence data}

In addition to the LC observations, FN~Lyr and AW~Dra were observed at 
short cadence (SC, 1 minute) in Q0 and Q5.  
The time baseline for the Q0 observations was 9.7~d (0.528~d to 10.254~d), while the  
the Q5 observations were made over 94.7~d (322.98~d to 417.67~d).
When combined the total number of data points amounted to $\sim$150,000 observations per star 
(14280 from Q0, and 136140 from Q5).
Processing of the data required removal of a small number of outliers ($\sigma>5$) and three spurious dips 
occuring in Q0 at times centered on 3.298~d, 6.261~d and 8.342~d.  The SC data were
detrended in the same way as the LC observations.

\section{PERIODS AND LIGHT CURVES}

\subsection{Previously published periods, risetimes, etc.}

All but one of the 19 stars in Table~1 (the exception being KIC~7021124) has been studied previously, either from
the ground or using the early {\it Kepler} photometry. In {\bf Table~2} some of this information is summarized. 

The GEOS RR~Lyr database (see LeBorgne {\it et al.} 2007) at http://rr-lyr.ast.obs-mip.fr/dbrr 
gives pulsation periods for eight of the 19 stars --
these are summarized in column~2 of Table~2. 
Much of the historical photometry was derived from photographic plates taken through various filters
and on different emulsions, often but not always approximating Johnson $V$ and $B$ magnitudes.  
The database also includes HJD times of maximum light, risetimes (RTs), and visual and CCD observations 
acquired over many decades.

Two historical studies of particular interest are the 
impressive 1953 and 1956 photographic investigations by W.J.~Miller, S.J.  The former study
includes an analysis of 
V715~Cyg (then known as Vatican Variable 14, or VV~14), while the latter includes
analyses of V782~Cyg (VV~22) and V784~Cyg (VV~24).  Both studies were based on 
photographic magnitudes ($m_{\rm pg}$) derived from Harvard and Vatican plates 
taken between 1927 (JD~2425148) and 1952 (JD~2434323).  The number of
individual $m_{\rm pg}$ observations were as follows: 840 for V715~Cyg,  735 for
V782~Cyg, and 741 for V784~Cyg.  Observed times of maximum light are given
for V782~Cyg (46 epochs) and V784~Cyg (42 epochs) but not for V715~Cyg.
For all three stars the GEOS catalog adopts the pulsation periods and times of maximum light 
from Miller's investigations.  

For the eight stars with periods given in the GEOS database agreement with the {\it Kepler} 
periods is very good, the average difference amounting to only $0.4\times10^{-6}$d.  The biggest period 
discrepancies are for V715~Cyg, V784~Cyg and AW~Dra.  The RT differences are discussed in $\S3.3$.
Possible slowly-changing periods that may not be detectable in the Q0-Q5 {\it Kepler} data 
were assessed from O$-$C diagrams and are discussed in $\S3.4$.

Nine of the {\it Kepler} non-Blazhko stars also were observed during the course of the 
ASAS-North survey (see Pigulski {\it et al.} 2009,  Szczygiel {\it et al.} 2009, and the ASAS
website at http://www.astrouw.edu.pl/asas).  
For the nine stars $\sim$80 $V$ and a similar number of $I$ CCD photometric measurements were 
made over $\sim$500 days in 2006 and 2007. 
A reanalysis of the on-line ASAS data was
made and risetimes and total amplitudes, Fourier-fitted $\langle V \rangle$ magnitudes, and $\langle V-I \rangle$ colours 
are given in the last four columns of Table~2. 
If the ASAS-South survey is any indication the ASAS-North RR~Lyr light curves will be ``close to the
standard $V$ system'' (Kovacs 2005).
The ASAS $V$ data are discussed further 
in $\S4.5$ below, and the cyclic behaviour of the stars in the HR-diagam is discussed in Appendix~B. 

Most recently, pulsation periods and preliminary Fourier analysis for 16 of the 19 stars  
were reported by B10.  They analyzed the long cadence {\it Kp} photometry from 
Q0 (9.7d interval), Q1 (33.5d) and Q2 (89d) and established the non-Blazhko behaviour of most of the stars. 
The periods (typical uncertainties of $\pm 3 \times 10^{-5}$~d) and Fourier $A_1$ amplitudes that they
derived are summarized in columns 3 and 4 of Table~2.

\begin{table*}
\begin{minipage}{150mm}
\caption[]{Other periods, amplitudes, risetimes and ASAS $\langle V \rangle$-magnitudes and $\langle V-I \rangle$-colours.}
\label{Table2}
\begin{flushleft}
\begin{tabular}{llcccccc}
\hline
\noalign{\smallskip}
\multicolumn{1}{c}{Star}  &\multicolumn{1}{c}{$P$\thinspace(GEOS)}  &$P\thinspace$(B10) & $A_1$({\it{Kp}}) & RT($V$) & $A_{\rm tot}$($V$) &
$\langle V \rangle $ & $\langle V-I \rangle $  \\
     &\multicolumn{1}{c}{[day]} &  [day]  &  [mag]  &[phase]  &[mag]   &[mag]   &[mag]  \\
\multicolumn{1}{c}{(1)}  &\multicolumn{1}{c}{(2)} &(3) &(4) &(5) &(6)  &(7) &(8) \\            
\noalign{\smallskip}
\hline
\noalign{\smallskip}
NR~Lyr          &0.6820287             & 0.68204 &0.266& 0.15 & 0.68 & 12.444(08) & 0.708(13) \\ 
V715~Cyg (VV14) &0.47067298(43)        & 0.47071 &0.340&(0.09)&  -   &   -        &  -        \\
V782~Cyg (VV22) &0.52363383(26)        & 0.52364 &0.195&(0.18)&  -   &   -        &  -        \\ 
V784~Cyg (VV24) &0.5341026(11)         & 0.53410 &0.229&(0.19)&  -   &   -        &  -        \\ 
KIC~6100702     &\multicolumn{1}{c}{-} & 0.48815 &0.206& 0.21 & 0.70 & 13.641(09) & 0.659(22) \\
NQ~Lyr          &0.5877888             & 0.58779 &0.280& 0.14 & 0.75 & 13.361(13) & 0.576(23) \\
FN~Lyr          &0.52739716            &  -      & -   & 0.13 & 1.27 & 12.779(09) & 0.597(14) \\
KIC~7030715     &\multicolumn{1}{c}{-} & 0.68362 &0.231& 0.15 & 0.81 & 13.241(14) & 0.667(25) \\
V349~Lyr        &\multicolumn{1}{c}{-} & 0.50708 &0.357&  -   &  -   &   -        &  -        \\
V368~Lyr        &\multicolumn{1}{c}{-} & 0.45649 &0.407&  -   &  -   &   -        &  -        \\ 
V1510~Cyg       &\multicolumn{1}{c}{-} & 0.58115 &0.341&  -   &  -   &   -        &  -        \\ 
V346~Lyr        &\multicolumn{1}{c}{-} & 0.57683 &0.322&  -   &  -   &   -        &  -        \\
V350~Lyr        &\multicolumn{1}{c}{-} & 0.59424 &0.339&  -   &  -   &   -        &  -        \\  
V894~Cyg        &\multicolumn{1}{c}{-} & 0.57139 &0.384& 0.15 & 1.07 & 12.923(11) & 0.585(17) \\
V2470~Cyg       &\multicolumn{1}{c}{-} & 0.54859 &0.219& 0.19 & 0.70 & 13.534(09) & 0.605(16) \\ 
V1107~Cyg       &\multicolumn{1}{c}{-} & 0.56579 &0.270&  -   &  -   &   -        &  -        \\
V838~Cyg        &0.4802795             & 0.48029 &0.390& 0.13 & 1.32 & 14.246(13) & 0.545(23) \\
AW~Dra          &0.6871941             &  -      & -   & 0.19 & 1.06 & 12.851(06) & 0.562(11) \\ 
\hline
\end{tabular}
\medskip
\end{flushleft}
\end{minipage}
\end{table*}


\subsection{Period estimation}

Pulsation periods for the non-Blazhko stars were derived mainly from analyses of 
the {\it Kp} data alone.  Computations were performed with  the 
package {\tt PERIOD04} (Lenz \& Breger 2005), which 
carries out multifrequency analyses with Fourier and least-squares algorithms, and
with a version of the CLEAN program (see Nemec, Walker \& Jeon 2009).
Details of the derived periods are given in {\bf Table~3}, where the 
best estimates of the periods and times of maximum light are in columns~3 and 4,
and column~5 describes the specific data that were analyzed.
The uncertainties in the periods (given in parentheses) are 
average values obtained using
the three methods available in {\tt PERIOD04}, {\it i.e.}, 
a theoretically expected uncertainty based on analytical considerations, a least-squares
method, and a 
Monte Carlo routine similar to the Fisher method of randomization described by 
Nemec \& Nemec (1985).
Typically the periods derived using the Q0-Q5 LC data alone 
are accurate to $1-2 \times 10^{-7}$~d.  The uncertainty is expected to decrease further as more
{\it Kepler} observations become available.    The most uncertain period
is that for KIC~7021124 which so far has been observed only in Q1.
Also given in Section~(a) of Table~3 are the mean {\it Kp} magnitudes obtained from the KIC.  In column~5 the abbreviations `LC' and `SC' stand for `long cadence' 
(30-min) and `short cadence' (1-min) {\it Kepler} data.  Columns 6-11 contain Fourier-based
results and are described below.  

In general, the new periods compare favourably with the GEOS and B10 periods (see Table~2), 
with the newer values being more precise and more accurate.  The largest differences are for 
V715~Cyg and AW~Dra, both of which have significantly longer periods than given in the GEOS catalog.  
In $\S3.4$ the periods derived directly from the {\it Kp} data are compared with the values from O$-$C diagrams.

\subsection{Light curves, amplitudes and risetimes}

Phased light curves for the {\it Kepler} non-Blazhko ab-type RR~Lyr stars are plotted in
the left panel of {\bf Figure~2},
assuming the pulsation periods and times of maximum light given in Table~3.   
The light curves have been ordered from smallest to largest amplitude (measured from minimum to 
maximum light), with each star offset from the next by 0.5 mag.  
This arrangement shows that there is a clear trend in light curve shape and risetime (RT), with the lowest amplitude stars
having the longest RTs and the most symmetric light curves, and the largest amplitude
stars having the shortest RTs and most asymmetric light curves.

The largest amplitude star is V368~Lyr, with $A_{\rm tot} = 1.133$ mag.  
This star also has the most asymmetric light curve and the shortest period, $P=0.456$~d. 
At the top of the stack of light curves the star with the smallest amplitude is V782~Cyg with $A_{\rm tot} = 0.523$ mag;
however, its period, $P=0.524$~d, is {\it not} the longest.  That distinction belongs to three stars with 
approximately equal periods: AW~Dra (0.687~d), NR~Lyr (0.682~d) and KIC~7030715 (0.684~d).  These
three stars all have intermediate amplitudes, and are shown below to have 
considerably lower metallicities than V782~Cyg (see $\S6.1$ and Figs.~7-9).   
 
The right panel of Figure~2 also shows a comparison of the total amplitudes, $A_{\rm tot}$ (Table~1, column~9), and the 
Fourier $A_{1}$ coefficients (see next $\S$), both derived from the {\it Kp}-photometry.
Although $A_{1}$ is approximately proportional to $A_{\rm tot}$ it is clear that the relationship is non-linear.
When a cubic equation was fitted to  the graph the resulting
standard error of the fit amounted to only 3 mmag.  It  is doubtful that this (slight) curvature would have been 
detected without the high precision of the {\it Kepler} data.
In general the relationship is roughly approximated by non-linear hydrodynamic models (see $\S$7,
in particular the top-left panel of Fig.~14).

\begin{figure*}
\begin{center}$
\begin{array}{cc}
\includegraphics[width=8cm]{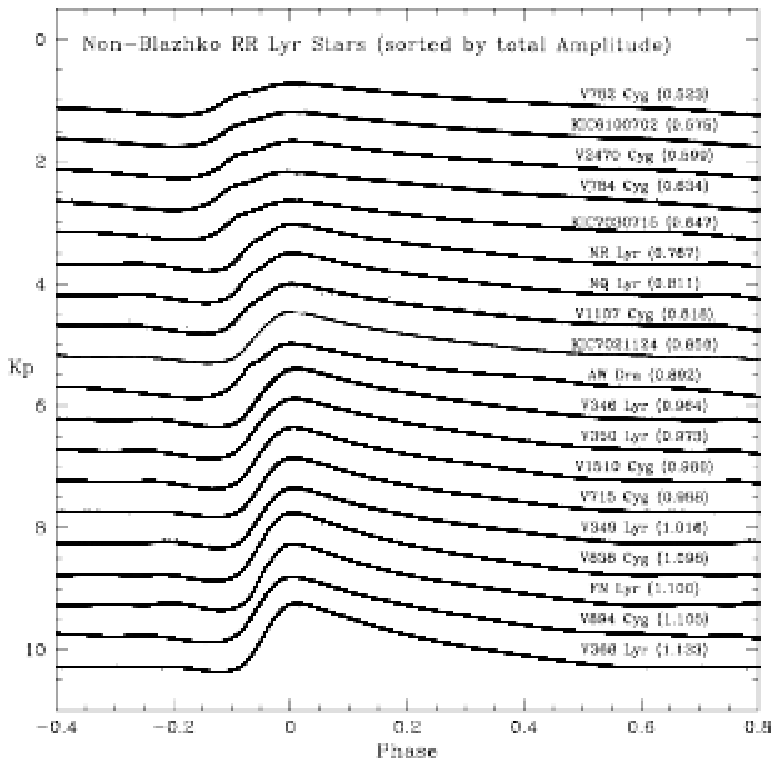}&
\includegraphics[width=8cm]{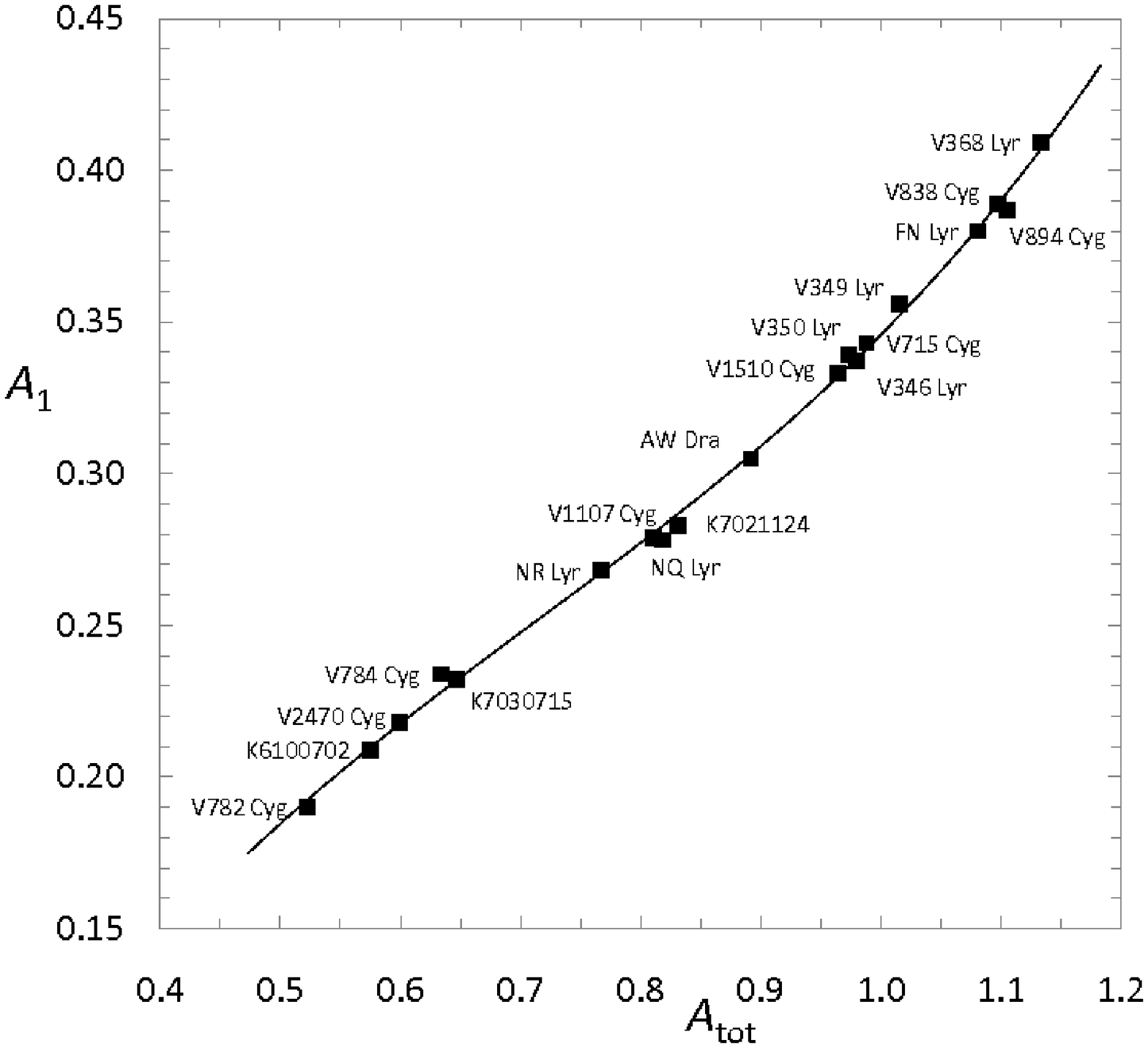} 
\end{array}$
\end{center}
\caption{(Left) Phased light curves for the 19 non-Blazhko ab-type RR~Lyr stars observed with the {\it Kepler}
telescope.  The quantities in parentheses following the star names are the total {\it Kp} amplitudes (Table~1, column~8), ranging
from $A_{\rm tot}=0.52$ to 1.13 mag.   Almost all of the light curves contain over 18000 points with a spread of 
less than one milli-magnitude. 
(Right) Graph comparing $A_{\rm tot}$ values and the 
Fourier $A_{1}$ coefficients (in column~8 of Table~3 below), all derived from the {\it Kp}-photometry.  The fit to the
points is described by the cubic equation $A_1 = 0.443 \thinspace A_{\rm tot}^3 - 0.950 \thinspace A_{\rm tot}^2 + 
0.973 \thinspace A_{\rm tot} - 0.120$,  with fit standard error $\sigma = 0.003$ mag. }
\label{nonBL_LCs_sorted_Atot}
\end{figure*}

The RTs, magnitude ranges, and total amplitudes given in Table~1 (columns 7-9) 
were calculated numerically from the fitted {\it Kp} light curves.  The magnitudes at maximum and minimum 
light are the values at which the slopes of the light curves were found to be zero, from which magnitude ranges 
and precise total amplitudes were calculated.  The phases at the light minima and maxima were 
used to calculate the RTs.   

It is informative to compare the {\it Kp}-based RTs with those given in the GEOS database and
with RTs based on the ASAS $V$-photometry.   Since RTs are not given at the ASAS on-line
website the values given in column~5 of Table $2$ were derived in the same way that the {\it Kp} values were 
computed.  For the nine stars with {\it Kp} and ASAS RTs the agreement is excellent, the mean difference  
being 0.002 (ASAS minus {\it Kp}, standard deviation 0.014).  For Miller's three Vatican variables (values 
are given in parentheses in column~5 of Table~2) there is reasonable agreement, with the mean difference 
being $-0.024$ (Miller minus {\it Kp}, standard deviation 0.021).  
The largest differences occur for NR~Lyr and V838~Cyg, where the GEOS database gives RT=0.27 and 0.40,
respectively, 
compared with the averages of the {\it Kp} and ASAS estimates of 0.147 and 0.132. 
An independent estimate of 0.141 for NR~Lyr comes from our analysis of the unpublished $V$ CCD photometry of 
Benk\H{o} and Nuspl, which agrees well with the {\it Kp} value.

\subsection{O$-$C diagrams and Period Change Rates}

O$-$C diagrams can be used to improve the precision of estimated pulsation periods, 
to detect slowly changing periods, and to derive 
accurate period change rates.  The longer the time baseline the greater the leverage on the 
period and the greater the probability of detecting curvature signifying a changing period.
For those {\it Kepler} stars with over one year of nearly continuous high-precision photometry (Q0-Q5) we have seen
that it is 
possible to derive periods accurate to $\sim$1$\times 10^{-7}$~d using only the {\it Kepler} photometry. 
When the {\it Kepler} data are combined with the long-baseline historical photometry (in some cases as many 
as 105 years -- see for example Le Borgne {\it et al.} 2007) even more accurate periods 
(and possible d$P$/d$t$ values) can be calculated.

\begin{figure*}
\begin{center}$
\begin{array}{ccc}
\includegraphics[width=5.5cm]{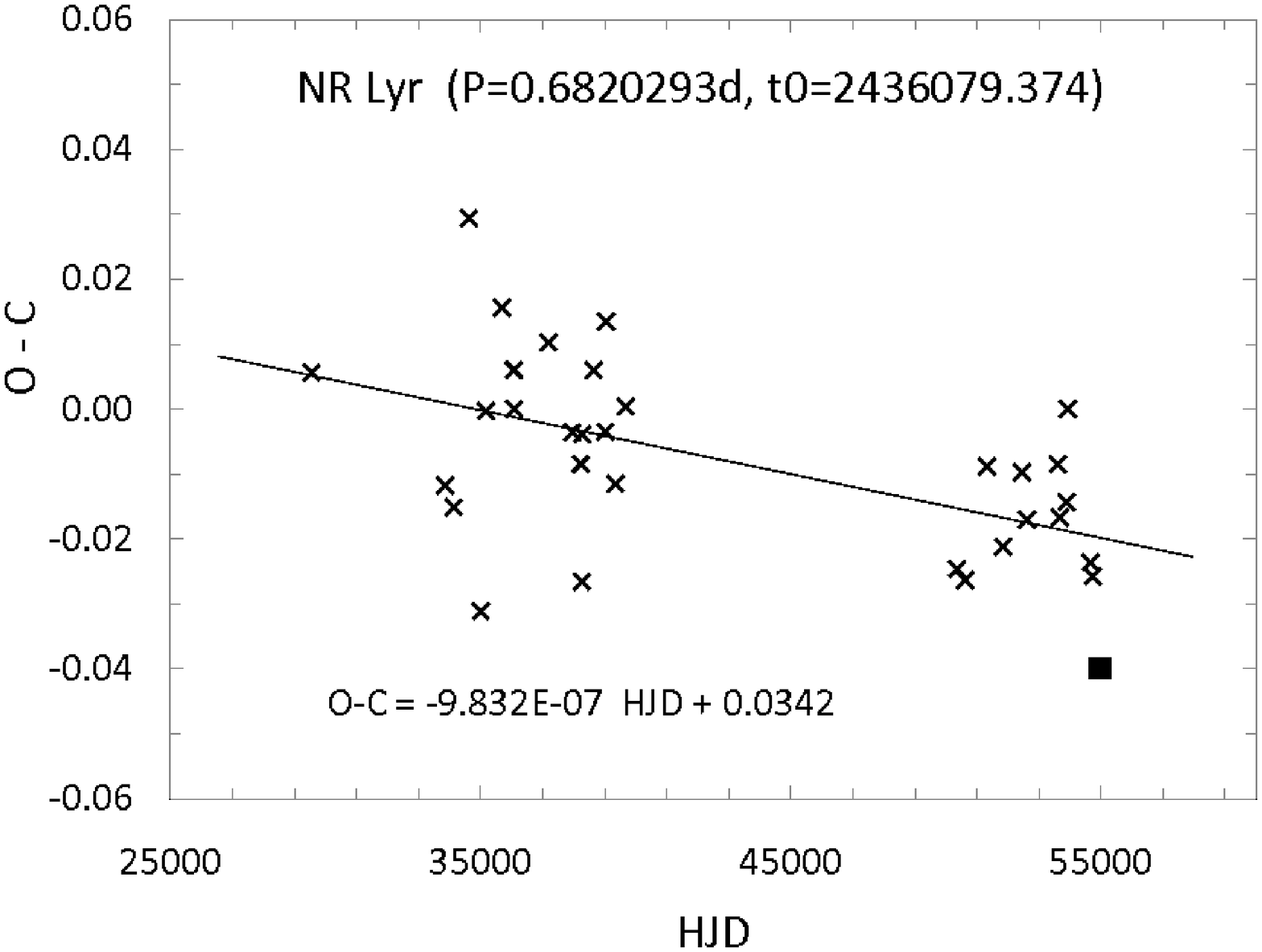} &
\includegraphics[width=5.5cm]{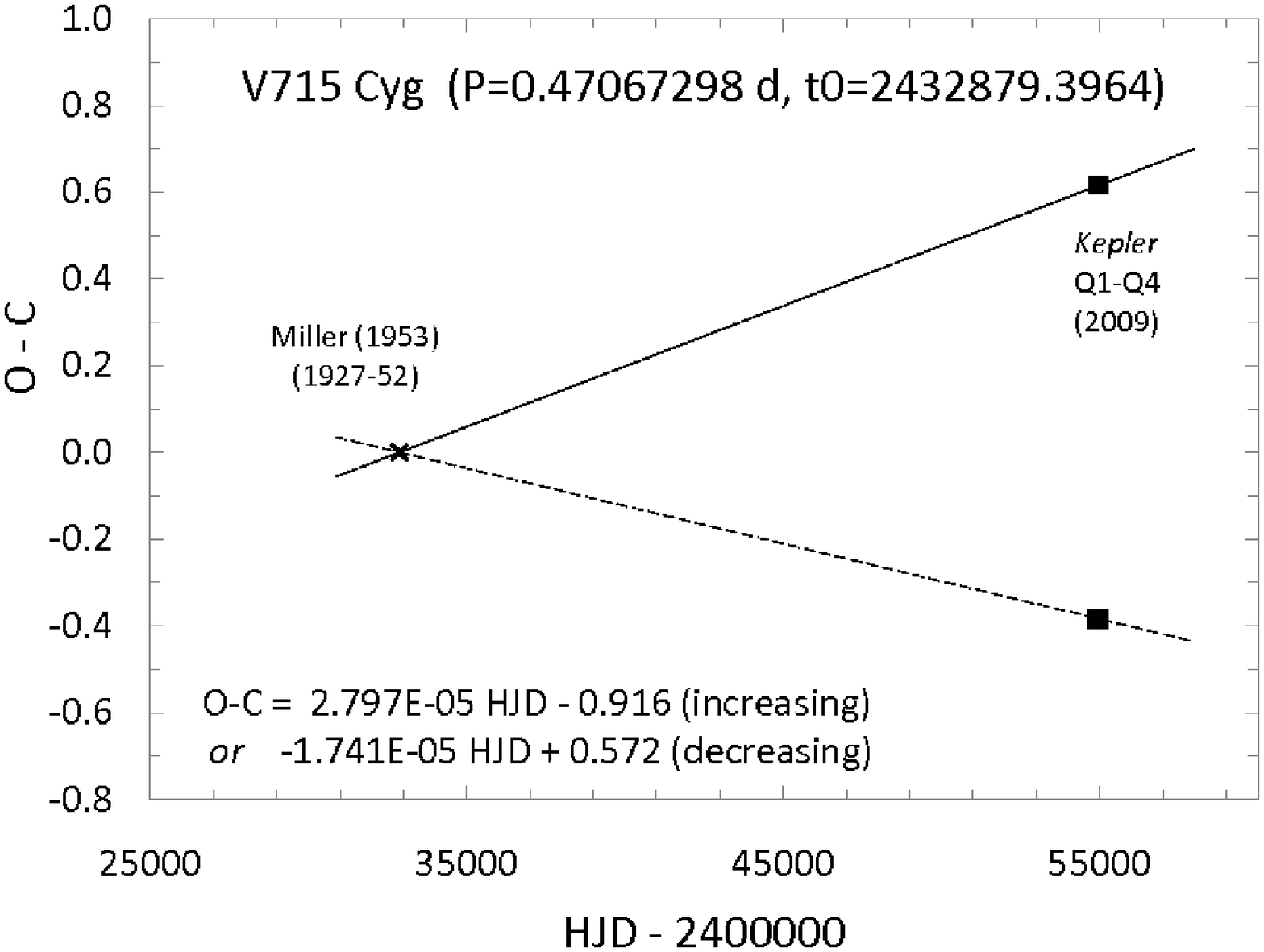} &
\includegraphics[width=5.5cm]{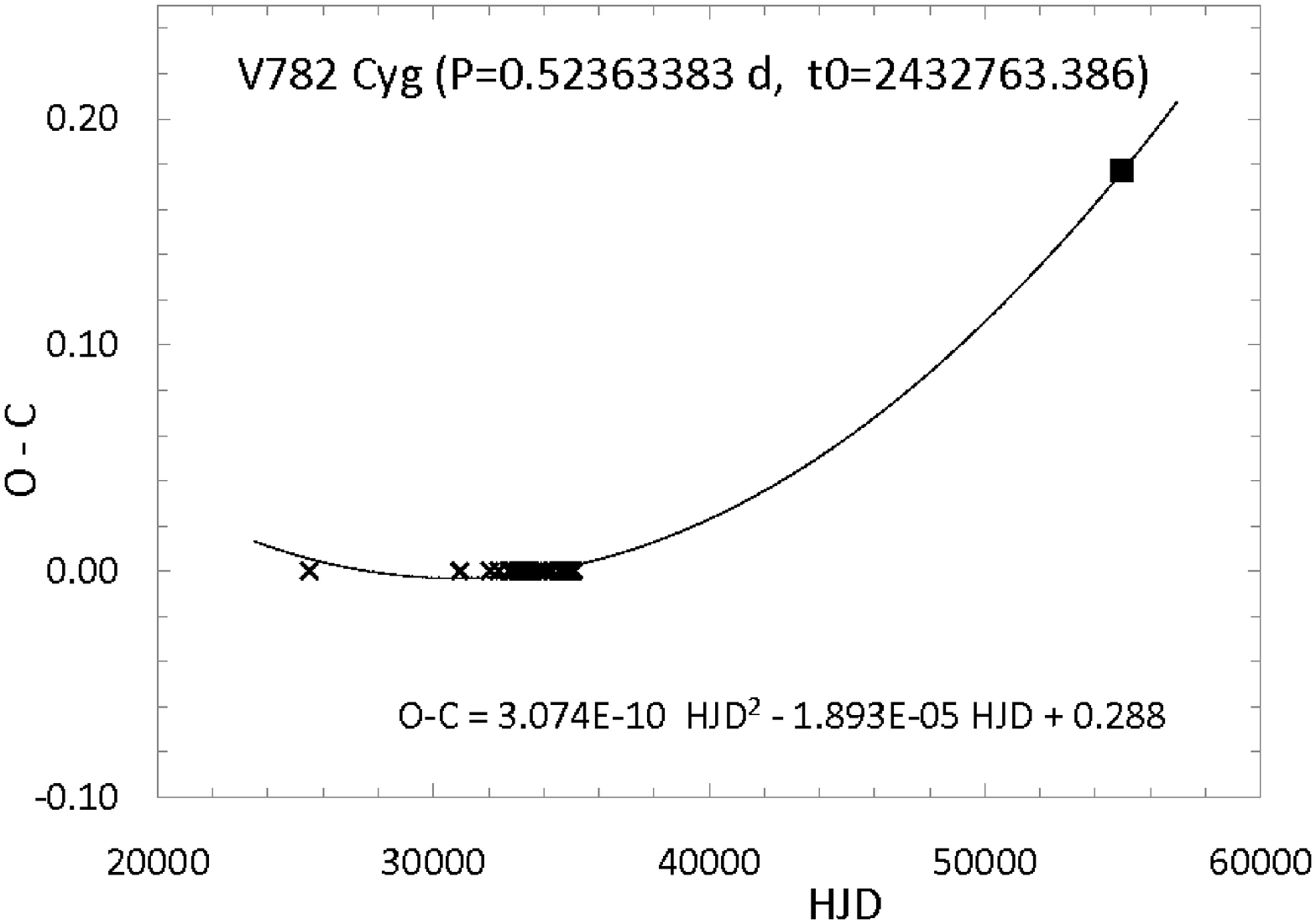} \\
\includegraphics[width=5.5cm]{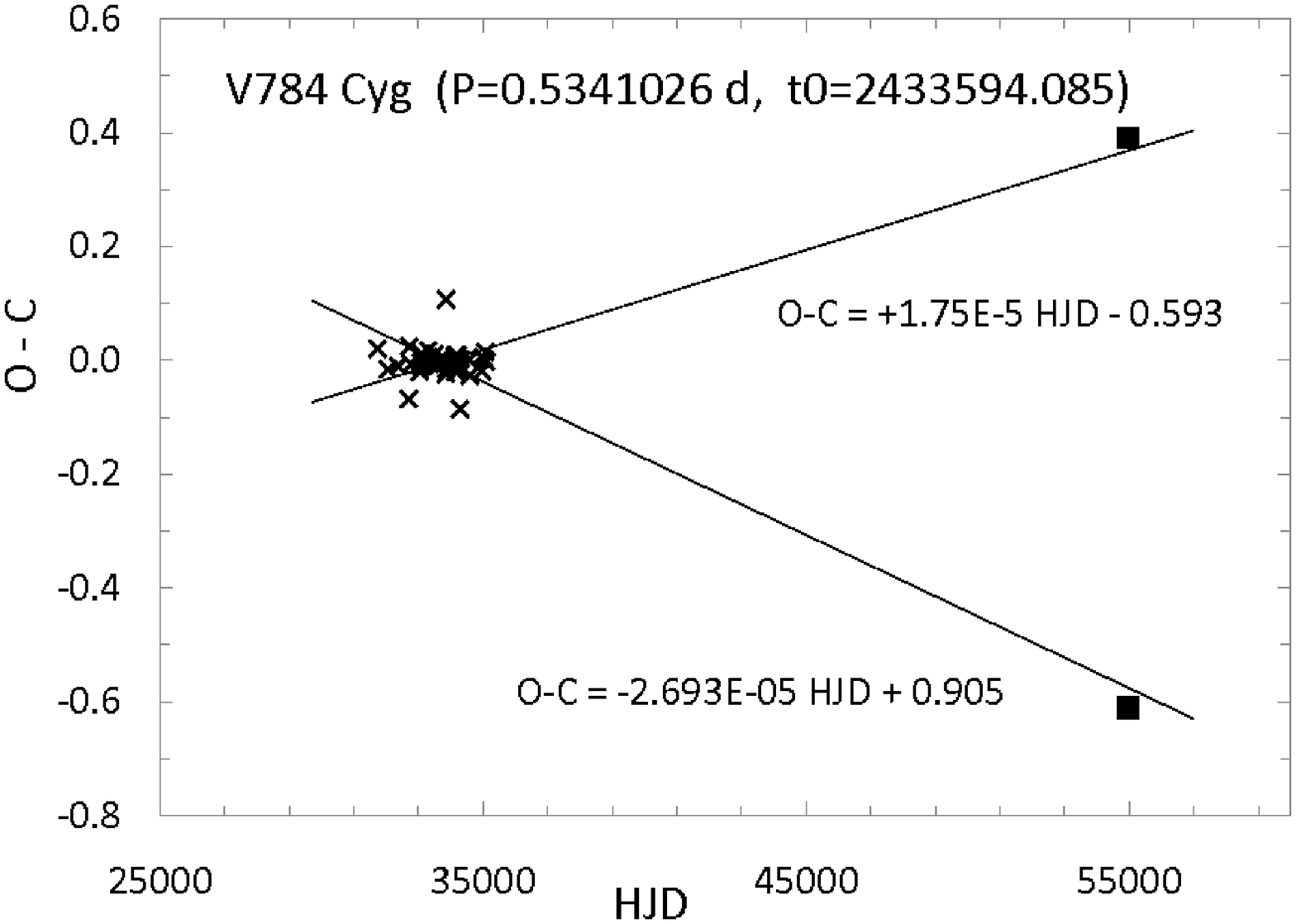}&
\includegraphics[width=5.5cm]{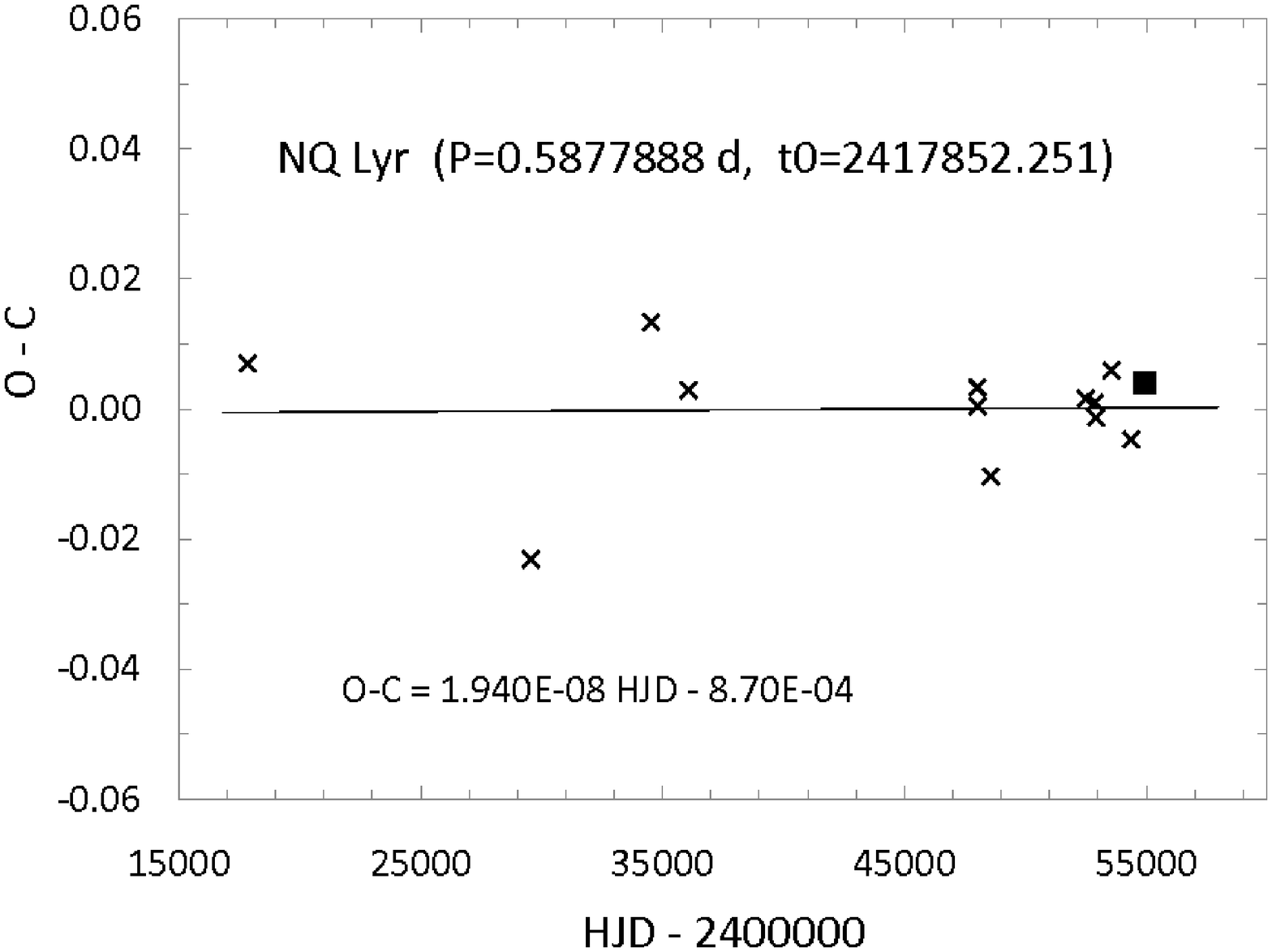} &
\includegraphics[width=5.5cm]{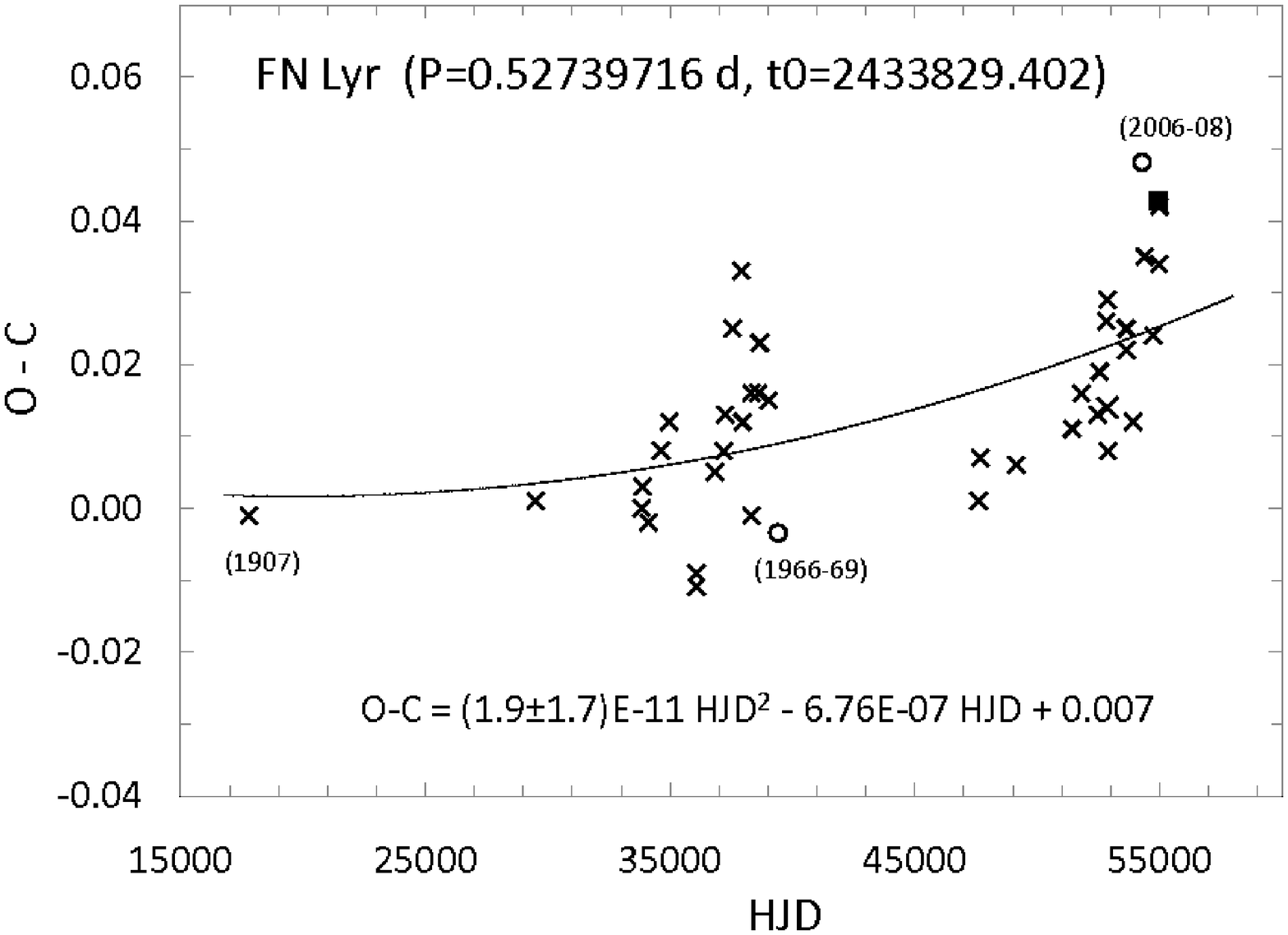} \\
\includegraphics[width=5.5cm]{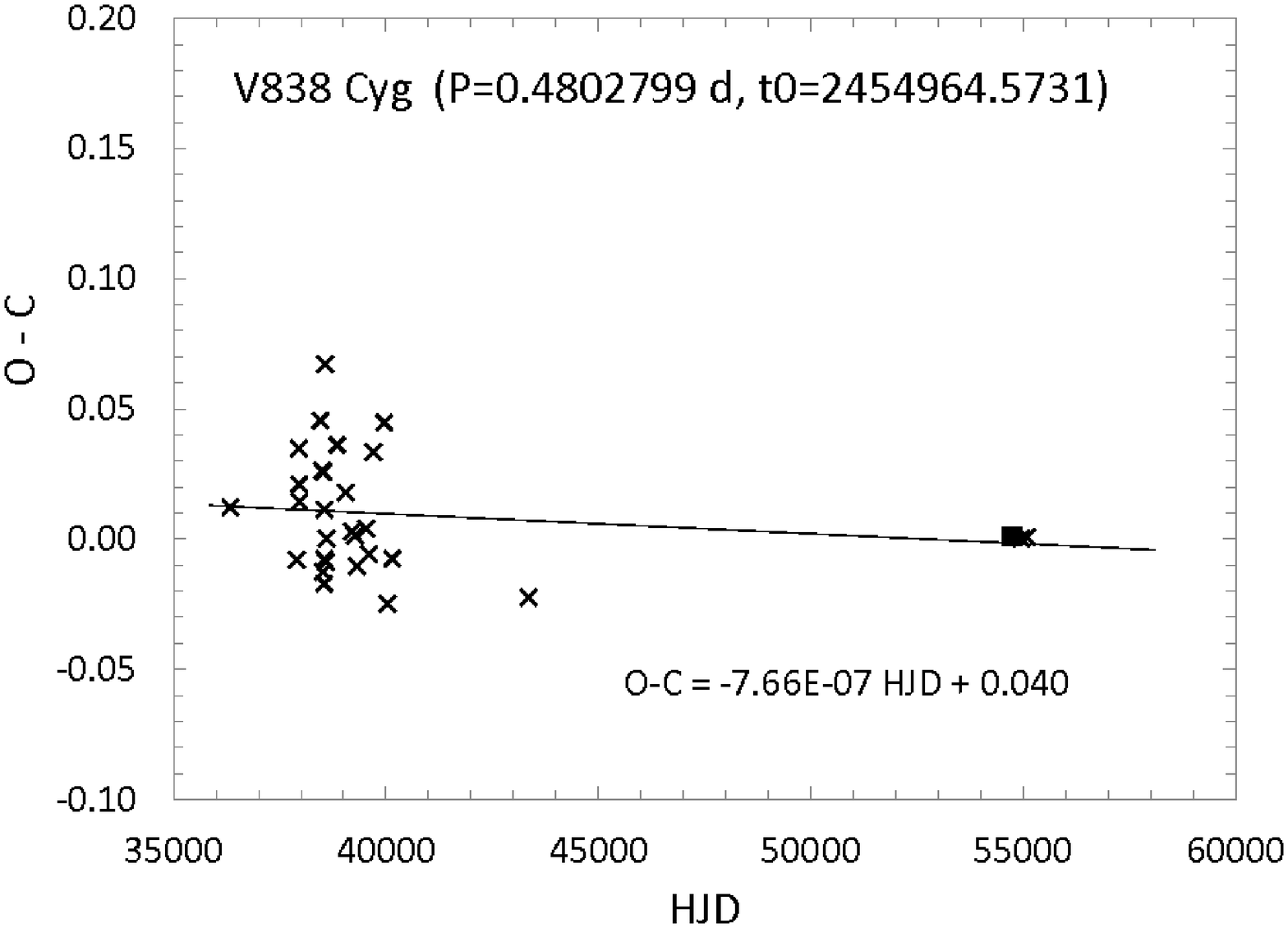} &
\includegraphics[width=5.5cm]{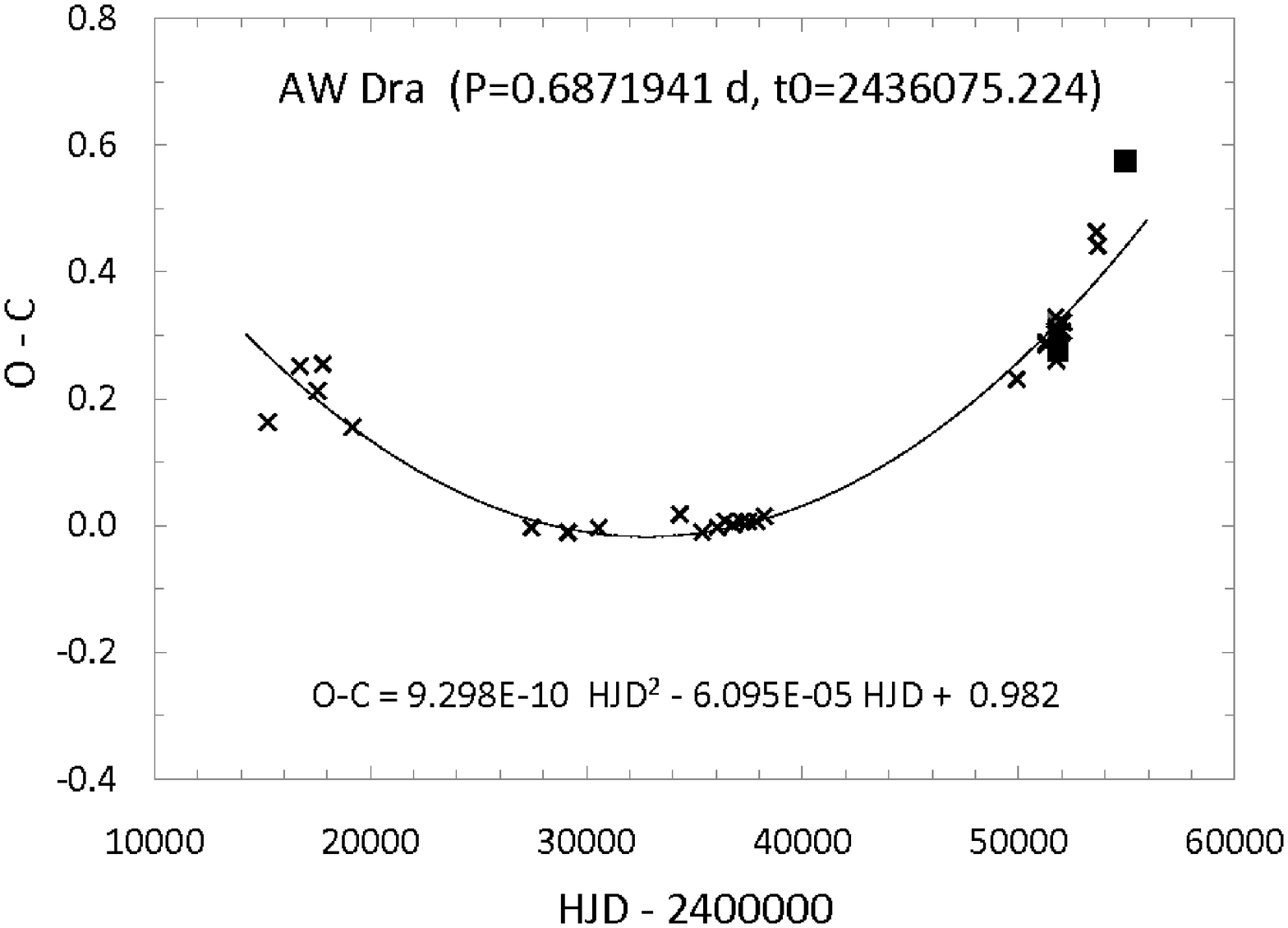} &
\includegraphics[width=5.5cm]{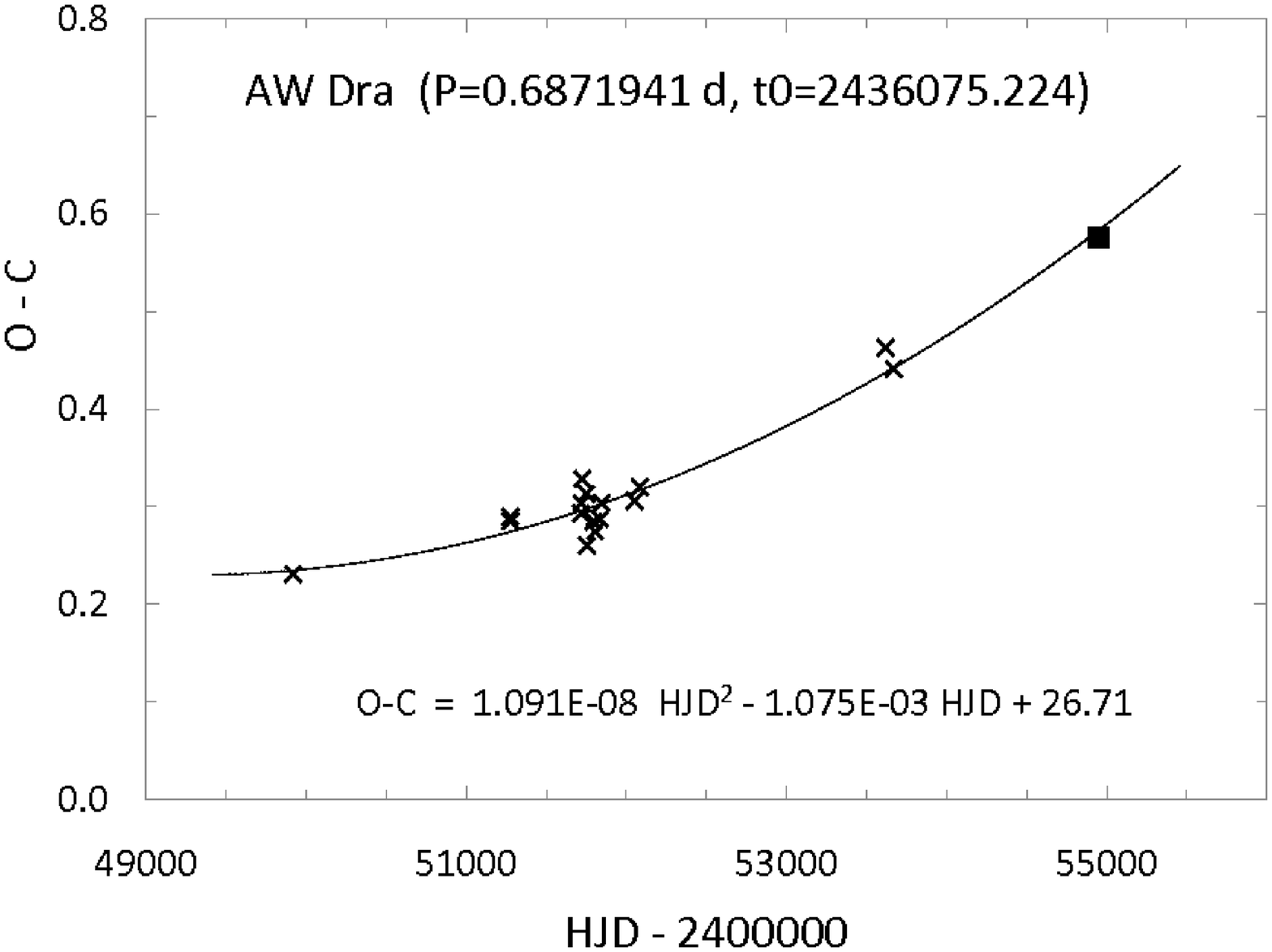}
\end{array}$
\end{center}
\caption{O$-$C diagrams for the eight {\it Kepler} non-Blazhko stars with previous photometry
in the GEOS atlas (AW~Dra is shown twice, with different time scales).  
In each panel the O$-$C value derived from the {\it Kepler} data is
plotted as a solid black square at HJD $\sim$55000, and the crosses 
are the O$-$C values at the various previous epochs.    
Error bars for the {\it Kepler} points are comparable to the size of the square and
have not been plotted.
The fitted lines (or curves) plotted in each panel, the equations of which are given 
in each graph, are discussed in the text. For V715~Cyg and V784~Cyg two, of many, 
possible solutions are plotted. }
\label{O$-$C_diagrams}
\end{figure*}

{\bf Figure 3} shows O$-$C diagrams for the eight {\it Kepler} non-Blazhko stars that have 
ephemeris information in the GEOS database. The ordinate represents the ``observed minus
calculated'' phase at which maximum light occurs, assuming
the GEOS period and time of maximum light given in parentheses at the top of each panel;  the 
abscissa is the Heliocentric Julian Date.\footnote[2]{Since the 
offset between Barycentric Julian Dates (BJDs) used by {\it Kepler} are very small compared to the 
HJDs usually used by ground-based observers, the two types of dates have been used interchangeably
in this paper.}   
At each epoch the O$-$C value was calculated by taking the observed time
of maximum light {\it minus} the assumed GEOS time of maximum light ($t_0$), and dividing the resultant 
by the assumed (constant) GEOS period ($P$).  This gave the number of pulsation cycles since $t_0$, and 
the remainder after subtracting an integer number of cycles (E) is the observed fractional phase shift 
from the expected value.
For five of the stars linear fits to the observations are given, a positive (negative) slope indicating that
the assumed period is too short (long).  Quadratic fits are shown for V782~Cyg, FN~Lyr and AW~Dra (shown twice,
the second graph representing the last $\sim$13 years), 
where the upwards curvature indicates an increasing period, the period change rate, d$P$/d$t$, being given by
$2 P c^2$ where $P$ is the assumed GEOS period, and $c$ is the curvature coefficient in the quadratic equation.  
None of the stars represented in Fig.~3 was studied by Le~Borgne {\it et al.} (2007).

For the three Vatican Variables (V715~Cyg, V782~Cyg, V784~Cyg)
there is an $\sim$60 year gap between the 2009-10 {\it Kepler}
data and the earlier observations.  Over such a long time interval one cannot hope to
keep track of cycle counts and there is little hope of using the O$-$C diagrams to either improve the period or 
calculate d$P$/d$t$.  To illustrate the cycle-count problem the panels for 
V715~Cyg and V784~Cyg show two (of many) possible linear solutions, where, in all the cases there is a 
large O$-$C shift between the {\it Kepler} and earlier 
epochs.\footnote[3]{Miller (1953) did not bin his 
V715~Cyg (VV~14) observations into separate epochs, and the GEOS database simply adopted for its
zero-phase epoch the time of maximum light given in the Miller ephemeris. Future period change
rate studies of this star might want to bin Miller's photometry into separate epochs rather than just let one epoch (1948) 
represent the extended 1927-1952 data. }  However, comparison of the assumed periods with the periods derived from the 
{\it Kepler} data do give some indication of either a period change or incorrect period.
For V715~Cyg, the {\it Kepler} period, 0.4707045(1)~d, is significantly longer than the GEOS period, 0.47067298(43)~d,
suggesting a period increase.     
And for V784~Cyg the {\it Kepler} period, 0.5340941(1)~d, is significantly shorter than that derived by 
Miller (1956), 0.5341026(11)~d.

Miller's (1956) data for V782~Cyg provides 46 times
of maximum light  at epochs between 1928 and 1954, and with these Miller derived   
$P=0.52363383$~d (adopted in the GEOS database).   
Our analysis of the $\sim$18419 Q1-Q5 {\it Kp} photometry gave $P=0.5236377$(1)~d, which is 3.9$\times 10^{-6}$d longer
than the Miller period (see column 3 of Table~2).  Using the time of maximum light given in column~5
of Table~1 the O$-$C phase shift of $\sim$0.20 (nearest) is undoubtedly significant.
The quadratic solution (shown in Fig.~3) gives d$P$/d$t$ = 0.06 d/Myr, which predicts, for 2010, the period
0.5236379~d, a value that is only 0.2$\times 10^{-6}$d longer than the measured {\it Kepler} period. 
While this agreement strongly supports the increasing period it is also possible that the period
changed abruptly in the 60-year gap.   

For V838~Cyg a total of 31 times of maximum light are available in the GEOS database: 26 epochs betwen 1958 and 1968, a single
epoch in 1977, and four recent epochs from 2008-09.  Although the 10- and 30-year gaps may be problematic, 
the {\it Kepler} ephemeris gives an O-C value near zero for every epoch, suggesting either a constant (or very slowly changing) 
period since 1958.  A linear fit to the O$-$C diagram (Fig.~3, bottom left panel) predicts a (constant) true period
$P=0.4802797$(2)~d, which matches, to within the uncertainties, the {\it Kepler}-based period $P=0.4802799(1)$~d
and the GEOS period $P=0.4802795$~d.  A parabolic fit suggests d$P$/dt = +0.05$\pm$0.04 d/Myr, with the assumed period
0.4802799~d occurring at BJD~48438.  The fit standard error in both cases is only 0.021 (phase units). 

For NR~Lyr (Fig.~3, top left panel) the 31 historical points and the new {\it Kepler} data are consistent with a 
shorter period than assumed in the GEOS database.  The negative slope suggests 
a period shorter by 0.46$\times10^{-6}$d than the assumed $P$(GEOS), {\it i.e.}, $P({\rm true}) = 0.6820288$~d;
this period is significantly longer than the estimate based on the 18358 {\it Kepler} Q1-Q5 data points,
0.6820264\thinspace(2)~d.  

The 13 historical points and the {\it Kepler} data for NQ~Lyr produce consistent period estimates, with a difference of only
0.1$\times10^{-6}$d between the two periods.   
Thus the assumed (GEOS) period of 0.5877888~d is to be compared with 0.5877887(1)~d  derived using
the 18759 {\it Kepler} Q0-Q5 data points.

The O$-$C diagram for FN~Lyr (middle right panel of Fig.~3) 
assumes $P=0.52739716$~d and $t_0 = 33829.402$ (GEOS values), and includes 34 O$-$C values from
as early as 1907 ($t=17793$) to as recent as 2009 ($t=54995$).  All but one of the 
phase shifts is near zero (a visual observation from 1937 was considered an outlier and excluded from our analysis). 
The diagram includes a point from the 1966 photoelectric $V$-photometry of 
Bookmeyer {\it et al.} (1977) at $t=39416.6457$, and a point from unpublished 2006-08 CCD $V$-photometry by Layden
at $t=54288.7454$  (see $\S$4.4 for further discussion of these data).  
These two points, which differ in time by $\sim$40 years, 
have been plotted as open circles and differ in phase by $\sim$0.055.  The O$-$C value derived for
FN~Lyr from the {\it Kepler} data (epoch 2009) is almost identical to the Layden O$-$C value (see Table~2).   
Considering that the total baseline is longer than 100 years the O$-$C phase shifts are 
very near zero. Nevertheless there appears to be a slight upward trend, and possibly  a small amount of curvature.  
A parabolic solution gives d$P$/d$t$=0.004$\pm$0.004~d/Myr, with the GEOS 
period occuring at t=18100, a residual standard error of the fit of 0.011, and a predicted 
period 0.5273975~d for the current epoch (2010-11). 
Since the curvature term does not differ significantly from zero ($p$-value$=0.27$) 
the question of a changing period remains unanswered. 
A linear solution is also consistent with the data, in which case 
a revised constant period of 0.52739741(5)~d is suggested by the data.
Again, the residual standard error of the fit was only 0.011.
When the 136140 SC observations made in Q5 were combined with the 14280 SC observations 
made in Q0, a period 0.52739845(1)~d with residual standard error 0.60 mmag, 
was derived, borderline consistent with both the 
constant period and small d$P$/d$t$ hypotheses.

Finally, the GEOS data for AW~Dra  have many epochs with O$-$C phases that appear to be
outliers.  These include a single point at $t\sim$20000 with O$-$C$\sim-0.20$, and a clump
of $\sim$20 points at $t\sim$52000 with O$-$C$\sim-0.30$.  If these values are correct then they present a 
challenge for modelling period variations.  We have chosen to discard them as outliers leaving 
the O$-$C diagram shown in the bottom middle panel of Fig.3.  The 
parabolic fit suggests an increasing period amounting to 0.32 d/Myr, and predicts 
a period 0.687214~d for the {\it Kepler} Q5 epoch.
However, close inspection of the diagram shows that the last three points, one of which
is that derived from the {\it Kepler} data, deviate substantially from the fitted curve.  
If the analysis is restricted to epochs 
since HJD~49000 (the first point is from the Castellani {\it et al.} 1998 study of AW~Dra) {\it and} if 
the suspected outliers identified above are excluded, then
the O$-$C diagram is as shown in the bottom right panel of Fig.~3.  In this case, the period increases
smoothly at the rate d$P$/d$t$ = 3.76 d/Myr, with the assumed period of 0.6871941~d having occurred at
HJD~49267.  For the {\it Kepler} Q1 and Q5 LC data (6117 points) the predicted period is 0.6872156(2)~d.   
And the period from the combined Q0 and Q5 SC data (149222 points) is similar, 0.6872162(1)~d, with a 
fit standard error 7.3 mmag.  Both periods agree, are significantly longer
than the GEOS period, and are consistent with AW~Dra having an increasing period.  Comparing this period with 
the predicted periods the new data favour the smaller d$P$/d$t$ value.

To summarize this section, periods based on O$-$C diagrams were calculated for NR~Lyr and NQ~Lyr, both of which 
seem to have constant periods, and d$P$/d$t$ values have been calculated for FN~Lyr and AW~Dra. 
AW~Dra is the only star for which we can at present be confident that the period is changing (assuming that the 
outliers really should be omitted from consideration) -- depending on the assumed
data set the rate is either d$P$/d$t$=3.786~d/Myr (for the more recent data), or 0.32~d/Myr (all data) with the possibility of
variations on top of this.  For FN~Lyr the period {\it either} is very nearly constant {\it or} is increasing at a very slow rate.
For the other four stars with GEOS times of maximum light the large gaps in the observation record prevent 
improvement over the GEOS or {\it Kepler} periods.  The ongoing high-precision {\it Kepler} observations should remedy 
this lack of knowledge (for the modern era), not only for the eight {\it Kepler} non-Blazhko stars in the GEOS database 
but also for all the {\it Kepler} non-Blazhko stars.

\section{FOURIER ANALYSIS}

The use of Fourier methods to characterize the shapes of RR~Lyr light curves 
began with the work of Simon and his collaborators (Simon $\&$ Lee 1981, Simon \& Teays 1982, Simon \& Clement 1993) and has 
continued through the more recent papers by  Kov\'acs, Jurcsik, Morgan and others.  
The goal of these studies has been to establish empirical correlations employing a minimum number of observed parameters
(periods, amplitude ratios, phase parameters) from which physical characteristics can be derived (see $\S6$).  
Calibration of the equations follows from independent observations (high dispersion spectroscopy, parallax measurements, {\it etc.}) 
and from theory ({\it e.g.}, hydrodynamic pulsation models, stellar evolution models).
After Simon (1988) discovered that the Fourier phase parameter $\phi_{\rm 21}^c$ decreases with decreasing metallicity (for field RRab stars with $P < 0.575$d), 
Kov\'acs \& Zsoldos (1995) proposed that correlations involving period and the Fourier phase parameter $\phi_{\rm 31}$ can be used to derive [Fe/H] values.
The first application of this method to field and globular cluster
RR~Lyr stars was the study by Jurcsik \& Kov\'acs (1995,1996), and subsequent calibrations have been made by Jurcsik (1998, hereafter J98) and 
Kov\'acs \& Walker (1999, 2001).   From the beginning Jurcsik \& Kov\'acs (1995) warned that the Fourier $\phi_{\rm 31}$ method ``is not applicable 
for the estimation of [Fe/H] in peculiar stars ({\it e.g.}, in Blazhko variables, highly evolved stars ...)'', and subsequently 
JK96 and Kovacs \& Kanbur (1998) introduced a `compatibility test' for identifying `peculiar' stars.   Indeed, when such stars were omitted from the ASAS-South database, 
agreement between Fourier-based metallicities and low-dispersion spectroscopic metallicities reached the 0.16 dex level (see Kovacs 2005).

\subsection{Coefficients and fitted light curves}

Fourier decomposition of the {\it Kepler} light curves was performed for each of the
19 non-Blazhko stars by fitting the following Fourier sine series to the observed photometry:  
\begin{equation}
m(t) = A_0 + \sum_{i=1}^F  A_i \sin  [  i \omega_0 (t - t_0)  + \phi_i ], \\
\end{equation}
where $m$($t$) is the apparent magnitude (either {\it Kp} for the {\it Kepler} data, or $m_V$($t$) for the ground based
$V$-photometry), $F$ is the number of fitted terms, $\omega_0$ is the angular pulsation frequency
of the star (=$2 \pi f_0$, where $f_0$=$P_0^{-1}$), $t$ is the observed time of the observation (BJD-54953 for the 
{\it Kepler} data, HJD for the ground-based $V$ photometry), $t_0$ is the
time of maximum light (used to phase the light curves so  that maximum light occurs at zero phase), 
and the $A_i$ and $\phi_i$ are the amplitude and phase coefficients for the individual Fourier terms.
The Fourier calculations were made with two FORTRAN programs, one kindly provided by Dr. G\'eza Kov\'acs, and
the other by Dr. Pawel Moskalik.  The assumed pulsation periods were those derived either from the {\it Kepler}
data or from the period change rate analysis ($\S3.4$), and the $t_0$ values were calculated numerically.
The derived periods and $t_0$ values are summarized in Table~1 (columns 4,5) and the details given in Table~3.

\subsection{{\it Kepler} {\it Kp}-photometry}

Initially the {\it Kepler} magnitudes were fitted with a Fourier series having 7-15 terms, the larger 
number being needed when the skewness is greatest, {\it i.e.}, the risetime is shortest.
Inspection of the residuals from the fitted light curves revealed that because most of the
stars show a sharp rise to maximum light, with additional detailed bumps and features, 
many more terms were needed.  {\bf Figure~4} illustrates 38-term Fourier fits for two non-Blazhko stars, 
AW~Dra on the left and FN~Lyr on the right.  These two stars  have 
both {\it Kp}-photometry (top diagrams) and high-precision ground-based $V$-band photometry (bottom diagrams).
For AW~Dra the {\it Kepler} data are Q5 SC observations, {\it i.e.}, brightness measurements every 
minute over 94.7 days, resulting in 135380 data points;  and for FN~Lyr the {\it Kepler} data are Q1-Q5 LC 
observations, {\it i.e.}, a measurement every 30 minutes over 417 days, resulting in 18338 points. 
In both cases the standard deviation of the fit to the {\it Kepler} data is less than one milli-magnitude (see column~6 of Table~3).   
However, the residuals still show systematic variations with the largest residuals occuring on the rise to maximum light where 
the slope of the light curve is steepest.  To achieve residuals showing normally distributed white noise 
perhaps 50-100 terms would be needed.

\begin{figure*}
\begin{center}$
\begin{array}{cc}
%
\includegraphics[width=8cm]{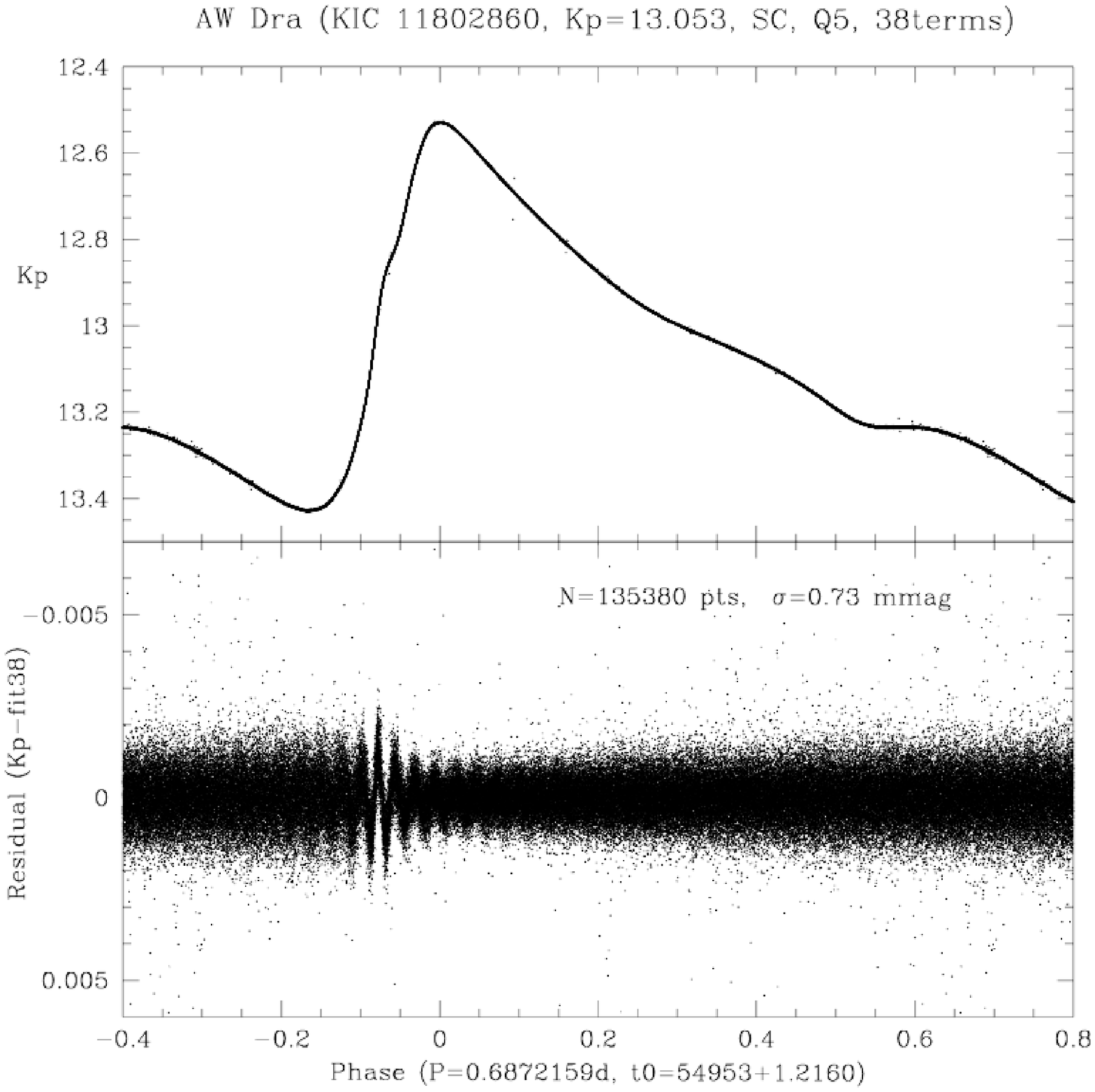}& 
\includegraphics[width=8cm]{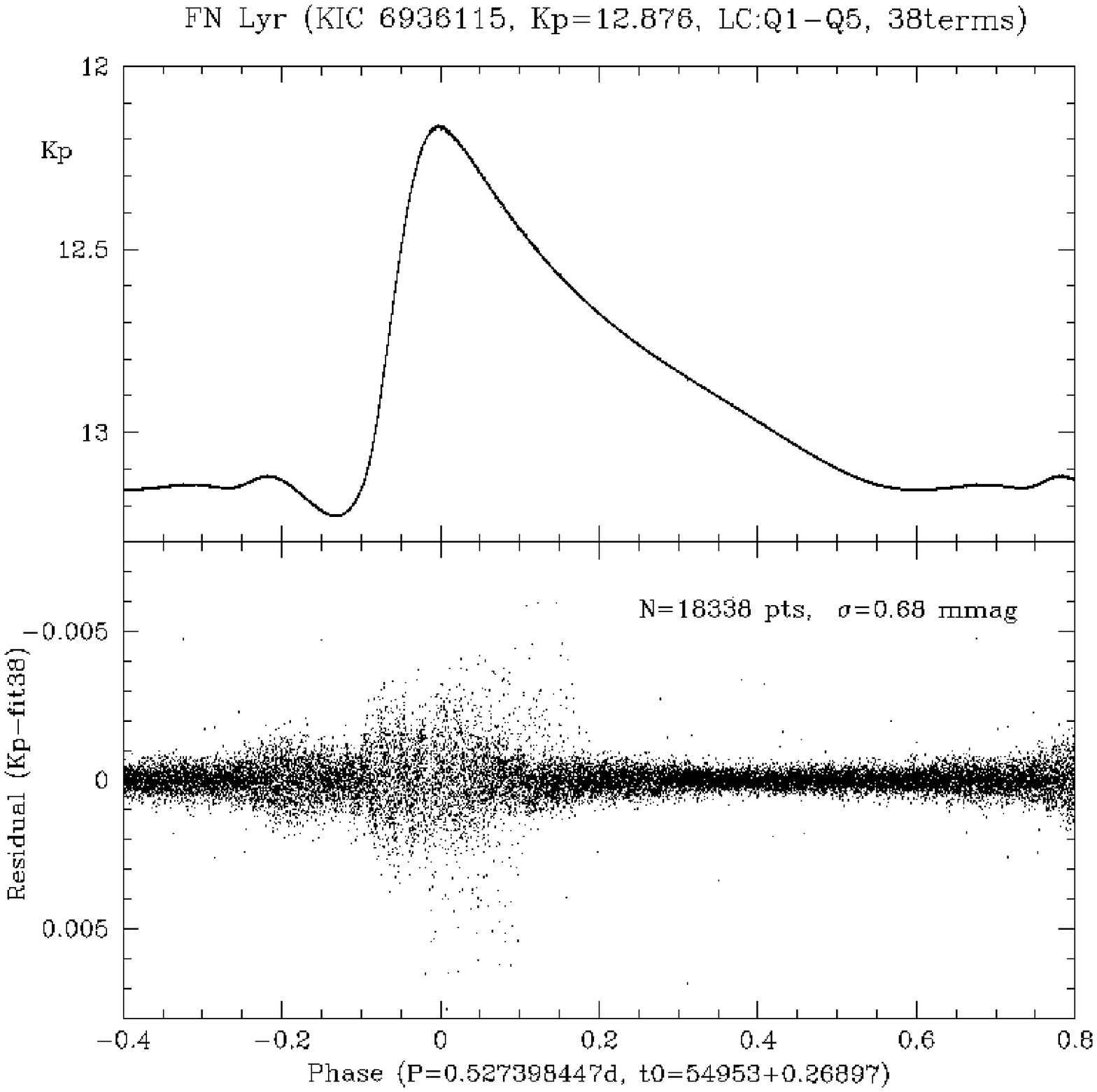} \\
\includegraphics[width=8cm]{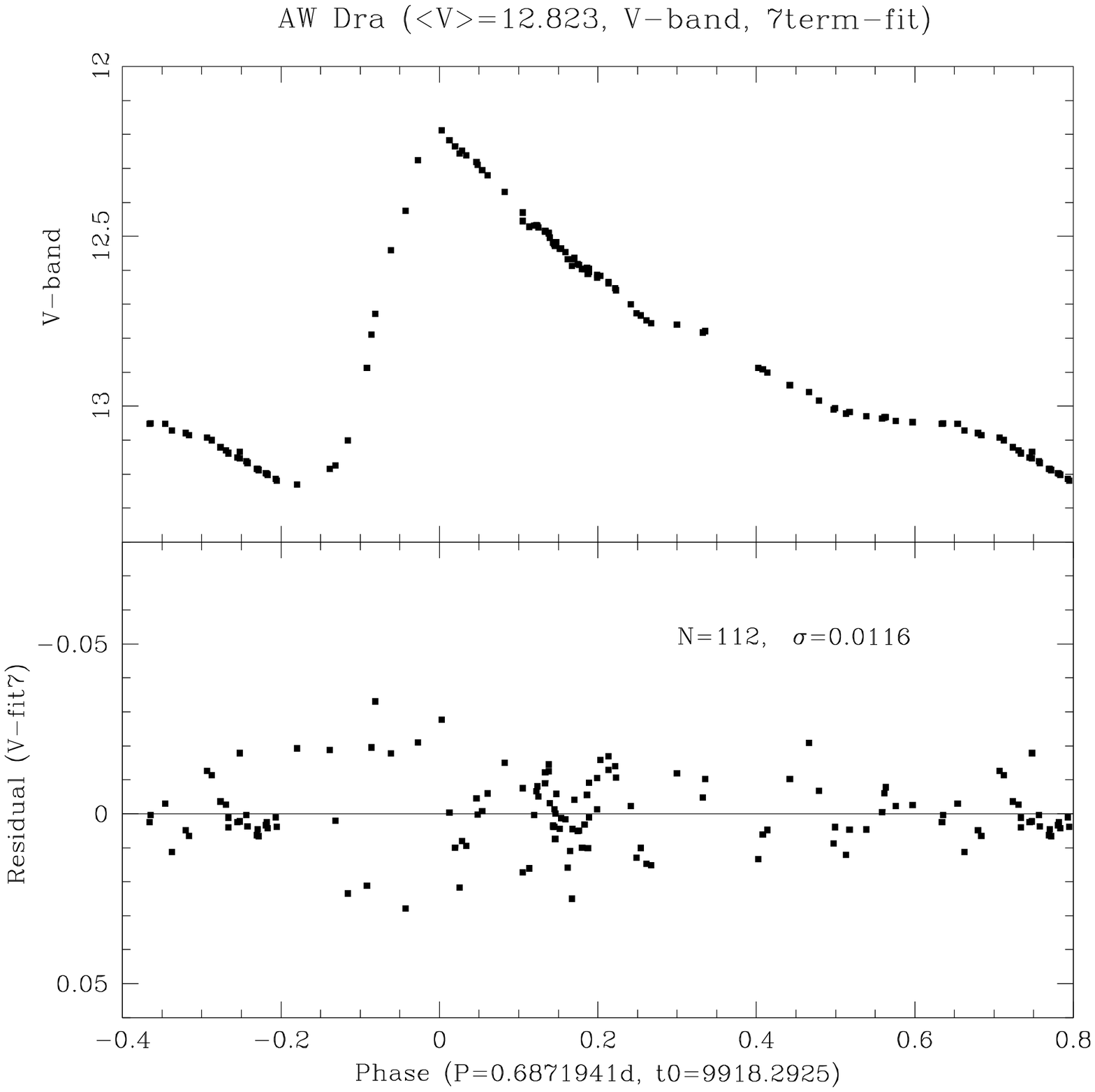}&
\includegraphics[width=8cm]{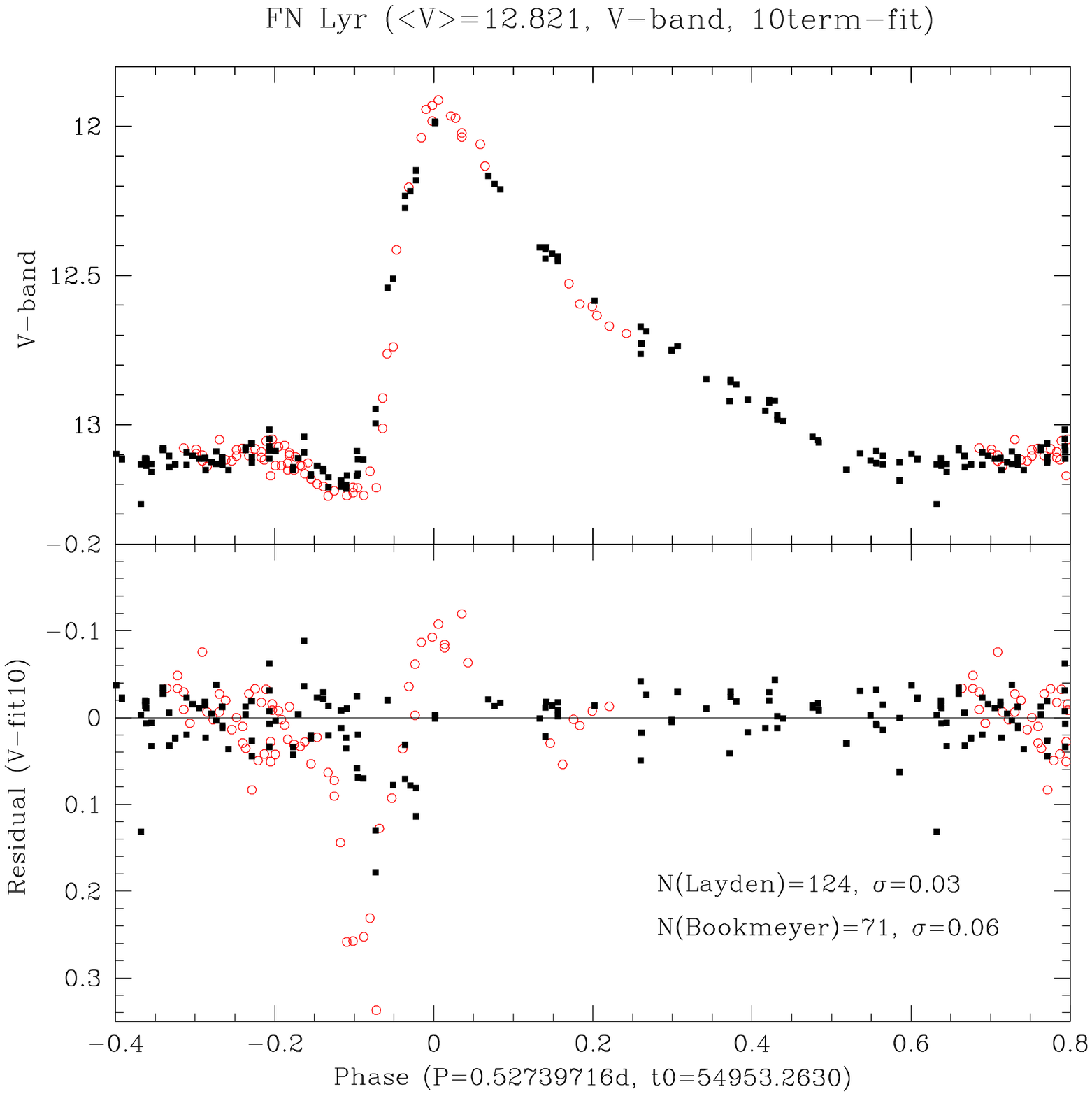} 
\end{array}$
\end{center}
\caption{Light curves and residual plots for AW~Dra (left) and FN~Lyr (right).  
The top diagrams show {\it Kp} photometry and the bottom diagrams $V$-band photometry,
and within each diagram the light curves are on top and the residual plots on the bottom.
The {\it Kp} photometry is SC(Q5) data for AW~Dra and LC(Q1-Q5) data for FN~Lyr. 
The $x$-axis labels give the assumed 
pulsation periods ($P_0$) and times of maximum light ($t_0$).  The diagram labels
indicate the data plotted and the number of terms in the fitted Fourier series, 
and the number of data points ($N$) and standard deviations of the fits ($\sigma$) are 
stated in each diagram.  }
\label{AWDra_fittedLCs}
\end{figure*}

For each of the program stars epoch-independent phase differences
$ \phi_{i1}^s =  \phi_i^s - i \phi_1^s $, and amplitude ratios 
$  R_{i1} =  { A_k / A_1 }$, were computed from the 
Fourier coefficients derived from the {\it Kp}-photometry.  The superscripts ``s''
and ``c'' signify phases and phase-parameters computed with sine and cosine series, respectively 
({\it e.g.} Fig.7 below).  The sine and cosine $\phi_{31}$ phase 
parameters differ by $\pi$ radians, that is,  $\phi_{31}^c  =  \phi_{31}^s - 3.14159$, and the
$\phi_{21}$ parameters differ by $\pi /2$ radians, {\it i.e.,} $\phi_{21}^c =  \phi_{21}^s + 1.5708$. 

Table~3 also contains the Fourier parameters needed for computing physical characteristics. 
In Section (a) results derived from various combinations of short- and long-cadence {\it Kp} data (see column 5) 
are given.  The $A_1$ coefficients (column~7) are practically identical to the $A_1$ coefficients 
calculated by B10 using the Q0-Q2 photometry (given in column~6 of Table~2).    
Columns 8 and 9 contain the amplitude ratios, $R_{21}$ and $R_{31}$;
and columns 10 and 11 contain the Fourier phase parameters, $\phi_{21}^s$ and 
$\phi_{31}^s$.  The accuracy of the {\it Kp}-based Fourier parameters is at least one part in 1000, 
and higher in many cases.  For example, the parameters derived for FN~Lyr from a 38-term fit of the 
18338 Q1-Q5 LC data points have formal values (including uncertainties) as follows: 
$R_{21}$ = 0.44208(2), $R_{31}$=0.34610(2), $\phi_{21}^s$=2.32417(7), and $\phi_{31}^s$=4.81916(9). 
Correlations and graphical representations of these parameters are discussed below.

\begin{table*}
\caption[]{Pulsation and Fourier parameters for the {\it Kepler} non-Blazhko ab-type RR~Lyr stars}
\label{Table3}
\begin{flushleft}
\begin{tabular}{lclclcccccc}
\hline
\noalign{\smallskip}
 Star &$\langle Kp \rangle $&\multicolumn{1}{c}{Period}  &     $t_0$ & \multicolumn{1}{c}{Observational}    & $\sigma$   & $A_1$  & $R_{21}$ & $R_{31}$&  $\phi_{21}^s$ & $\phi_{31}^s$  \\
                &  [mag] & \multicolumn{1}{c}{ [day]}      &    [BJD]     &\multicolumn{1}{c}{data (No.pts)}    & [mmag]  & [mag]   &          &         &   [rad]   & [rad]   \\
  (1)           &  (2)   & \multicolumn{1}{c}{(3)}        &     (4)    &\multicolumn{1}{c}{(5)}            &  (6)   & (7)   &    (8) & (9)  & (10)    &     (11)       \\
\noalign{\smallskip}
\hline
\noalign{\smallskip}
\multicolumn{11}{c}{(a) Results from analysis of the Q0-Q5 {\it Kp} photometry} \\
\noalign{\smallskip}
NR~Lyr          & 12.683 & 0.6820264(2)  & 54964.7381 & LC:Q1-Q5 (18333)    & 0.69   & 0.266 & 0.456 & 0.352 & 2.416 & 5.115 \\ 
V715~Cyg        & 16.265 & 0.47070494(4) & 54964.6037 & LC:Q1-Q5 (18374)    & 1.74   & 0.338 & 0.479 & 0.358 & 2.314 & 4.901 \\                      
V782~Cyg        & 15.392 & 0.5236377(1)  & 54964.5059 & LC:Q1-Q5 (18381)    & 0.79   & 0.190 & 0.488 & 0.279 & 2.777 & 5.808 \\
V784~Cyg        & 15.370 & 0.5340941(1)  & 54964.8067 & LC:Q1-Q5 (18364)    & 0.96   & 0.234 & 0.487 & 0.253 & 2.904 & 6.084 \\
KIC~6100702     & 13.458 & 0.4881457(2)  & 54953.8399 & LC:Q0-Q4 (14404)    & 0.66   & 0.209 & 0.493 & 0.279 & 2.743 & 5.747 \\
NQ~Lyr          & 13.075 & 0.5877887(1)  & 54954.0702 & LC:Q0-Q5 (18759)    & 0.65   & 0.280 & 0.471 & 0.356 & 2.389 & 5.096 \\
FN~Lyr          & 12.876 & 0.52739845(1) & 54953.2690 & SC:Q0+Q5 (149925)   & 0.60   & 0.380 & 0.445 & 0.354 & 2.321 & 4.817 \\
                & 12.876 & 0.527398471(4)& 54953.2690 & LC:Q1-Q5 (18338)    & 0.68   & 0.379 & 0.442 & 0.346 & 2.324 & 4.819 \\
KIC~7021124     & 13.550 & 0.6224926(7)  & 54965.6471 & LC:Q1 (1595)        & 1.10   & 0.283 & 0.512 & 0.351 & 2.372 & 5.060 \\
KIC~7030715     & 13.452 & 0.6836137(2)  & 54953.8434 & LC:Q0-Q5 (18802)    & 0.71   & 0.231 & 0.494 & 0.303 & 2.683 & 5.606 \\
V349~Lyr        & 17.433 & 0.5070740(2)  & 54964.9555 & LC:Q1-Q5 (18314)    & 3.24   & 0.346 & 0.450 & 0.352 & 2.328 & 4.845 \\
V368~Lyr        & 16.002 & 0.4564851(1)  & 54964.7828 & LC:Q1-Q5 (18273)    & 1.57   & 0.405 & 0.464 & 0.341 & 2.272 & 4.784 \\
V1510~Cyg       & 14.494 & 0.5811436(1)  & 54964.6695 & LC:Q1-Q5 (18394)    & 0.82   & 0.345 & 0.473 & 0.355 & 2.389 & 5.068 \\
V346~Lyr        & 16.421 & 0.5768281(1)  & 54964.9211 & LC:Q1-Q5 (18362)    & 2.83   & 0.330 & 0.473 & 0.352 & 2.372 & 5.060 \\
V350~Lyr        & 15.696 & 0.5942369(1)  & 54964.7795 & LC:Q1-Q5 (18326)    & 1.67   & 0.340 & 0.485 & 0.342 & 2.389 & 5.124 \\
V894~Cyg        & 13.293 & 0.5713866(2)  & 54953.5627 & LC:Q1-Q5 (18362)    & 0.91   & 0.377 & 0.490 & 0.338 & 2.364 & 5.067 \\
V2470~Cyg       & 13.300 & 0.5485894(1)  & 54953.7808 & LC:Q0-Q5 (18794)    & 0.79   & 0.220 & 0.488 & 0.282 & 2.745 & 5.737 \\
V1107~Cyg       & 15.648 & 0.5657781(1)  & 54964.7532 & LC:Q1-Q5 (18373)    & 0.99   & 0.280 & 0.495 & 0.350 & 2.421 & 5.196 \\
V838~Cyg        & 13.770 & 0.4802799(1)  & 54964.5731 & LC:Q1-Q5 (18241)    & 1.22   & 0.393 & 0.465 & 0.349 & 2.300 & 4.853 \\
AW~Dra          & 13.057 & 0.6872158(6)  & 54954.2160 & SC:Q0 (14240)       & 0.55   & 0.305 & 0.527 & 0.348 & 2.731 & 5.563 \\
                & 13.053 & 0.687217(1)   & 54954.2160 & LC:Q1 (1614)        & 0.69   & 0.306 & 0.524 & 0.342 & 2.730 & 5.561 \\     
                & 13.053 & 0.6872158(2)  & 54954.2160 & LC:Q5 (4474)        & 0.79   & 0.306 & 0.524 & 0.343 & 2.728 & 5.557 \\
                & 13.053 & 0.68721632(3) & 54954.2160 & SC:Q5 (135380)      & 0.71   & 0.308 & 0.527 & 0.347 & 2.729 & 5.558 \\
\noalign{\smallskip}
\multicolumn{11}{c}{(b) Results from (re-)analysis of high-precision ground-based $V$ photometry} \\
\noalign{\smallskip}
NR~Lyr          &   -    & 0.6820264    & 54964.7381 & Benk\H{o} \& Nuspl (183)         &  5.7   & 0.322 & 0.445 & 0.343 & 2.32  & 4.94  \\
FN~Lyr          & 12.810 & 0.52739716   & 54953.2630 & Layden unpub. (124)              & 30.1   & 0.436 & 0.433 & 0.344 & 2.25  & 4.67  \\
AW~Dra          & 12.823 & 0.6871941    & 49918.2925 & Castellani {\it et al.} (112)    & 11.6   & 0.357 & 0.510 & 0.353 & 2.63  & 5.43  \\
\noalign{\smallskip}
\multicolumn{11}{c}{(c) Results from re-analysis of the ASAS $V$ photometry (http://www.astrouw.edu.pl/asas)} \\
\noalign{\smallskip}
NR~Lyr          & 12.44  & 0.68200(5)   & 54964.7383 & 190827+3848.8 (82)   &  61    & 0.24  & 0.45  & 0.35  & 2.2   & 4.9  \\
KIC~6100702     & 13.64  & 0.48817(2)   & 54953.8398 & 185038+4125.4 (79)   &  71    & 0.23  & 0.59  & 0.32  & 2.4   & 4.8  \\
NQ~Lyr          & 13.36  & 0.58779(3)   & 54954.0702 & 190749+4217.9 (68)   &  68    & 0.30  & 0.56  & 0.40  & 2.4   & 5.2  \\   
FN~Lyr          & 12.78  & 0.52740(3)   & 54953.2690 & 191022+4227.6 (82)   &  76    & 0.41  & 0.49  & 0.36  & 2.3   & 4.8  \\
KIC~7030715     & 13.24  & 0.68362(3)   & 54953.8434 & 192325+4231.7 (84)   & 103    & 0.26  & 0.56  & 0.35  & 2.4   & 5.4  \\
V894~Cyg        & 12.92  & 0.57140(2)   & 54953.5625 & 193301+4614.3 (76)   & 108    & 0.37  & 0.41  & 0.38  & 2.2   & 4.8  \\                 
V2470~Cyg       & 13.53  & 0.54857(1)   & 54953.7808 & 191958+4653.3 (78)   &  62    & 0.25  & 0.42  & 0.33  & 2.7   & 5.6  \\
V838~Cyg        & 14.25  & 0.48030(2)   & 54964.5731 & 191404+4812.1 (68)   & 105    & 0.44  & 0.39  & 0.38  & 2.2   & 4.7  \\
AW~Dra          & 12.85  & 0.68721(1)   & 54954.2158 & 190048+5005.5 (86)   &  49    & 0.37  & 0.53  & 0.36  & 2.5   & 5.4  \\ 
\hline
\end{tabular}
\end{flushleft}
\end{table*}

\subsection{Cycle-to-Cycle variations and stationarity}

\begin{figure*}
\begin{center}$
\begin{array}{cc}
\includegraphics[width=8cm]{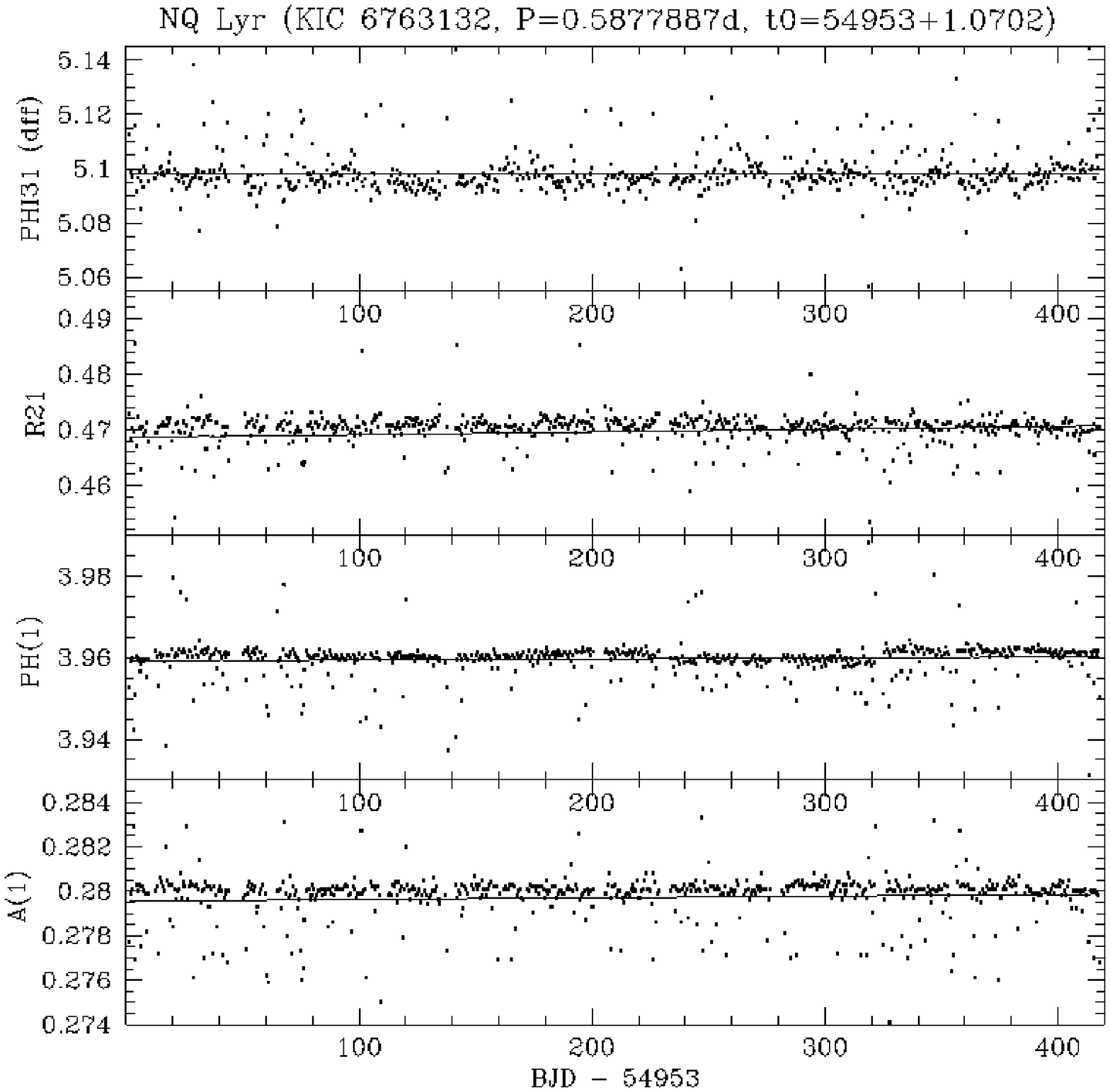} &
\includegraphics[width=8cm]{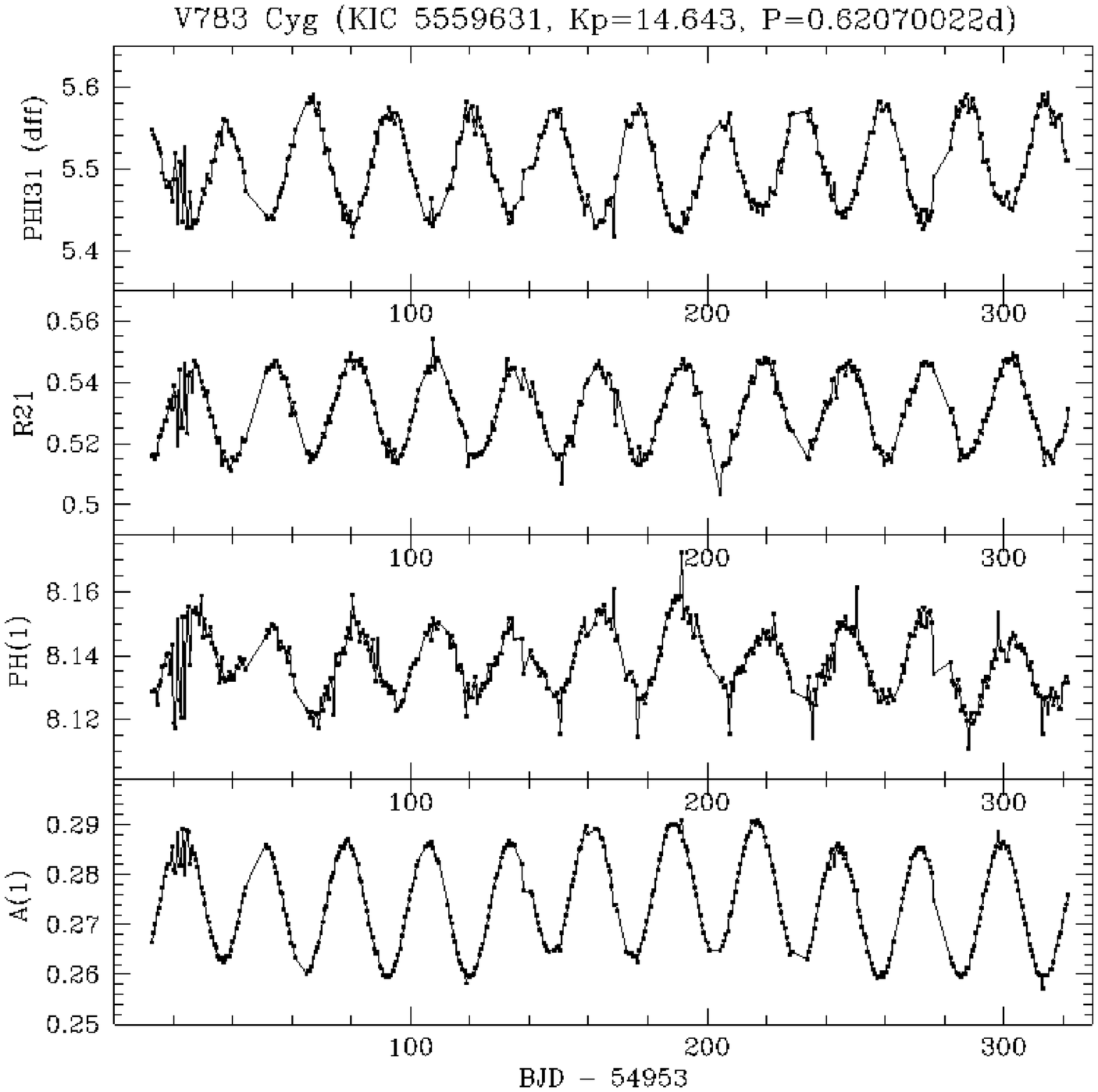} 
\end{array}$
\end{center}
\caption{Variations in time of the Fourier parameters  $\phi_{31}^s$, $R_{21}$, $\phi_1^s$ and 
$A_1$ (all sine-series) derived for the typical non-Blazhko star NQ~Lyr (left panel) and for 
the known low-amplitude Blazhko star V783~Cyg (right panel).  Each point corresponds to the variable of interest 
as determined from a single pulsation cycle.  }
\label{cf_nonBL_BL}
\end{figure*}

To examine cycle-to-cycle variations in the {\it Kepler} light curves the Fourier 
parameters were re-calculated after binning the data so that each bin included 
data for one pulsation period ({\it i.e.}, each bin represented a single 
pulsation cycle).   For the long cadence (30 min) data there were $\sim$24 observations per bin for an RRab star
with a period of 12 hours, and fewer (more) observations per cycle for stars with shorter (longer) periods.   
None of the 19 stars exhibits the recently discovered ``period doubling'' effect, thus confirming the
result found earlier by Szab\'o {\it et al.} (2010).  

For the Fourier calculations 
`direct Fourier fitting' (dff) rather than `template Fourier fitting' (tff) methods were used (see Kov\'acs \& Kupi 2007),
and for each star the resulting time series were plotted for four Fourier parameters: 
$A_1$, the first term in the Fourier series (see right panel of Fig.~2);  $\phi_1$, the phase of the first term; 
$R_{21}$, the amplitude ratio $A_2/A_1$; and $\phi_{31}^s$, the Fourier parameter found by 
Simon, Kov\'acs and others to be one of the most significant variables for deriving physical characteristics.
Typical time series are illustrated in {\bf Figure~5}, on the left for the unmodulated RRab star NQ~Lyr having 
intermediate amplitude ($A_1$=0.279 mag) and intermediate brightness ($\langle Kp \rangle =13.075$~mag), and 
on the right for V783~Cyg, a low-amplitude 
Blazhko star with the shortest Blazhko period ($P_B = 27.7\pm0.04$~d -- see B10)
among the stars in our sample. 

For NQ~Lyr linear trend lines were fit to each time series and in almost every case the slope is
zero to within the systematic and random uncertainties.  The mean values for the four parameters are
as follows:  
$\phi_{31}^s = 5.0958 \pm 0.0023$, with residual standard error, $\sigma =0.002$;
$R_{21} = 0.4710 \pm 0.0006$, with $\sigma = 0.0006$;
$\phi_1^s = 3.961 \pm 0.001$, with $\sigma = 0.001$; and
$A_1 = 0.28016 \pm 0.00014$ mag, with $\sigma = 0.0001$.
Since the other non-Blazhko stars show random variations of the Fourier parameters similar to those shown
here for NQ~Lyr these means and errors provide a measure of the typical uncertainties.       
  
Inspection of the NQ~Lyr time series in Fig.~5 shows that the stability over the $\sim$420~d interval (Q0-Q5) is exceptional, 
not only for each Fourier parameter but for the ensemble of parameters.  To better illustrate the remarkable
stationarity of the light curves for all 19 sample stars a set of `animated gif' light curves has been 
prepared and these are available in the electronic version of the Journal.

\begin{figure*}
\begin{center}$
\begin{array}{cc}
\includegraphics[width=8cm] {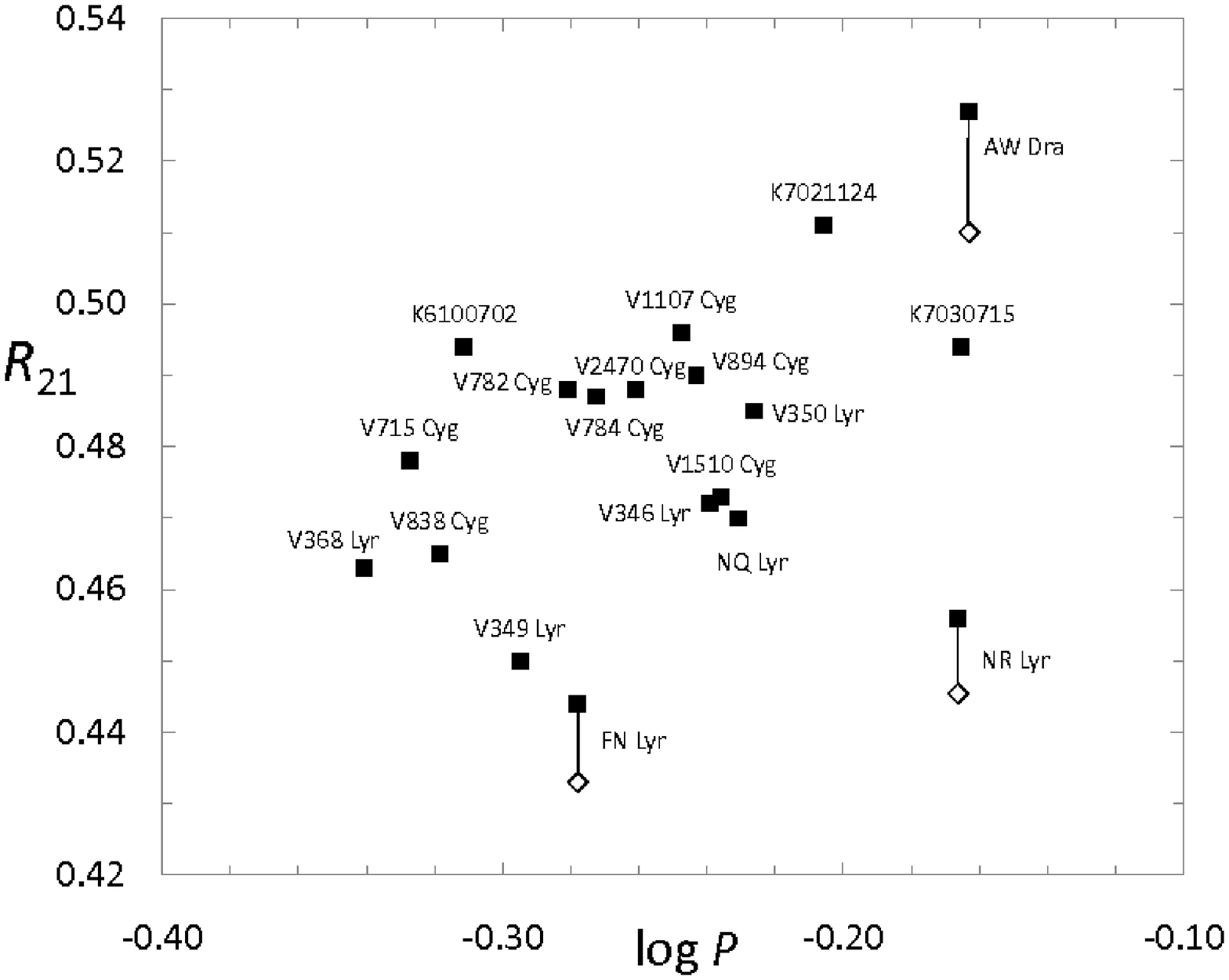}&
\includegraphics[width=8cm] {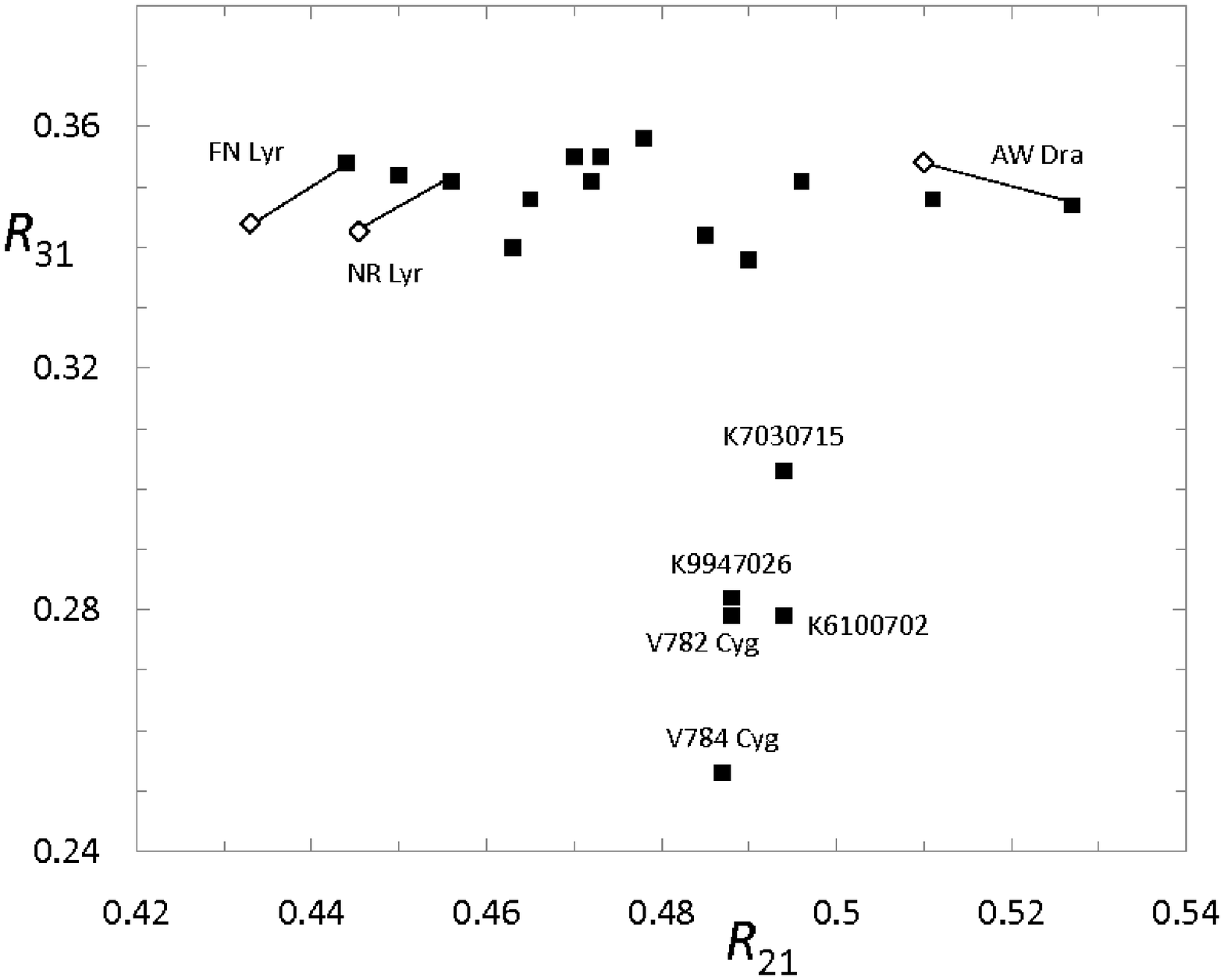} \\
\includegraphics[width=8cm] {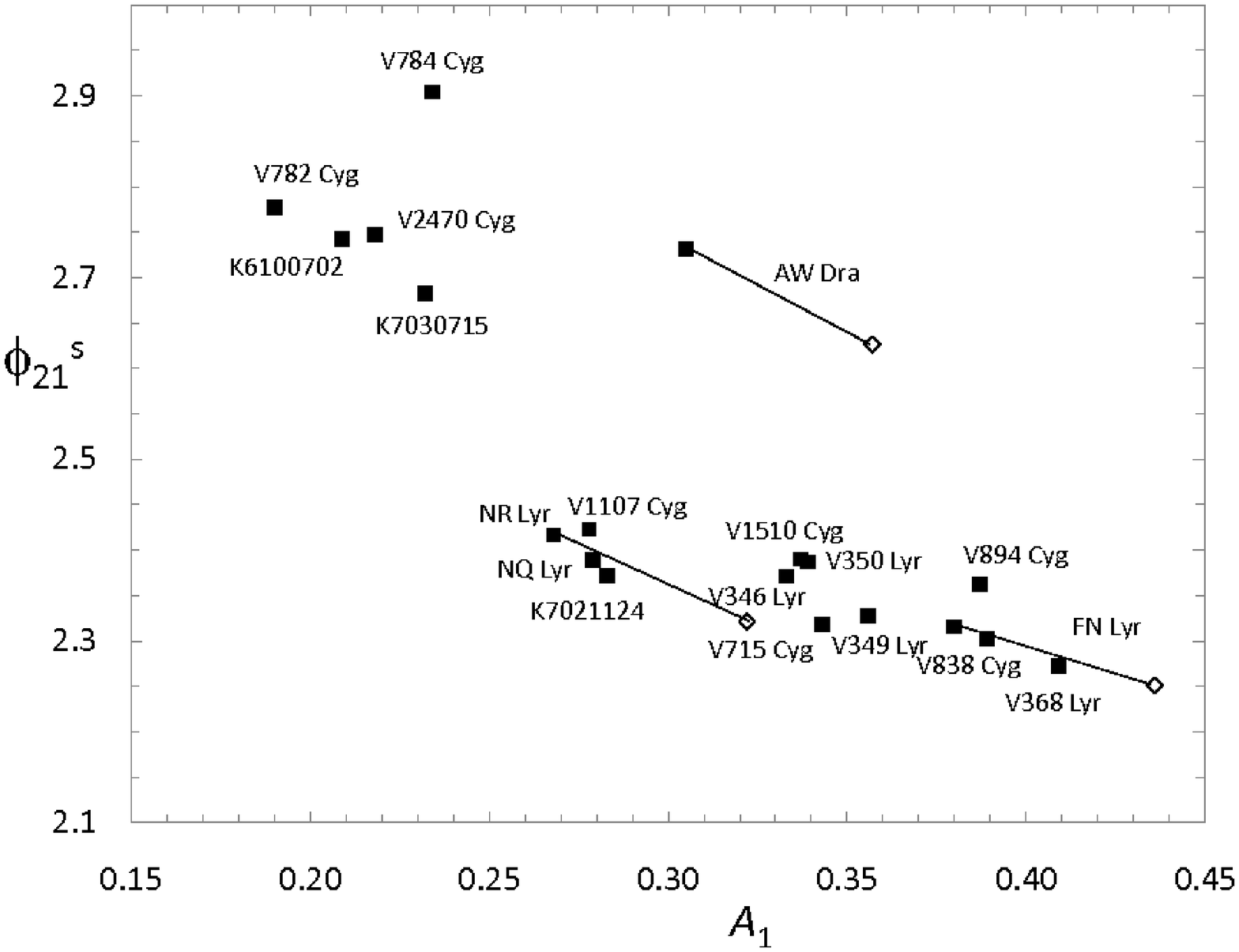} &
\includegraphics[width=8cm] {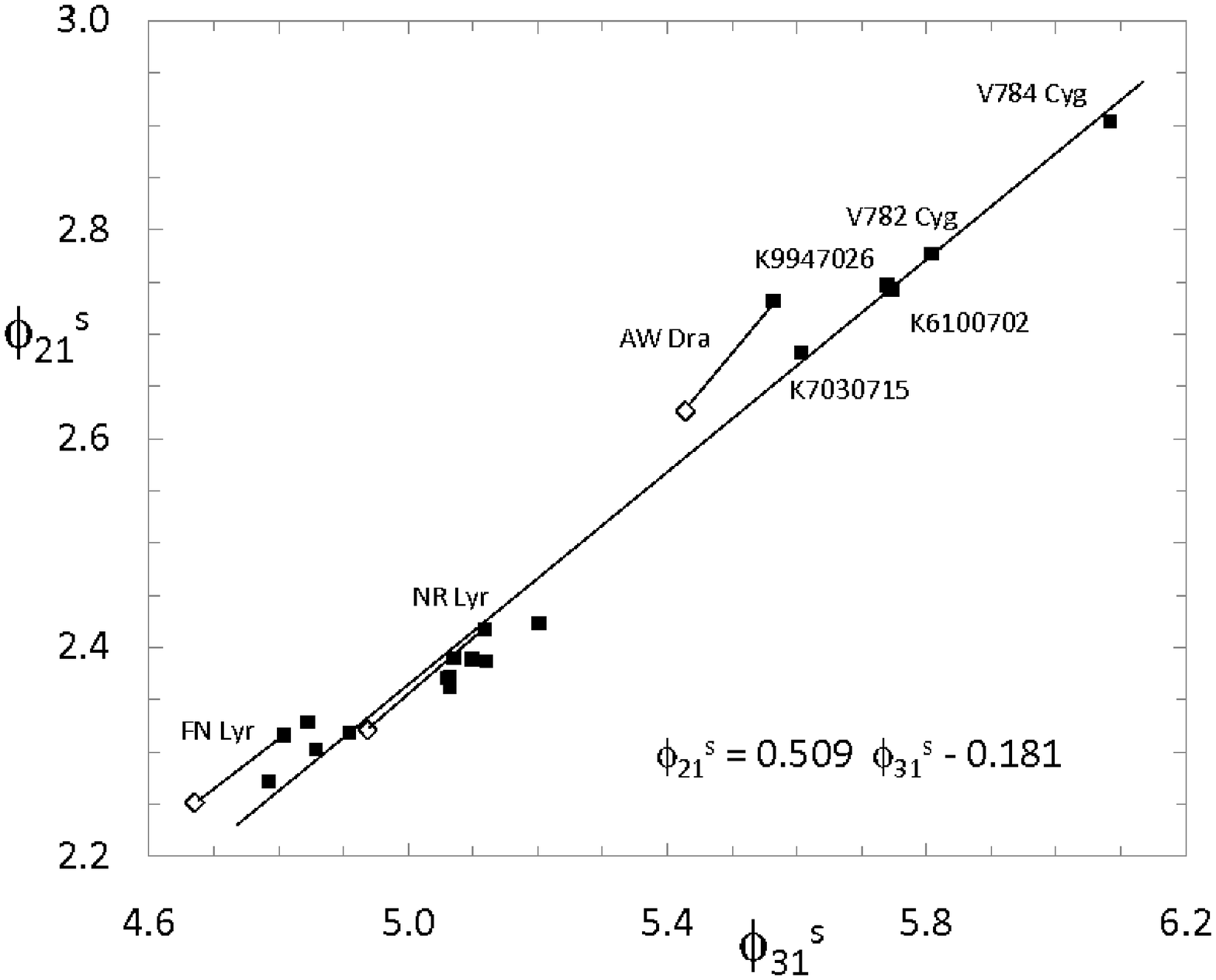}  
\end{array}$
\end{center}
\caption {Four panels comparing various Fourier parameters derived from the {\it Kp}-photometry 
for the 19 {\it Kepler} non-Blazhko stars.  The open diamonds for AW~Dra, FN~Lyr and NR~Lyr were derived 
from the $V$-photometry of Castellani {\it et al.} (1998), Layden (2010, unpublished) and 
Benk\H{o} \& Nuspl (2010, unpublished), respectively.  }
\label{nonBL_PHI21s_vs_PHI31}
\end{figure*}

For the Blazhko star V783~Cyg the time plots of the four Fourier parameters 
are in  striking contrast
to the stochastic noise exhibited by NQ~Lyr and by 
the candidate non-Blazhko star V349~Lyr. 
For V783~Cyg pronounced 
periodic variations are seen over the 27.7~d Blazhko cycles.   There 
does appear to be a slight discontinuity at around 140~d, but this is certainly a data
processing defect.  Over the 11 Blazhko cycles seen there is very little
difference from cycle-to-cycle; this is not always the case for Blazhko stars where
large cycle-to-cycle differences often are seen.

\subsection{High-precision $V$-photometry}

By comparing the light curves and Fourier parameters  for the {\it Kp} and $V$ photometric systems 
it is possible to test (in a limited way) the
hypothesis that the two broad-band  
filters give similar results.    If this hypothesis is true then the Fourier parameter  
correlations  established for the $V$-band system
by Simon, Kov\'acs, Jurcsik, Sandage, and others {\it might} be applied to the {\it Kp} mags and 
be used to derive first-order physical characteristics (see $\S6$).
Unfortunately there is very little {\it high-precision} published ground-based $V$-band photometry
for our {\it Kepler} stars.  Three exceptions are NR~Lyr, AW~Dra and FN~Lyr.  Using the available $V$-photometry 
for these stars Fourier parameters were derived and compared with the {\it Kp} parameters.      
The results are given in the middle section of Table~3 (Section b).

\subsubsection{AW~Dra}

The light curve and residuals derived from the $V$-photometry of
Castellani {\it et al.} (1998) are plotted in the lower left diagram of Figure~4.
The 112 points are from CCD observations made in 1995 with the 72-cm Teramo/Scuola Normale Telescope, 
modelled with a 7-term Fourier series fit.  
The shape of the $V$ light curve is very similar to the light curve based on the {\it Kepler} data, and the Fourier
parameters (Table~3) support this observation.   Notice also 
that,  although the $V$-data have a residual standard error of only 0.0116~mag, the residuals are $\sim16\times$ larger 
than the {\it Kp}-residuals (Figure~4, upper left panel).     
On the other hand, the risetimes are quite similar: 0.165 for the {\it Kp}-photometry and 0.174 for the $V$-photometry.
The joined points in the RT graphs below (lower left panel of Figure~9, and left panel of Figure~10)  
give a visual representation of these differences.

\subsubsection{FN~Lyr}

For FN~Lyr, the $V$ photometry shown in the lower right panels of Figure~4 are from two sources: 
Bookmeyer {\it et al.} (1977) and Layden (2010, unpublished).  
The Bookmeyer {\it et al.} data comprise 71 $UBV$ photoelectric observations
made in 1966 (21) and 1969 (50), and the Layden $V$-photometry constitute 124 $VR$ CCD observations made in 2006-08.
When the two data sets, which are separated by $\sim$40 years, were combined and plotted as single light curve 
({\it i.e.},  the same period and time of maximum light) a phase shift 
amounting to 5.5\% of the period (0.029~d) was seen, with the Layden data having the larger phases.  
The shift corresponds to the offset of the two O$-$C values seen in Fig.~3 (middle right panel, 
where the O$-$C values for the two data sets are plotted as open circles
and are labeled `1966-69' and `2006-08').   Our period change rate analysis suggested that the shift is 
probably due {\it either}  to the assumed 
period (0.52739716~d) being too small  {\it or} to a slowly increasing period (d$P$/d$t$$=0.004\pm0.004$~d/Myr).

Figure~4 (lower right) shows the composite $V$ light curve for FN~Lyr, where the Layden data (solid black squares)
are plotted alongside `the Bookmeyer data shifted in phase by $+0.055$' (red open circles).   
There is a reasonably good match  between the Layden and Bookmeyer light curves, the chief differences 
being: (1) only two (very close) Layden data points were acquired near maximum light,
and they are fainter than the Bookmeyer observations near maximum light;  
(2) the slope of the rise to maximum light is shallower for the Layden data than for the Bookmeyer data; and 
(3) the Layden data are uniformly distributed in phase while the Bookmeyer data
show two large gaps on the descent to minimum light. 

Fourier coefficients and parameters were calculated using the composite `Layden {\it plus} shifted-Bookmeyer' data, 
and using the 124 Layden $V$ data points only (the former were found to be less reliable than the latter).  
Owing to the large gaps it was not possible to obtain a reliable  
Fourier fit using only the Bookmeyer data. 
Table~3 gives the Fourier decomposition results based on the Layden-only $V$-photometry.  
The $V$-residuals shown in the bottom-right panel of Figure~4  were calculated with respect to the Layden-only fit.  
The Layden data have a residual standard error of 0.03 mag, while the fit for the Bookmeyer data is poorer ($\sim$0.06 mag).  
In both cases (as for the {\it Kepler} data) the largest residuals occur on the rise to maximum light.
The observed smaller total amplitude and longer risetime for the Layden $V$ photometry 
compared with the Bookmeyer photometry might be expected if FN~Lyr is a low-amplitude 
Blazhko variable.  The available data are insufficient to check this possibility but as
additional  {\it Kepler} observations are acquired  the answer may be evident. 

\subsubsection{NR~Lyr}

In 2008 $B,V,R,I$ CCD observations of NR~Lyr were made by J.~Benk\H{o} and J.~Nuspl 
using the Konkoly telescope.  At present the 183 data points per filter
are on the instrumental system of the telescope; however, the photometric colour constant is very small and the
$V$ data differ from the Johnson $B,V$ system by only a zero point shift.  Because the data completely 
cover the phase range and are of high precision the $V$ data were Fourier analyzed (using the
period and $t_0$-value favoured by the {\it Kepler} data) and the 
results summarized in Table~3.   When the Fourier  
parameters are compared with those for FN~Lyr and AW~Dra (see Figures 6, 9 and 10) all three sets
of observed offsets are in excellent agreement (see $\S5.2$ below).

\subsection{ASAS $V$-photometry}

For the nine non-Blazhko RR~Lyr stars in the ASAS-North survey (see $\S 3.1$) the on-line $V$-photometry was re-analyzed and 
the results reported in section (c) of Table~3.     
The derived $\langle V \rangle$ magnitudes (column 2) have uncertainties of $\pm 0.01$ and range from 12.44 to 14.24.
A comparison of the $\langle V \rangle$  and $\langle Kp \rangle$ mags
shows the following trend over the observed range:  $\langle V \rangle  = (1.45\pm0.24) \langle {\it Kp} \rangle - (5.97\pm3.20)$.
Until further $V$ photometry fainter than $14^{\rm th}$ mag is obtained it is unclear
whether this trend of $V$ mags fainter than the {\it Kp} mags will continue.
Pulsation periods (with uncertainties in parentheses) derived from our re-analysis of the $V$ data are given in column (3);  
in general these agree with the periods given at the ASAS website and with those derived from the {\it Kepler} data (see Tables~2-3).
Columns~4-6 contain, respectively, the assumed time of maximum light, the ASAS name of the star, and the spread of the Fourier fit about the
mean light curve (found to range from 49 to 108 mmag).
Fourier parameters derived using the on-line $V$ data are given on the right side of Table~3.  These were computed using the 
accurate {\it Kepler} periods and times of maximum light.  The uncertainties
typically are $\pm 0.04$ for $R_{21}$, $\pm 0.01$ for $R_{31}$, $\pm 0.08$ for $\phi_{21}$,
and $\pm 0.1$ for $\phi_{31}$, and in several cases are larger.

\section{FOURIER PARAMETER CORRELATIONS}

\subsection{{\it Kp} Correlations}

{\bf Figure~6} shows four graphs relating the Fourier parameters $A_1$, $R_{31}$, $R_{21}$, $\phi_{21}^s$ 
and $\phi_{31}^s$, derived from the sine-series decomposition of the {\it Kp}-photometry of the 
19 {\it Kepler} non-Blazhko stars.  
Each star is represented by a solid black square, and is labelled for easy identification.
Also plotted in Fig.~6 are the $V$-band Fourier parameters for 
AW~Dra, FN~Lyr and NR~Lyr ($\S4.3$) which are shown as open diamonds connected to the corresponding {\it Kp} points by lines.  
The offsets between the {\it Kp} and $V$ points appear to be systematic (i.e., generally go in one direction)
and are not very large,  even in the case of $R_{31}$ which is the least certain of the parameters.
Perhaps this relationship between $V$ and {\it Kp} is not surprising because both passbands are quite wide. 

The upper-left panel of Fig.~6 shows that there is very little correlation between $R_{21}$ and period
(except possibly a slight increase in $R_{21}$ with increasing period).
The $R_{21}$ values from the $V$ photometry are all slightly smaller than from the {\it Kp} photometry, and  
the  $R_{21}$ values for FN~Lyr and NR~Lyr are significantly smaller than that for AW~Dra.  

The upper-right panel shows 14 stars (including AW~Dra, FN~Lyr and NR~Lyr) with $R_{31}$ between 0.34 and 0.36, 
and five stars with $R_{31}$ values smaller than 0.31.  
In particular, AW~Dra, FN~Lyr and NR~Lyr are seen to have similar $R_{31}$ values even though the 
$R_{21}$ values for FN~Lyr and NR~Lyr are significantly smaller than that for AW~Dra.  Although the overall bimodal distribution 
is striking, we shall see below (upper right panels of Figs.~7 and 8) 
that RR~Lyr stars with relatively low $R_{31}$ values are 
quite common and can be simulated with hydrodynamic models ($\S7$). 

In the lower-left panel the stars separate into high- and low-$\phi_{21}^s$ stars, with a discontinuity
at $A_1$$\sim$0.25 mag.  AW~Dra and V784~Cyg (suspected of having a relatively high metallicity) appear to stand out from the other stars;  even when
compared with  RR~Lyr stars in globular clusters ($\S5.3$) these two stars appear extreme.  

Finally, the lower-right panel shows that $\phi_{21}^s$ is roughly proportional 
to $\phi_{31}^s$, where the linear relationship is given by $\phi_{21}^s = 0.509 \thinspace \phi_{31}^s - 0.181$.
For AW~Dra, FN~Lyr and NR~Lyr the phase parameters are smaller for the $V$ data than the {\it Kp} data, and the offsets 
are parallel to the fitted line.     

\begin{figure*}
\begin{center}$
\begin{array}{cc}
\includegraphics[width=8cm] {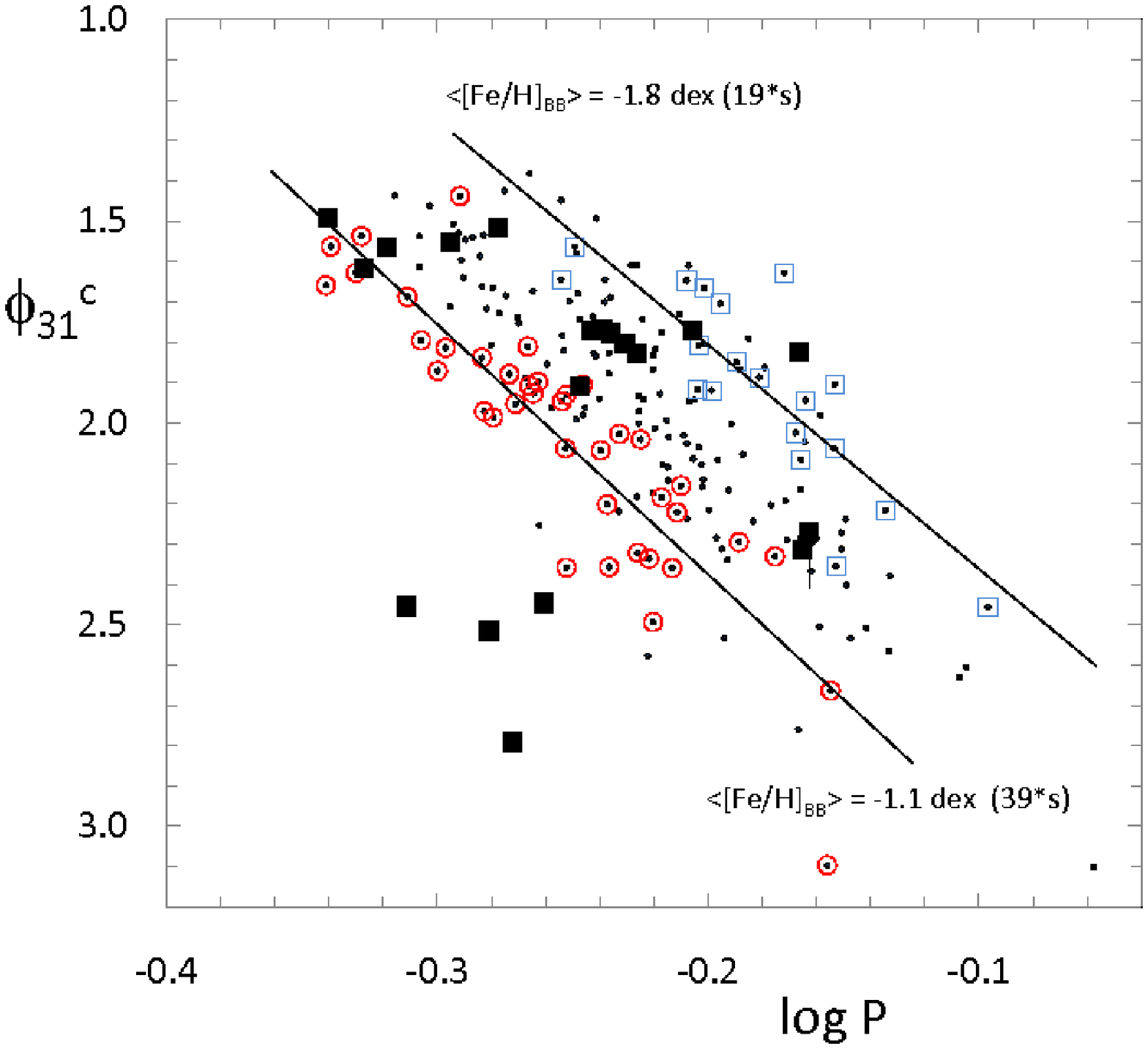} &
\includegraphics[width=8cm] {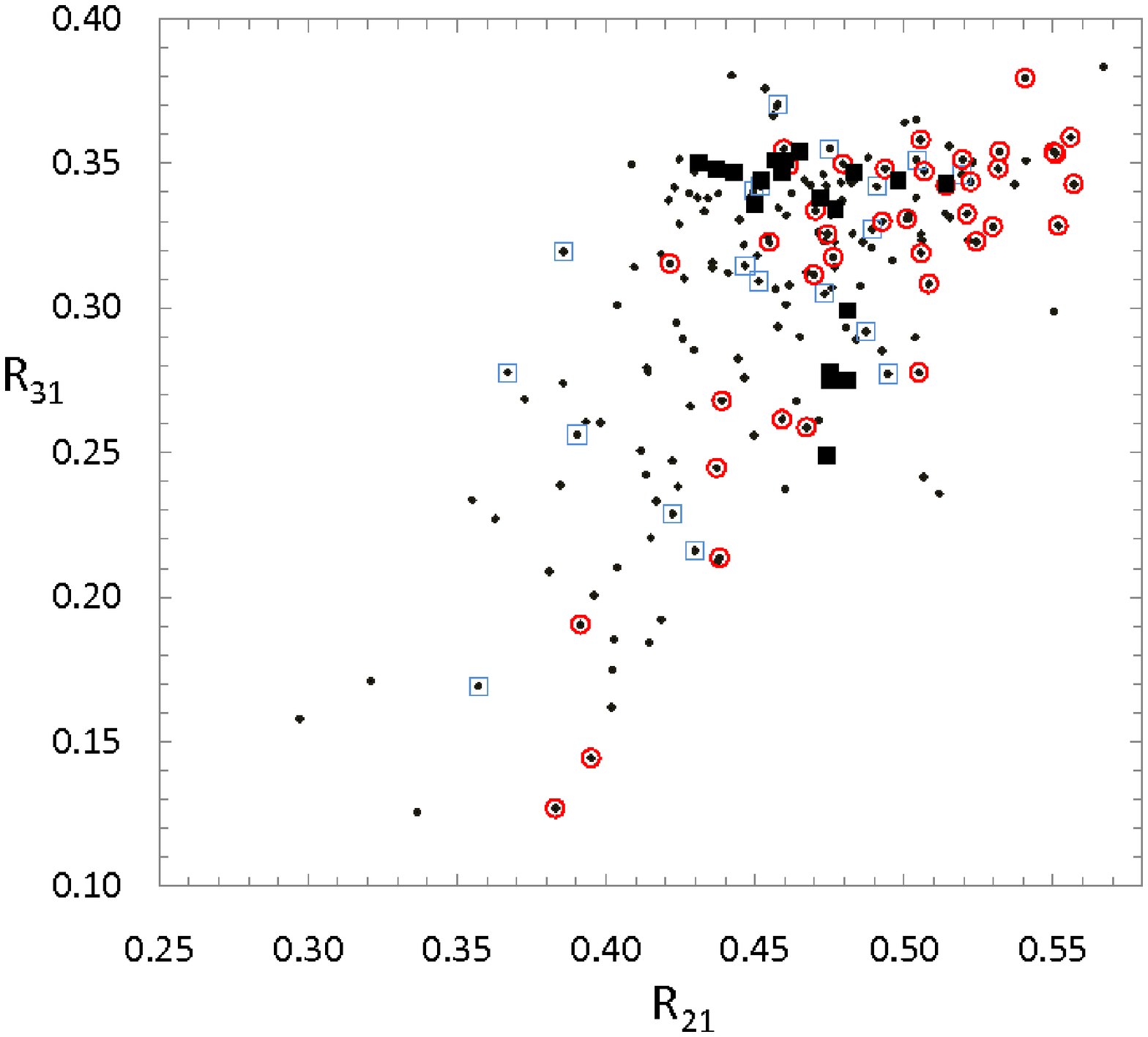} \\
\includegraphics[width=8cm] {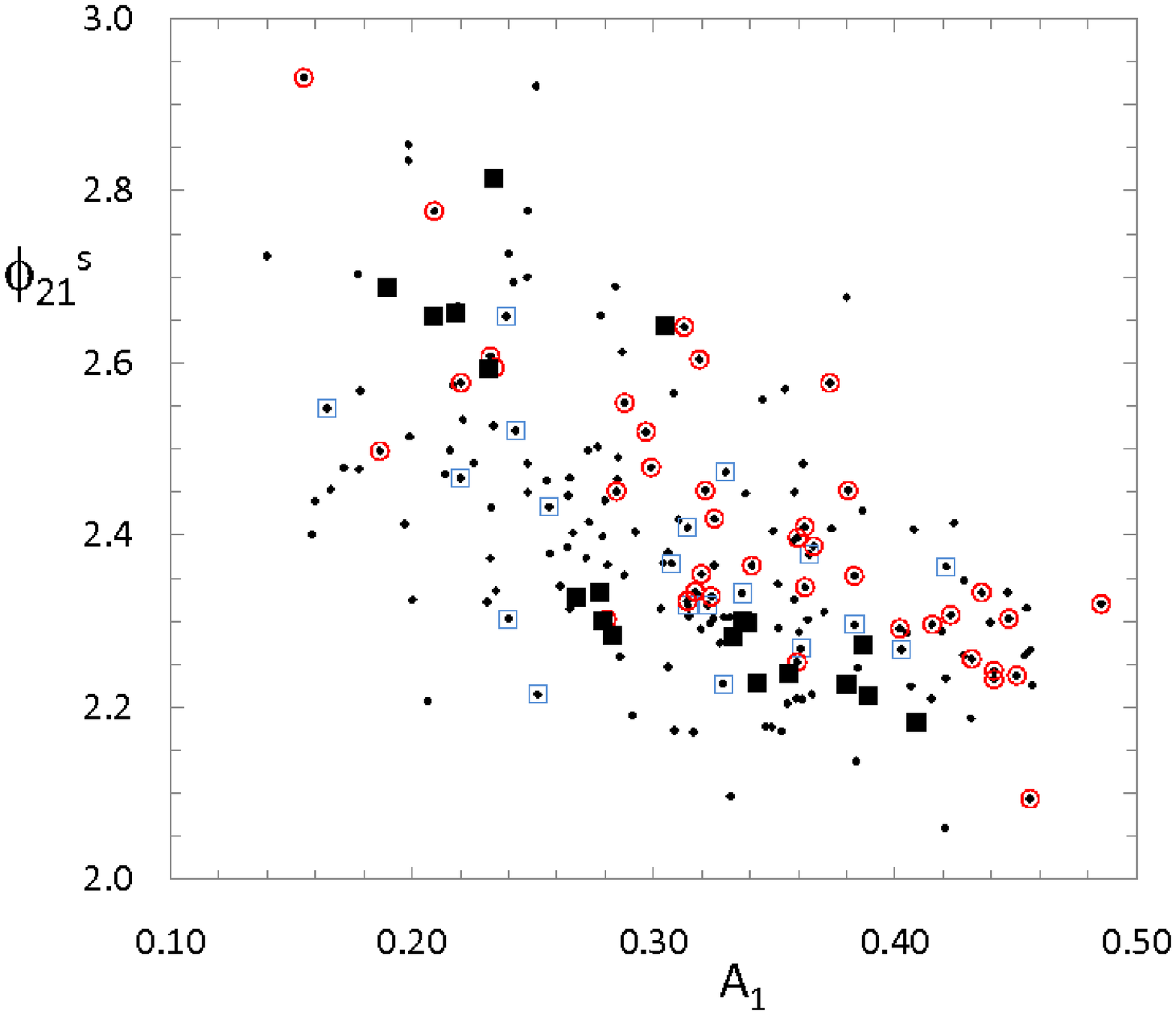} &
\includegraphics[width=8cm] {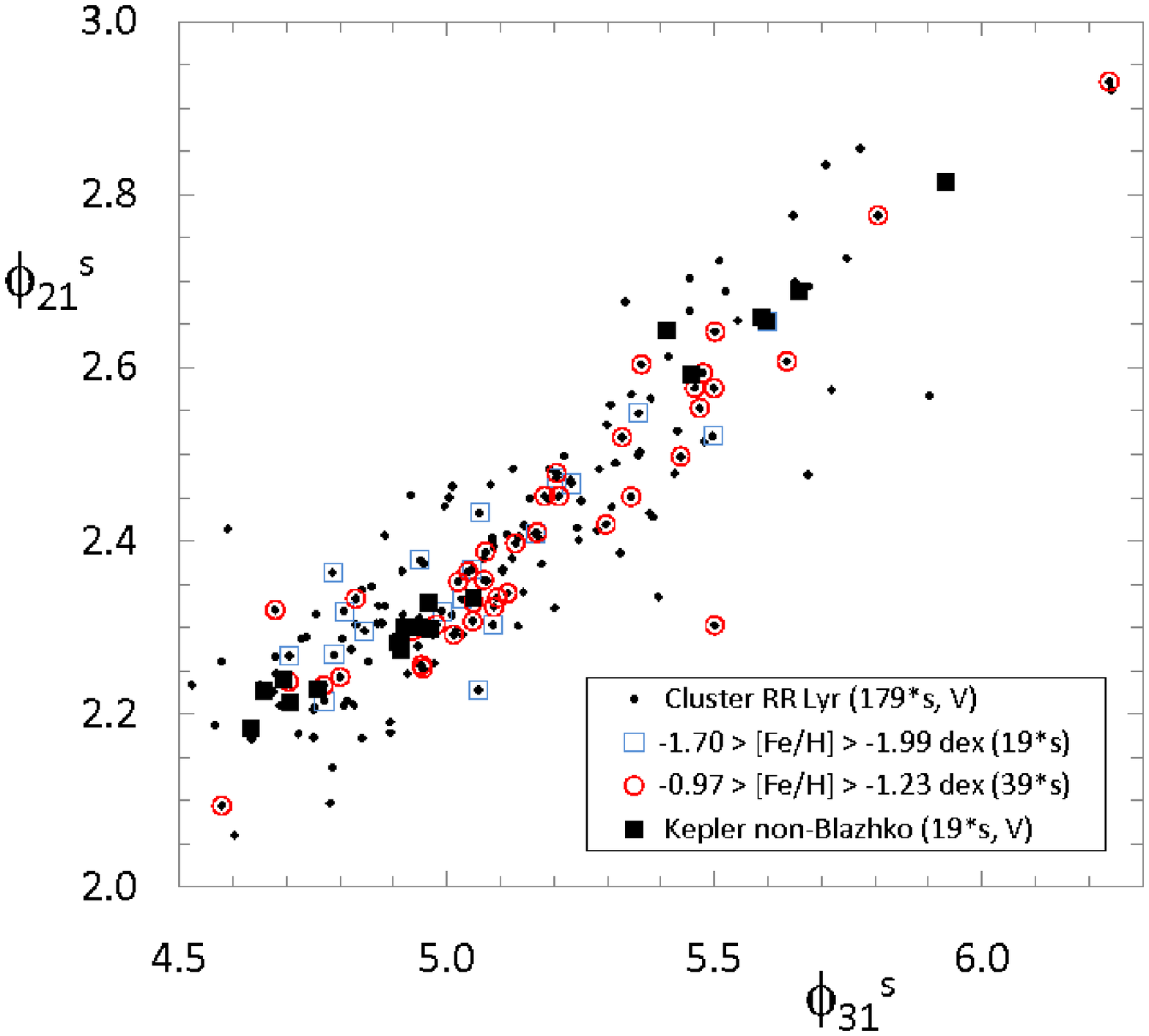}
\end{array}$
\end{center}
\caption{Four panels comparing the Fourier parameters (transformed to the $V$-band) for 
the {\it Kepler} non-Blazhko stars (large black squares) and 
for the 177 RR~Lyr stars located in several galactic and LMC globular clusters (small black dots).
The cluster RR~Lyr stars (from Kov\'acs \& Walker 2001) are in the 
globular clusters M2 (12 stars), M4 (4), M5 (12), M55 (4), M68 (5), M92 (5), NGC~1851 (11), 
NGC~5466 (6), NGC~6362 (12), NGC~6981 (20), IC~4499 (49), Ruprecht~106 (12), NGC~1466 (8), Reticulum (8) and NGC~1841 (9). 
The points for the 19 most metal-poor globular cluster stars are surrounded 
by blue squares, and the points for the 39 most metal-rich globular cluster stars are circled
in red.  }
\label{PHI21vsA1and PHI31}
\end{figure*}

\subsection{ {\rm {\it Kp}}-$V$ offsets}

Since the offsets between the {\it Kp} and $V$ Fourier parameters for AW~Dra, FN~Lyr and NR~Lyr appear 
to be small and systematic (Figs.~6,9,10; Table~3), transformation from {\it Kp} to $V$ should be possible
using the high-precision $V$ parameters (inclusion of the ASAS parameters would not have been helpful owing to 
their relatively large uncertainties).
Transformation to the $V$ system allows derivation of physical characteristics through the 
application of well-established $V$-band Fourier relations (see $\S6$ below). 
Because the three stars exhibit a range of periods (0.53~d to 0.69~d) and  
light curve shapes ($\phi_{31}^s$ = 4.8 to 5.6) and mean colours (see Fig.~B1 below),
confounding effects that depend, for example, on location in the `instability strip', are likely to 
have been revealed.  On the other hand, if the offsets depend on [Fe/H] such an effect might 
not have been detected because all three stars appear to have low metal abundances 
(see column~3 of Table~4).

\begin{figure*}
\begin{center}$
\begin{array}{cc}
\includegraphics[width=8cm] {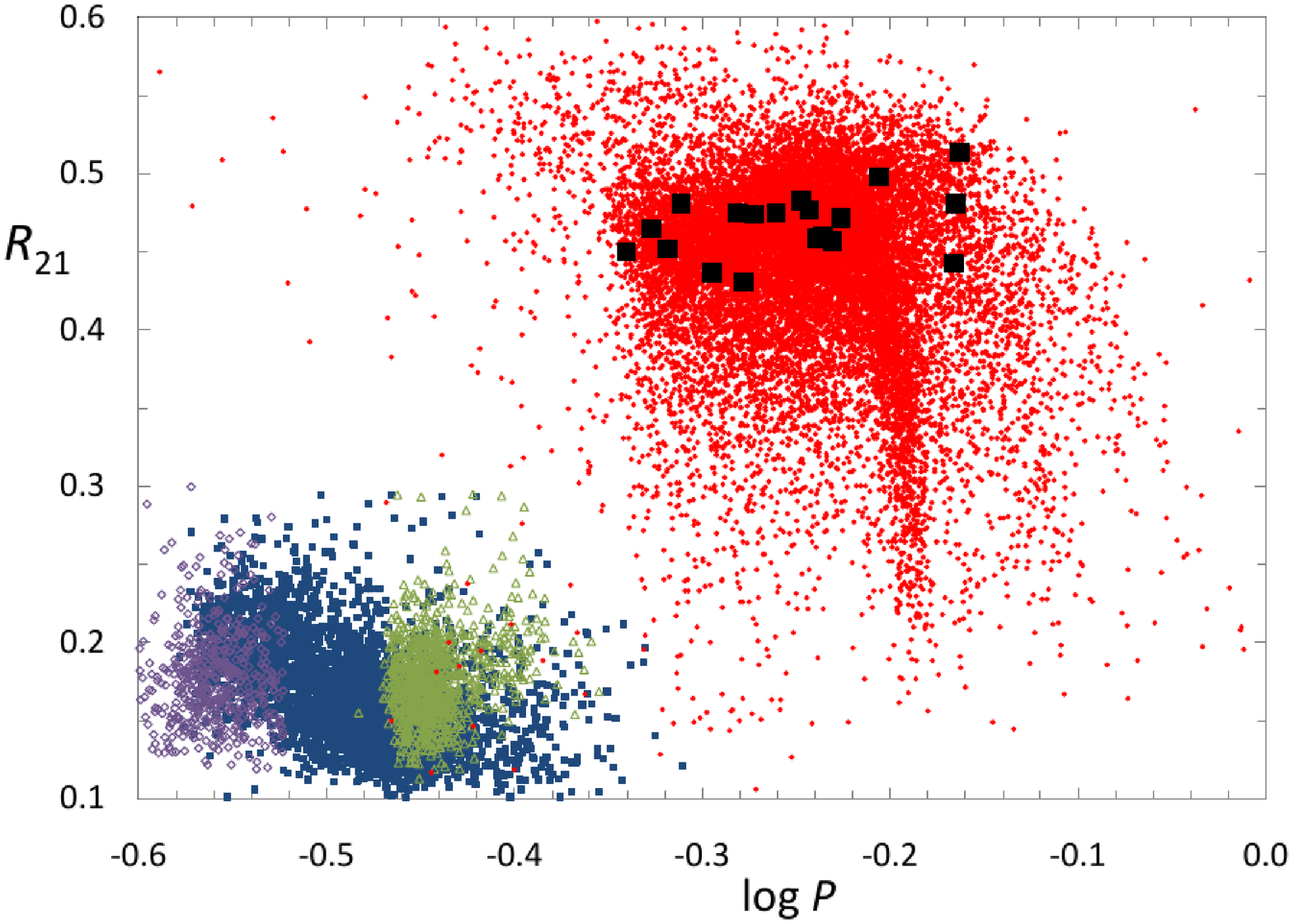} &
\includegraphics[width=8cm] {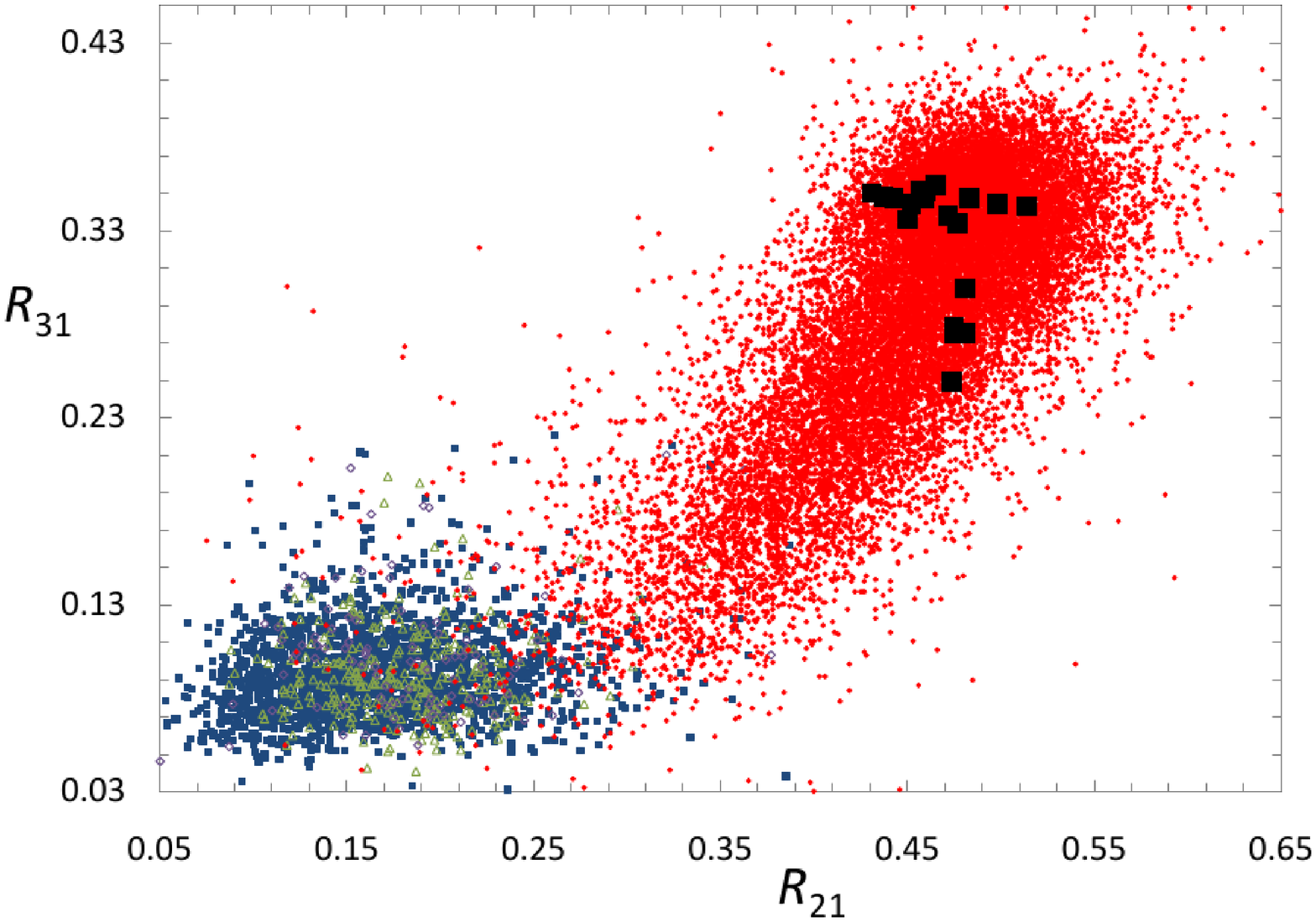} \\
\includegraphics[width=8cm] {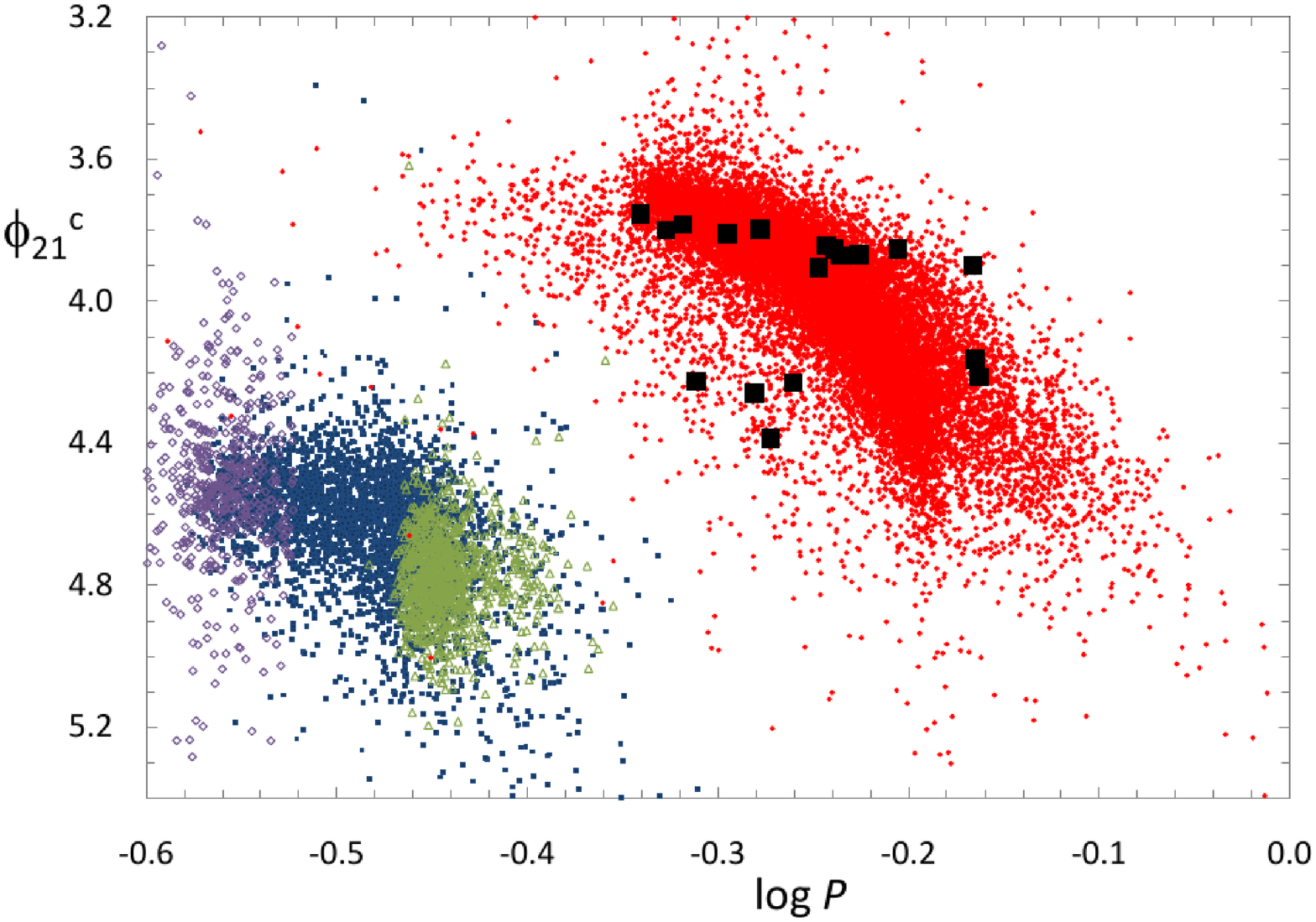} &
\includegraphics[width=8cm] {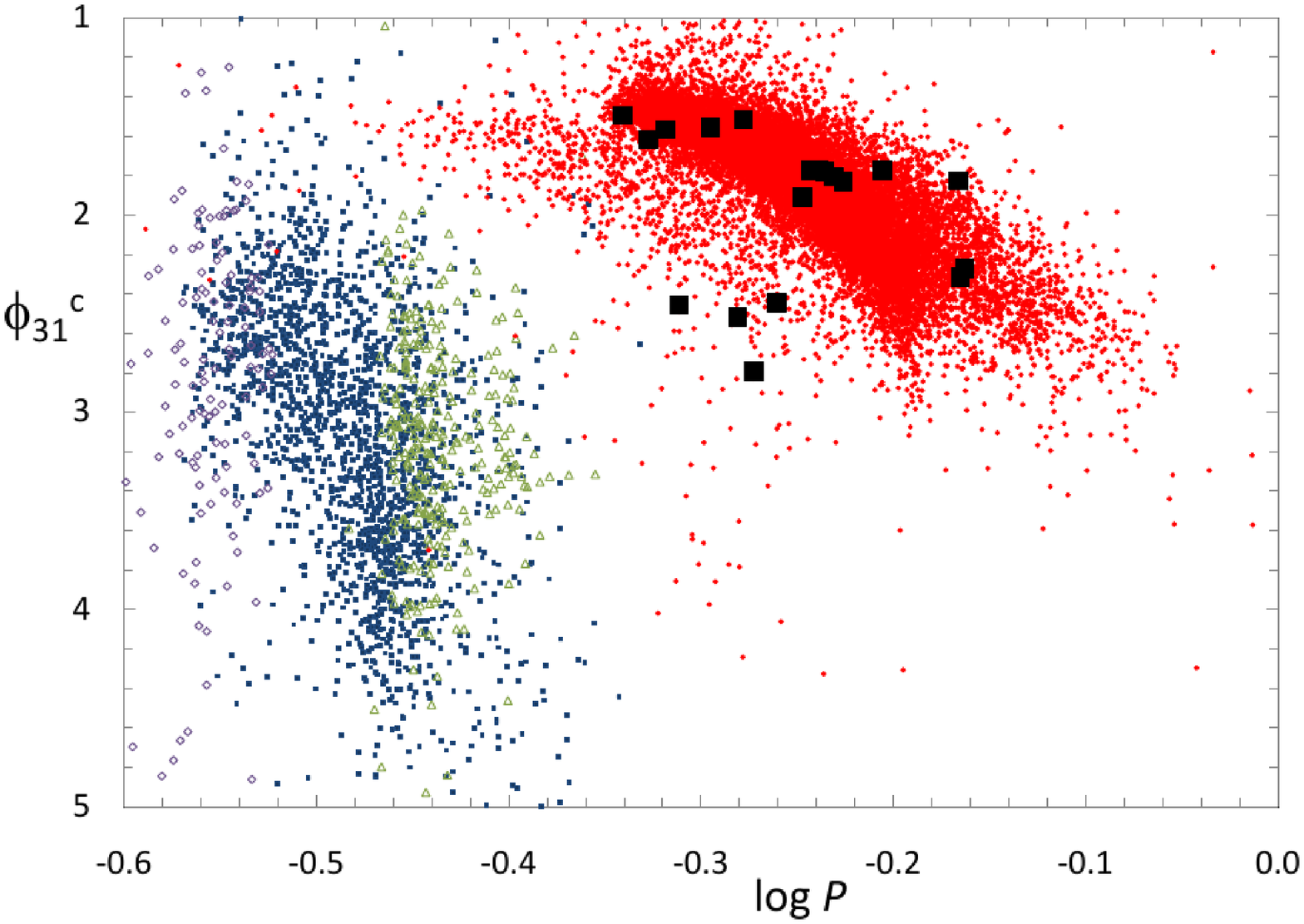}
\end{array}$
\end{center}
\caption{Four panels comparing Fourier parameters ($V$-band) for the {\it Kepler} 
non-Blazhko ab-type RR~Lyr stars (large black squares) and for
24905 RR~Lyr stars in the central regions of the Large Magellanic Cloud.  
The LMC data are 
from the Soszynski {\it et al.} (2009) OGLE-III study, with colour and symbol coding as follows: 
red filled dots (17693 RRab stars);  blue filled squares (4957 RRc stars);  green open triangles (986 RRd stars), 
and purple open diamonds (1269 RRe stars). 
Because of the limited $x$- and $y$-ranges not all of the LMC RR~Lyr stars are represented in the graphs.  }
\label{OGLE-III}
\end{figure*}

Our approach to estimating the {\it Kp}-$V$ offsets was simply to average the 
observed differences for AW~Dra, FN~Lyr and NR~Lyr,
taking the standard deviation as a measure of the uncertainty. 
In this way we arrived at the following {\it Kp}-$V$ transformations: 
\begin{equation}
\renewcommand{\arraystretch}{1.2}\begin{array}{l}
      A_{\rm tot}(V) = A_{\rm tot}({Kp}) + (0.14\pm0.01), \\
      {\rm RT}(V) = {\rm RT}({Kp}) + (0.002\pm0.006), \\
      A_1(V) = A_1({Kp}) + (0.054\pm0.002),   \\
      A_3(V) = A_3({Kp}) + (0.018\pm0.003),   \\
      R_{21}(V) = R_{21}({Kp}) - (0.013\pm0.004),       \\
      R_{31}(V) = R_{31}({Kp}) - (0.004\pm0.009),       \\
      \phi_{21}(V) = \phi_{21}({Kp}) - (0.089\pm0.021), \\
      \phi_{31}(V) = \phi_{31}({Kp}) - (0.151\pm0.026). \\
\label{eq}
\end{array}
\end{equation} 
The most uncertain of these equations are those for RT and $R_{31}$.
While the percent uncertainties are large the offsets are small; and besides,
these two quantities are rarely used in correlations to derive physical properties.  
 
A comparison of the ASAS $A_{\rm tot}(V)$ values (given in Table~2, column~7) with the 
{\it Kp} values (given in column~8 of Table~1) is not helpful since several of 
the stars show larger than expected differences (for example, V838~Cyg).
The accuracy of the above transformation equations is expected to improve 
with future $BVI$ photometry.  In $\S6$ these offset equations are used to convert 
the well-established $V$-band correlations relating
Fourier parameters and physical characteristics to {\it Kp} correlations.

\subsection{{\it Kepler} vs. globular cluster RR Lyr Stars}

In {\bf Figure~7} the Fourier parameters 
for the 19 non-Blazhko stars in the {\it Kepler} field (derived using the {\it Kp} photometry
but transformed to $V$ values using the offsets given in Eq.~2) 
are compared with parameters derived from $V$-photometry for 177 RR~Lyr stars in several well-studied globular clusters (from Kov\'acs \& Walker 2001). 
The same symbols and colour coding are used in all four panels.

The upper-left panel is analogous to the $P$-$A_{\rm tot}$ diagram and can be used to derive metallicities (see $\S$6.1 below).  
The two diagonal lines represent mean relations when the globular cluster data (small black dots) are sorted into two [Fe/H]$_{\rm BB}$ bins (here `BB' refers to the 
Butler-Blanco system, which approximates the Carretta-Gratton system - see below): 
a metal-poor bin consisting of 19 stars (surrounded by blue squares) with metallicities between $-1.70$ and $-1.99$ dex and average $-1.8$ dex (BB-scale);  and 
an intermediate-metallicity bin consisting of 39 stars (circled in red) with [Fe/H]$_{\rm BB}$ between $-0.97$ and $-1.23$ dex and average $-1.1$ dex.
The equations of the lines are: $\phi_{31}^c = 5.556 \thinspace {\rm log}P + 0.2920$ (upper metal-poor bin)
and  $\phi_{31}^c = 6.200 \thinspace {\rm log}P + 3.615$ (lower intermediate metallicity bin).
Four of the {\it Kepler} non-Blazhko stars are clearly richer than [Fe/H]$_{\rm BB}$=$-1.1$ dex, 
while the remainder are apparently more metal poor.

The axes of the other three panels in Fig.~7 are similar to those in Fig.~6.
The upper-right panel shows that the {\it Kepler} stars appear to be drawn from a distribution similar to
that of the globular clusters.  In particular, the {\it Kepler} stars with low $R_{31}$ values are not unusual,
except possibly that they all have relatively high $R_{21}$ values.
Since the globular cluster stars of higher metallicity (red open circles) all tend to reside on the right side of the 
diagram this separation is probably a metallicity effect,  supporting our conclusion that 
V784~Cyg, V782~Cyg, KIC~6100702 and V2470~Cyg are metal rich.
Likewise, the lower-left panel shows that the majority of the {\it Kepler} stars do not differ 
from the stars in globular clusters.
Note too that there is very little metallicity discrimination in this plane.
The two stars located at the extreme upper edge of the envelope of the globular cluster distribution are AW~Dra and V784~Cyg.
Finally, the lower-right panel shows close agreement between the phase parameters of the {\it Kepler} and globular 
cluster RR~Lyr stars, which supports the conclusion  drawn earlier (lower-right panel in Figure~6) that
there is a strong approximately-linear correlation between $\phi_{21}^s$ and $\phi_{31}^s$.
The  diagram also shows very little  dependence on metallicity.

\subsection{{\it Kepler} vs. LMC field RR Lyr stars}

In {\bf Figure~8} we compare the  $R_{21}$,  $R_{31}$,  $\phi_{21}^c$ and $\phi_{31}^c$ 
Fourier parameters for the 19 {\it Kepler} non-Blazhko RR~Lyr stars 
with those for the field RR~Lyr stars in the central regions of the Large Magellanic Cloud.  
The LMC data are from the massive OGLE-III Catalogue of Variable Stars by Soszynski {\it et al.} (2009), which comprises 
almost 25000 RR~Lyr stars.  All the parameters are $V$-band values, the LMC values having been
transformed from $I$ to $V$ using the transformation equations given by 
Morgan, Simet \& Bargenquast (1998), and the values for the  
{\it Kepler} RR~Lyr stars from {\it Kp} to $V$ using the Eq.~2 offsets.  

The upper-left panel shows that the {\it Kepler} stars have $R_{21}$ values in the narrow range 0.43 to 0.51, which
is much smaller than the range seen for the LMC stars, 0.25 to 0.55.   
In particular, none of the {\it Kepler} RRab stars are among the LMC RRab 
stars in the `$R_{21}$ dropdown' at ${\rm log}P \sim -0.19$, {\it i.e.,} $P \sim 0.65$~d. 
Note that the upper-left panel of Fig.~6 can be used to identify individual {\it Kepler} stars.  

The $R_{21}$ and $R_{31}$ amplitude ratios are compared in the upper-right panel of Fig.~8.
This graph, which is similar to the upper-right panels of Figs.~6 and 7 (and again, Fig.6 can be used to
identify individual {\it Kepler} stars), further puts into perspective the small ranges of $R_{21}$ and $R_{31}$ seen for the {\it Kepler} 
RRab stars, both relative to that for the LMC RRab stars, and relative to the ranges seen for the  
other types of RR~Lyr stars (RRc, RRd and RRe). 
  
The bottom panels of Fig.~8 are both `metallicity diagnostic' diagrams, such as those discussed
in $\S6.1$ below.  We have already seen (upper-left panel of Fig.~7) that the {\it Kepler} stars sort into 
four metal-rich and 15 metal-poor stars, that the metal-poor stars are similar to the 
RRab stars found in both Oosterhoff type I and type II galactic globular clusters, and that 
the metal-rich stars have no counterparts among the 
Kov\'acs \& Walker (2001) sample of globular cluster RR~Lyr stars.   Here we see 
that both the metal-rich {\it Kepler} stars (the clump of four stars at log$P$$\sim$$-0.3$ and $\phi_{21}^c$$\sim$$4.3$)
and the metal-poor {\it Kepler} stars have counterparts in the LMC, and that the metal-rich stars appear to have relatively
long periods for metal-rich RRab stars ({\it i.e.}, the subgroup of LMC RRab stars with shorter periods at a given
$\phi_{31}^c$).

\section{PHYSICAL CHARACTERISTICS} 

RR~Lyrae stars provide fundamental insight into the late evolution of low mass stars, and in particular
the instability strip (IS) region of the horizontal branch (HB).  Being present in many globular clusters (and galaxies) 
exhibiting different HB types\footnote[4]{The HB type of a globular cluster is defined as (B$-$R)/(B+V+R), where B, R and V are the
number of blue HB stars in the cluster, the number of red HB stars, and the number of RR~Lyr stars, respectively (Lee 1989).
A table of values is given in Lee, Demarque \& Zinn (1994).} 
their properties are related to the evolutionary and chemical histories of these systems.  Thus it is desireable 
to derive (from photometry, spectroscopy and detailed pulsation and stellar evolution models) their physical 
characteristics, such as the mass $\cal M$, luminosity $L$, effective temperature $T_{\rm eff}$, 
iron abundance [Fe/H], chemical composition (X,Y,Z), age, {\it etc.} 
An equally important goal is to understand the interdependencies of these quantities.
Significant early contributions are summarized in the papers by Christy (1966), Iben (1971), Smith (1995), 
Sandage \& Tammann (2006, hereafter ST6), and Catelan (2009).  This knowledge becomes critical    
when RR~Lyr stars (and Cepheids) are used to derive {\it accurate} distances within and beyond our Galaxy (see Tammann, Sandage \& Reindl 2008, hereafter TSR8; 
Sandage \& Tammann 2008, hereafter ST8).

Of course high-dispersion spectroscopy permits direct measurements to be made of some of these quantities;  
however, such observations are time-consuming and often impractical.  
Thus efforts have been on-going to derive physical characteristics from photometry only, in particular from 
pulsation periods, mean magnitudes and colours, quantities that characterize the shapes of the light-curves
({\it e.g.}, amplitudes, risetimes, Fourier phase parameters), and most recently, from fitting detailed light curve shapes
to nonlinear convective pulsation models (see Bono, Castellani \& Marconi 2000, Marconi \& Clementini 2005, 
Marconi \& Degl'Innocenti, 2007).

A potential difficulty for the present Fourier investigation is that almost all the available calibration
relations use empirical correlations derived from $V$-band photometry,  while for the stars studied here we 
have {\it Kp} photometry, and only a limited amount of high-precision $V$ photometry and colour information. 
Furthermore, the mean {\it Kp} magnitudes from 
the KIC are somewhat uncertain for giant stars.
Despite these limitations we have shown in the previous section, at least for AW~Dra, FN~Lyr and NR~Lyr, three 
RR~Lyr stars with quite different periods and light curve shapes, that differences between the  $V$ and {\it Kp} 
Fourier parameters appear to be small and systematic. For this reason we have chosen to present 
approximate physical characteristics based on the extant $V$-band correlations applied 
to {\it Kp} correlations transformed using the observed {\it Kp}-$V$ parameter offsets.  
Our results are summarized in Tables~4-6 and their derivation is discussed next.

\subsection{Iron Abundances}

\subsubsection{Background: {\rm [Fe/H]} from P-shifts in P-A diagram}

Starting with the papers by Oosterhoff (1939, 1944), Arp (1955) and Preston (1959), the
period-amplitude ($P$-$A$) diagram, also known as the `Bailey diagram', has been found to be a useful 
tool for deriving iron-to-hydrogen ratios for field and cluster/galaxy RR~Lyr stars. 
The amplitude usually employed is the $B$ or $V$ total amplitude, $A_{\rm tot}$, and often the abscissa 
is log$P$.  The basic $P$-$A_{\rm tot}$-[Fe/H] correlation is such that
for a given value of $A_{\rm tot}$ lower-metallicity stars tend to have longer periods than higher-metallicity stars.
Sandage (2004, hereafter S04) has called this the `OAP period-metallicity correlation'.
Other recent discussions of the $P$-$A$ diagram and period shifts are given by
Di Criscienzo {\it et al.} (2004), Bono {\it et al.} (2007, hereafter BCD7) and Sandage (2006 and 2010, hereafter S06 and S10).

The basic explanation for  $P$-shifts in the Bailey diagram follows from the `stacked HB luminosity levels' model 
first proposed by Sandage (1958), the latest refinement of which is given in S10 (see his figs.~1 and 3).   
Owing to the Ritter (1879) pulsation relation,  $P \sqrt \rho = Q$ (where Q is a
constant over a wide range of $L$, $T_{\rm eff}$ and [Fe/H]), lines 
of constant density ($\rho$), and therefore period, cut diagonally across HBs of different luminosity
(see fig.13 of Sandage, Katem \& Sandage 1981, hereafter SKS).  When coupled with a monotonic $A_{\rm tot}$-$T_{\rm eff}$ correlation (in the 
sense that $A_{\rm tot}$ increases with distance from the red edge of the IS -- see, for example, fig.12 of SKS, and fig.12 of Di Criscienzo {\it et al.} 2004)   
one expects to see at a given $A_{\rm tot}$ more luminous HB stars having longer periods.  In fact, $P$-shifts and correlations such 
as the log$P$-$A_{\rm tot}$-[Fe/H] correlation are expected in all diagrams where the
abscissa is pulsation period and the ordinate is a light curve descriptor that correlates with $A_{\rm tot}$.  Two such descriptors 
are the risetime (RT) and the Fourier $\phi_{\rm 31}$ phase parameter (defined in $\S4.2$) -- see Fig.10 below. 
For example, in the RT {\it vs.} log$P$ diagram the $P$-shift at a given RT between M3 and M15 amounts to $\Delta$log$P = 0.055$ (see fig.10 of SKS).
Period shifting can also be seen between M3 and M2 in the $\phi_{\rm 31}$ {\it vs.} period diagram (see fig.2 of Jurcsik {\it et al.} 2003), and 
between metal-rich and metal-poor field RR~Lyr stars (see fig.3 of S04). 
Thus, such correlations have the potential for measuring [Fe/H] (S04; Kovacs 2005;
Cacciari, Corwin \& Carney 2005, hereafter CCC5;  S06,S10; Sodor, Jurcsik \& Szeidl 2009). 

Identifying the less luminous HB in the Sandage model with that of the unevolved zero-age horizontal branch (ZAHB) stars 
in the IS of a relatively metal-rich system, and the more luminous HB with the same near-ZAHB stars in 
the IS of a more metal-poor system, leads directly to the OAP period-metallicity correlation. 
Indeed, ZAHB stellar evolution models clearly show that the mean luminosity levels in the IS for metal-poor ZAHB stars are higher than 
those for more metal-rich ZAHB stars  ({\it e.g.}, Faulkner 1966;  Faulkner \& Iben 1966;  Iben \& Faulkner 1968;  
Hartwick, H\"{a}rm \& Schwarzschild 1968;  Iben \& Rood 1970;  Iben 1971; Sweigart \& Gross 1976; 
Lee, Demarque \& Zinn 1990,1994, hereafter LDZ90 and LDZ94;  Dorman 1992;  VandenBerg {\it et al.} 2000;  Demarque {\it et al.} 2000, hereafter D00; 
Cassisi {\it et al.} 2004;  Pietrinferni {\it et al.} 2004, 2009; Dotter {\it et al.} 2007, {\it etc.}). 
Indeed, in the IS at log$T_{\rm eff}=3.85$ the M15 HB is $\sim$0.22 mag brighter than that for M3 (SKS). 
And, for [Fe/H] values ranging from $-2.3$ to $-0.5$ the mean absolute magnitudes for the RR~Lyr stars
range from $\langle M_{\rm V}{\rm (RR)} \rangle \sim 0.40$ to 0.85 mag,
with, for all but the most extreme clusters, little dependence on HB type (see figs.1,2 of D00;
and fig.3 of Cassisi {\it et al.} 2004).
The models also show that in the instability strip at the luminosity level of RR~Lyr stars the variation in $L$(ZAHB) is 
due primarily to a difference in mean mass, with  $\langle {\cal M} \rangle$ being higher in more metal-poor systems\footnote[5]{Based on the $A_{\rm tot}$-$T_{\rm eff}$ correlation
one expects also to see period shifting in the log$T_{\rm eff}$ {\it vs.} log$P$ diagram, and indeed fig.11 of SKS 
shows at `any given temperature' an observed shift between M3 and M15 of $\Delta$log$P$=0.070. 
However, when comparing model predictions with observations one must `let the temperature  be cooler for more metal-poor RR~Lyr
stars at their higher luminosities' (Simon \& Clement 1993, S93, S06).}.

\subsubsection{Period shifts caused by post-ZAHB evolution}

Despite the general success of the Bailey diagram as a tool for deriving [Fe/H] values 
it has become apparent since the early 1980's that simple correlations of 
period-shift with metallicity do not exist in all cases and factors other than metal abundance  
play a role in the interpretation of these diagrams and correlations. Chief among these effects
is evolution away from the ZAHB.

All HB stellar evolution models show tracks leading away from the ZAHB toward the asymptotic giant branch,
with HB lifetimes $\sim$50-100 Myr (core helium burning phase). 
While the evolution can be either redward or blueward it almost always is in the direction of higher luminosity, with 
the pace quickening as the evolution progresses.  As $L$ increases the  
period becomes longer and larger $P$-shifts are seen in the $P$-$A$ diagram. 
However, most HB stars are expected to be located near to 
the ZAHB, with fewer stars in advanced evolutionary phases (CCC5 put the 
fraction of evolved RR~Lyr stars in M3 at $\sim$14\%).  Indeed skewed HB {\it and} RR~Lyrae luminosity distributions are observed 
in many globular clusters ({\it e.g.}, fig.15 of Sandage 1990, hereafter S90).
Such skewed distributions are also seen in simulations of HB and RR~Lyr stars (see fig.10 of Marconi {\it et al.} 2003, and fig.3 of Catelan 2004),
and in period shift diagrams (see fig.6 of Kunder \& Chaboyer 2009).

As a first step towards identifying and correcting for such evolutionary effects 
Sandage (1981a,b; hereafter S81a,b) introduced the concept of `reduced' period,  log$P'$ = log$P$+0.336$\Delta$$m_{\rm bol}$,
where $\Delta m_{\rm bol} = m_{\rm bol}$ minus the mean apparent magnitude of the bulk of the RR~Lyr
stars in a given cluster, and the multiplicative factor follows from the van Albada \& Baker (1971,1973) version of the 
$P \sqrt \rho = Q$ relation, and demonstrated the existence of a period-luminosity-amplitude relation for equal metallicity 
RR~Lyrae stars.  The net effect of making this correction is to reduce the period shifts of the evolved RR~Lyr stars,
resulting in tighter correlations in the $A_{\rm tot}$ vs. log$P'$ diagram.  In this way S81a,b identified several stars in 
Messier~3 that are extra-luminous with longer periods at a given $A_B$ than the bulk of the RRab stars\footnote[6]{Sandage's candidate evolved stars in
M3 include V24, V60, V65, V96 and V124, all of which can be seen (labelled) in the S81a fig.2 $m_{\rm bol}$ 
vs. log$T_{\rm eff}$ diagram and in the S81a fig.3 period-amplitude diagrams.  Subsequent studies of M3 by Kaluzny {\it et al.} (1998) and  
Clement \& Shelton (1999) confirmed V65 as an evolved star and also have identified V14 and V104 as post-ZAHB stars.
Post-ZAHB evolution in M3 was also investigated by Marconi {\it et al.} (2003) and by Jurcsik {\it et al.} (2003).
Jurcsik {\it et al.} used  ``luminosity bins'' and Fourier parameters and found that RR~Lyr ``stars close to the ZAHB show OoI type properties, while the
brightest stars have Oo~II statistics regarding their mean periods and RRab/RRc number ratios'', confirming the earlier result of Clement \& Shelton (1999). 
Thus, again, 
the Oosterhoff dichotomy seems to be connected with evolutionary effects. 
More recently, CCC5 confirmed V24 and V14 as evolved stars and identified three more post-ZAHB stars based on their long periods and relatively high luminosities (V3, V35, V67).
Most recently, Valcarce \& Catelan (2008) identified in M3 $\sim$15 non-Blazhko and another $\sim$10 Blazhko RR~Lyr stars in 
an advanced post-ZAHB evolutionary state (see their Fig.2).}. 
Many other globular clusters are now known to contain extra-luminous large-period-shift stars, including
47~Tuc (V9 -- see Storm {\it et al.} 1994), NGC~6388 and NGC~6441 (Pritzl {\it et al.} 2000, 2001, 2002; Catelan {\it et al.} 2006),
Messier~5 (Kaluzny {\it et al.} 2000, CCC5),  NGC~5896 (Alves {\it et al.} 2001), IC~4499 (V9 and V54 -- see CCC5; Kunder {\it et al.} 2011; Walker {\it et al.} 2011), 
and Messier~13 (Sandquist {\it et al.} 2011).
  
The reduced-period concept is useful when deriving mean metallicities for ensemble populations
of RR~Lyr stars in globular clusters and galaxies ({\it e.g.}, Dall'Ora {\it et al.} 2003), and for field stars
that are near the ZAHB (which are assumed to be in the majority), but the method fails for evolved RR~Lyrae stars:
what looks in the Bailey diagram like a metal-poor ZAHB star might well be an evolved metal-rich 
star, {\it i.e.}, an evolved OoI star masquerading as an unevolved OoII star.   

Thus, the amount of $P$-shifting is determined by the amount of $L$-evolution, which in turn  
depends on the the HB morphology of the parent population (LDZ90, Lee 1990)\footnote[7]{The main parameter controlling the distribution of stars along the HBs of
globular clusters (GCs) is metal abundance (Sandage \& Wallerstein 1960), with metal-rich clusters (e.g., NGC~6356, 47~Tuc) 
having predominantly red HBs and metal-poor systems (e.g., M15, M9) having predominantly blue HBs. 
Since the discovery by Sandage \& Wildey (1967) and van den Bergh (1967) that there is not a unique 
correlation between [Fe/H] and HB morphology, and the recognition that the GCs with the bluest HBs are not the 
most metal poor clusters (Renzini 1983),
the search has been on for the so-called `second parameter'.
The two leading candidates are differences in age (see Rood \& Iben 1968; Lee, Demarque \& Zinn 1994; 
Lee {\it et al.} 1999, 2007; Rey {\it et al.} 2001; Dotter {\it et al.} 2010), and variations in the 
initial helium abundance (see Sandage \& Wildey 1967, 
van den Bergh 1967, Hartwick 1968, Busso {\it et al.} 2007).
Other candidates include variations in the amount of mass loss along the red giant branch (RGB),
rotation rates, and the amount of helium mixing deep inside progenitor RGB stars (see Sweigart \& Catelan 1998). }.
When RR~Lyr stars originate from a position on the ZAHB that is {\it within} the IS,
which occurs in OoI clusters with HB type near zero, such as M3 and M4 with HB types (from LDZ94) of $0.08\pm0.04$ and $-0.07\pm0.10$, respectively,   
the evolution manifests itself in the IS region of the HR diagram 
as a vertical widening of the HB with RR~Lyr stars of similar mass spread over a range of $L$ and $T_{\rm eff}$ (see S90; 
and fig.1 of Jurscik {\it et al.} 2003, hereafter J03; and fig.2 of Valcarce \& Catelan 2008). 
On the other hand, when the original location on the ZAHB was {\it blueward and outside of } the IS, 
which occurs in OoII clusters with HB types that are large and positive, such as M15
(see Sobeck {\it et al.} 2011),  M92 (see Roederer \& Sneden 2011) and M2 (Lee \& Carney 1999) with 
HB types of $0.72\pm0.10$, $0.88\pm0.08$ and $0.96\pm0.10$ (all from LDZ94), respectively,   
most if not all the RR~Lyr stars are evolved and few if any lower-luminosity ZAHB stars are 
present (see LDZ90, D00).  In the case of M2  Lee \& Carney (1999) concluded that its RR~Lyrae 
stars are $\sim$0.2 mag more luminous than those in M3, owing to all the
M2 stars having evolved away from the blue side of the HB, whilst the bulk of the M3 stars lie near the ZAHB. 
Additional observational support for this idea comes from the clear correlation 
of metallicity and HB intrinsic vertical width, with more metal-rich systems having greater vertical widths
(see Fig.16 of S90).
Schematic tracks (along with grid lines of constant period and amplitude) comparing the post-ZAHB evolution of stars in M2 (and M15) with the 
tracks for M3 (and NGC6441) are given in Fig.~3 of S10.
Completing the picture, in clusters with red HBs ({\it i.e.}, HB types that are large and negative, such as 47~Tuc and M107, 
with HB types of $-1.0\pm0.03$ and $-0.76\pm0.08$, respectively) the evolution occurs mainly redward
of the IS and few, if any, RR~Lyr stars are expected 
(exceptions to this rule include systems with red HBs and extended blue HBs, such as NGC~6388 and NGC~6441, 
-- see references above).   
Thus, the luminosity of an individual RR~Lyr star in a given cluster depends not only on its metallicity
(usually assumed to be that of the parent cluster) but on its evolutionary state, which in turn is determined
by the HB-type of the parent cluster and the amount of evolution away from the ZAHB.

In closing, a potential solution to the evolution degeneracy problem in the Bailey diagram is to identify the evolved RR~Lyrae stars using  
some sort of light curve index, such as the JK96 `compatibility criterion' (see Kovacs \& Kanbur 1998, Clement \& Shelton 1999, Nemec 2004,
CCC5).  Of particular interest in this regard is fig.15 of CCC5, which
shows large $P$-shifts for several stars in the $A_{\rm B}$ {\it vs.} log$P$ diagram, but no such offset in the
$\phi_{\rm 31}$ {\it vs.} log$P$ diagram;  whether post-ZAHB stars stand out more in the $P$-$A$ diagram
than in the period-$\phi_{\rm 31}$ diagram remains an open question.  In any case it is important to
excercise caution when using the individual Fourier-based metallicities derived here.

\begin{figure*}
\begin{center}$
\begin{array}{cc}
\includegraphics[width=8cm] {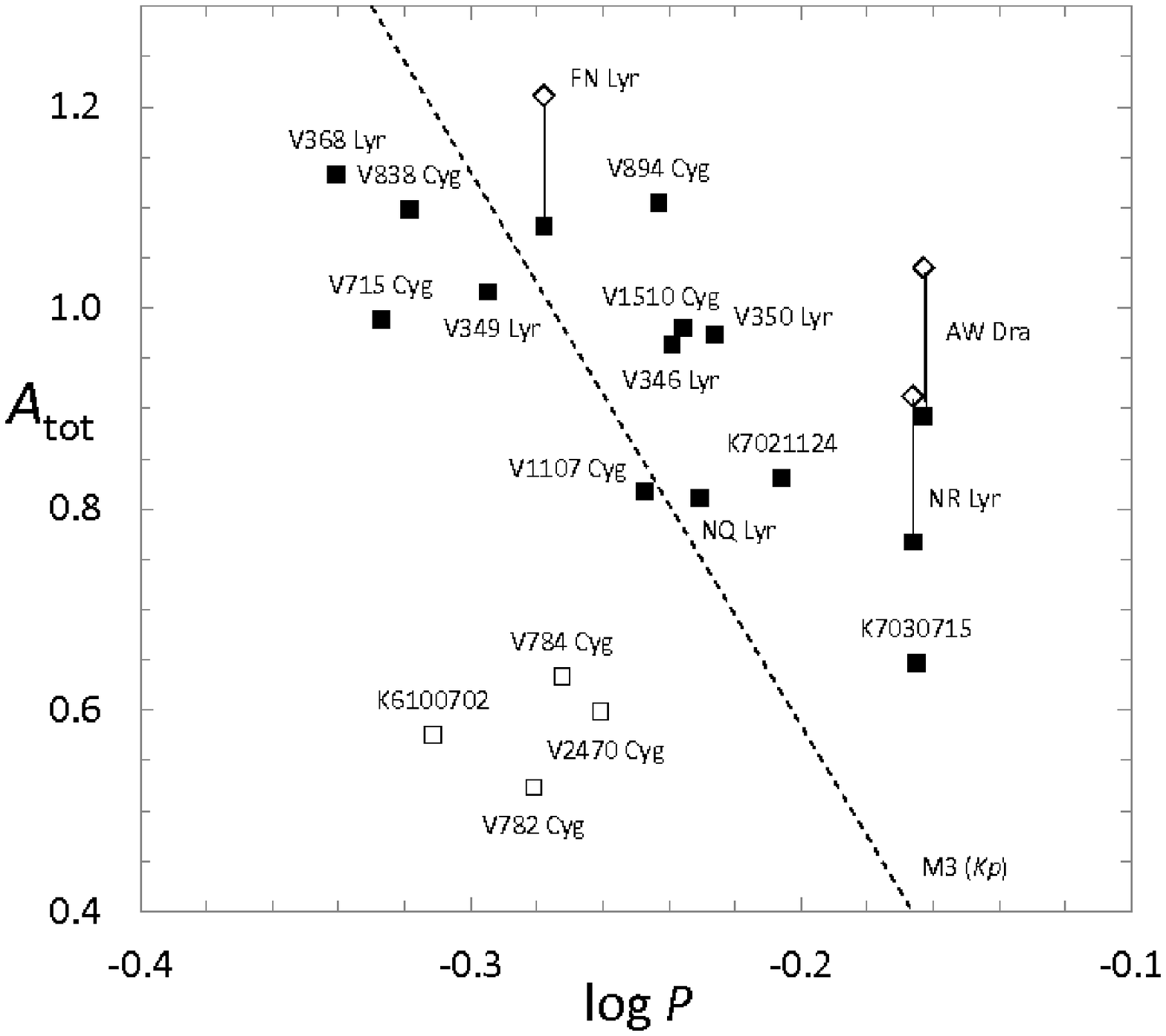} &
\includegraphics[width=8cm] {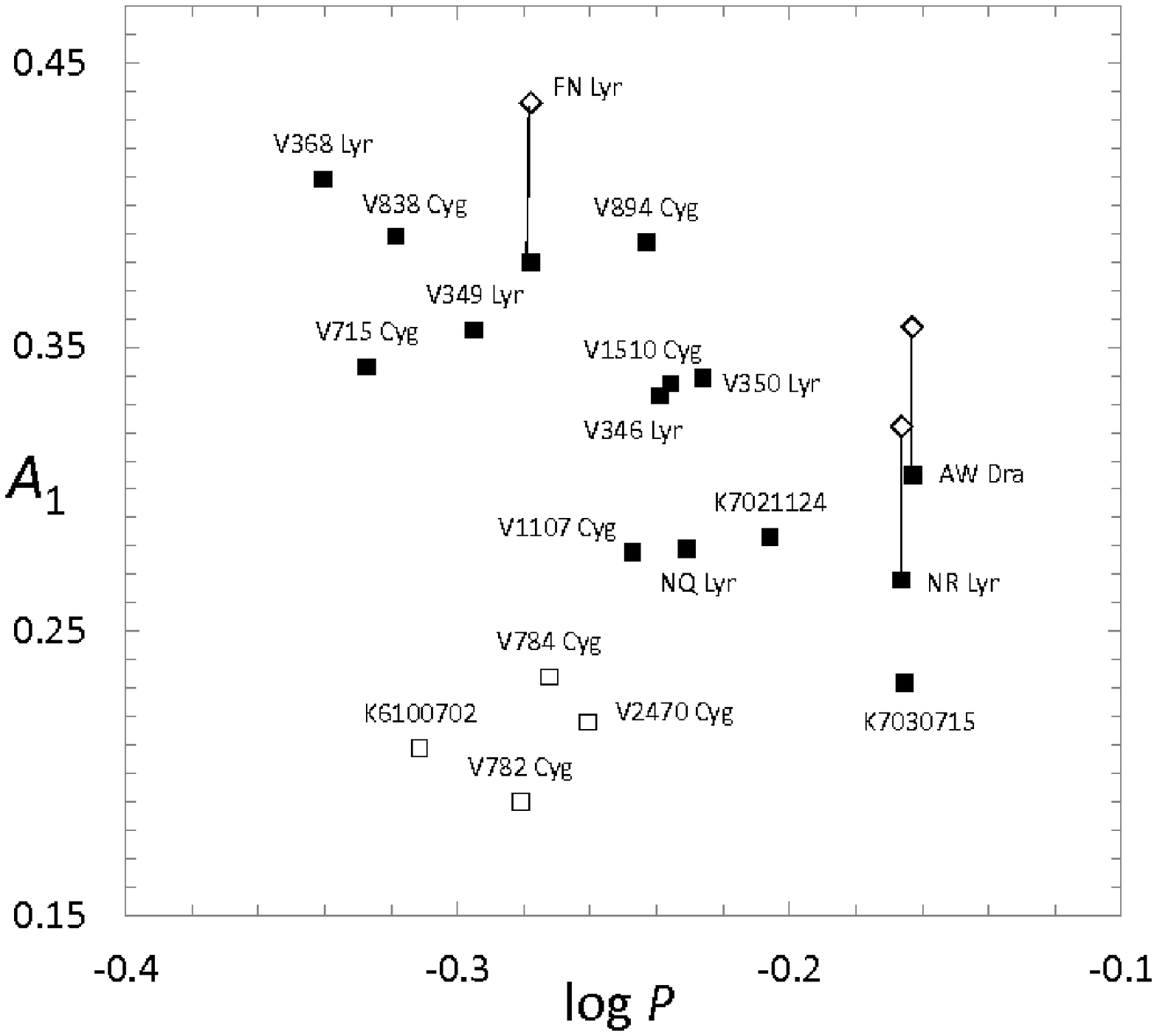} \\
\includegraphics[width=8cm] {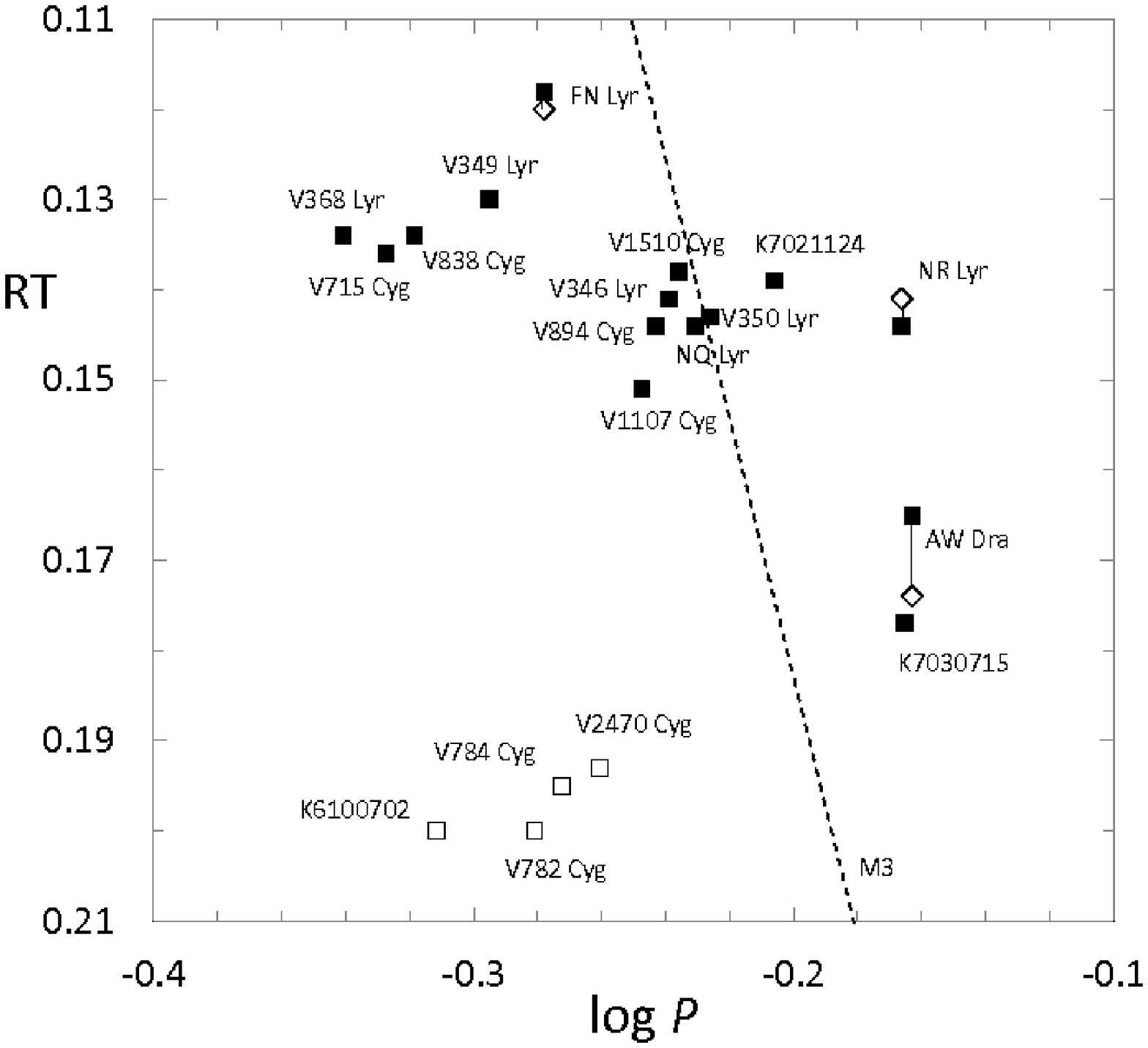} &
\includegraphics[width=8cm] {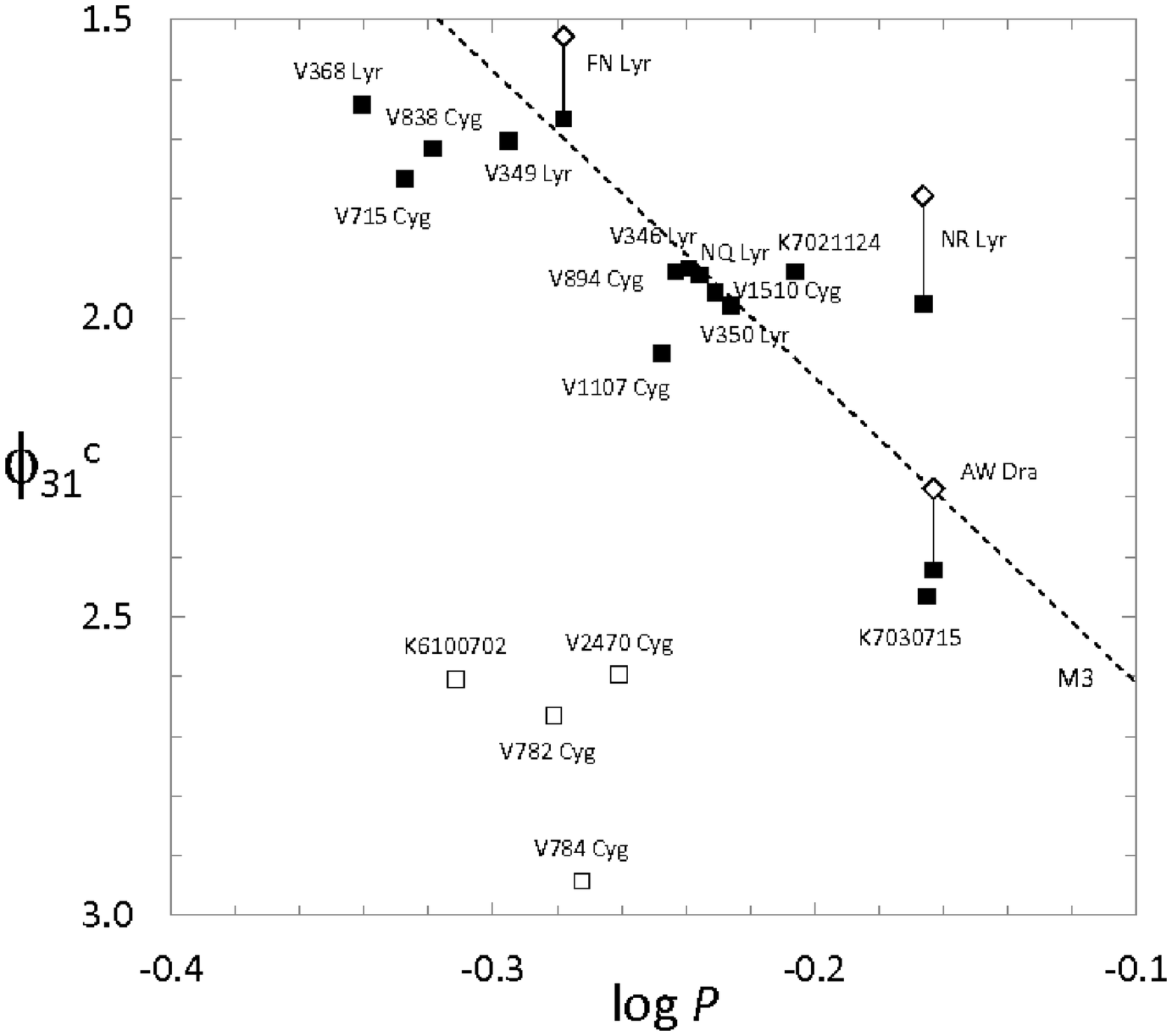}\\
\end{array}$
\end{center}
\caption{Period-amplitude and three other metallicity diagnostic diagrams for the 
{\it Kepler} non-Blazhko RRab stars.  For all four graphs the abscissa is log$P$,  
all the stars have been labelled, and the four candidate metal-rich stars are plotted
as open squares.  AW~Dra, FN~Lyr and NR~Lyr are plotted twice in each panel, with the points derived from the
{\it Kp} photometry (solid black squares) connected to the points derived from the  
high-precision $V$-band photometry (open diamonds) -- the consistency of the {\it Kp}-$V$ 
offsets seen here (and in Figs.6 and 10) is quite striking.
The diagonal dashed lines,  representing the relations for the M3 RR~Lyr stars  
(assumed to have [Fe/H]$_{\rm ZW} = -1.6$ dex), are from Equation~5 (S04eq6, top left),  Equation~6  (S04eq7, bottom left) 
and fig.15 of CCC5 (bottom right).  }
\label{MetallicityDiagnostics}
\end{figure*}

\subsubsection{Metallicity Scales}

For the {\it Kepler} non-Blazhko RRab stars we have used the correlations involving all three light curve descriptors to estimate [Fe/H] values
(see $\S6.1.4$).  The resulting metallicities are reported in {\bf Table~4}. 
The metal abundances based on the S04 equations are on the Zinn \& West (1984, hereafter ZW) scale, 
and those based on J98's equation~1 are on the Carretta \& Gratton (1997, hereafter CG) scale. 
For comparison purposes the latter ({\it i.e.}, the  $\phi_{31}^s$-metallicities) have been 
transformed to the ZW scale {\it and} to the Carretta {\it et al.} (2009, hereafter C9) scale.
Although several other metallicity systems have been proposed ({\it e.g.}, Layden 1994, Jurcsik 1995,  Butler \& Blanco - see S04, and Kraft \& Ivans 2003), 
the C9 scale is most extensive.  It is based on a homogeneous data set derived from $\sim$2000 modern spectra, 
and is gaining considerable favour -- see, for example, its adoption by Harris (1996) for his on-line globular cluster 
catalog at 'http://www.physics.mcmaster.ca/$\sim$harris/mwgc.dat'.
In addition, equations are provided for transformation between the various scales.  
When transforming between the ZW and CG systems the following equation (from S04, footnote~1) has been used:  
$ [{\rm Fe/H}]_{\rm ZW} = 1.05 \thinspace [{\rm Fe/H}]_{\rm CG} - 0.20 $.
And when transforming from the ZW and CG scales to the C9 scale we have used (from C9):
$ [{\rm Fe/H}]_{\rm C9} = 1.137 \thinspace [{\rm Fe/H}]_{\rm CG} - 0.003 $;
and
$ [{\rm Fe/H}]_{\rm C9} = -4.13 + 0.130 \thinspace [{\rm Fe/H}]_{\rm ZW} - 0.356 \thinspace [{\rm Fe/H}]_{\rm ZW}^2 $.

\subsubsection{Metallicities for the {\it Kepler} stars}

{\bf Figure~9} contains, for the {\it Kepler} non-Blazhko RRab stars, four such 
`metallicity diagnostic diagrams'.  The principal observation here is the similarity, to first order, of the distribution 
of the points in the four graphs.   
The top-left panel is the classical $P$-$A_{\rm tot}$ diagram, with total {\it Kp} amplitude plotted along the ordinate.  
Traditionally, lines of constant [Fe/H] serve as metallicity calibration lines; these run diagonally from the upper-left 
to the lower-right corner (as seen in three of the panels).  

The top-right panel has $A_1$ along the ordinate.  As expected, because $A_1$ and $A_{\rm tot}$ are highly correlated 
(see Fig.~2), the distribution of the points closely resembles that in the $A_{\rm tot}$ {\it vs.} log$P$ diagram.
In principle the small range of $A_1$ (only $\sim$0.2~mag) compared with that of $A_{\rm tot}$ 
(nearly 0.7~mag here, but often larger) makes the $P$-$A_1$ diagram less useful for estimating [Fe/H];  but with such high-precision data
as provided by {\it Kepler} the derived [Fe/H] results should, in principle, be similar. 

\begin{figure*}
\begin{center}$
\begin{array}{cc}
\includegraphics[width=8cm] {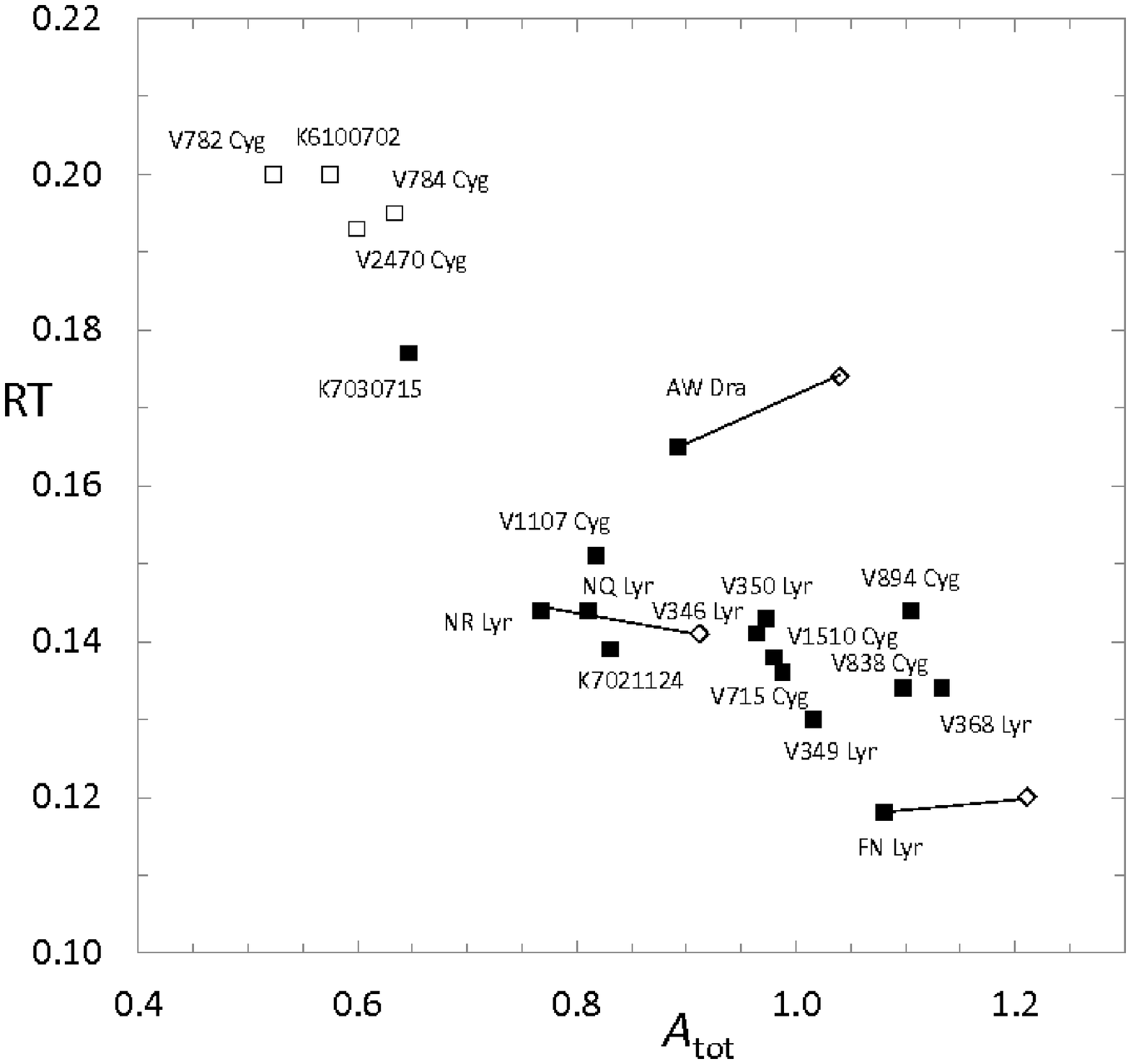} &
\includegraphics[width=8cm] {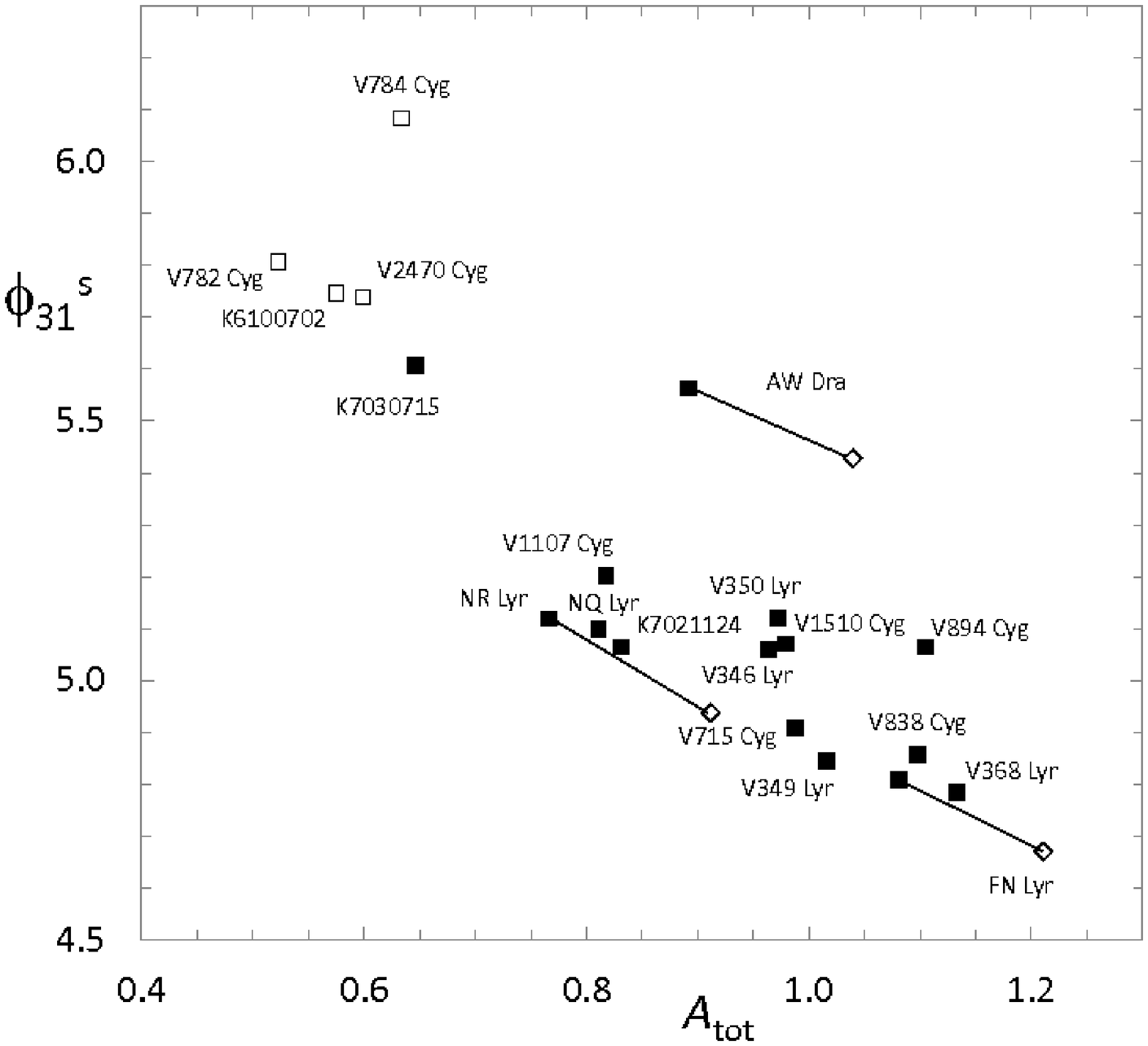}
\end{array}$
\end{center}
\caption{Risetime vs. total amplitude diagram (left panel) and Fourier phase parameter $\phi_{31}^s$ vs. total amplitude diagram (right panel)
for the {\it Kepler} non-Blazhko RRab stars.
The symbols are the same as in the bottom-right panel of Fig.9 ({\it i.e.}, the four metal-rich stars have been plotted with
open squares), and again AW~Dra, FN~Lyr and NR~Lyr are plotted twice in each panel, 
indicating the magnitudes and directions of the (systematic) $V$-{\it Kp} offsets.  }
\label{CorrelationDiagrams}
\end{figure*}

The bottom panels of Fig.~9 show risetime (left) and $\phi_{31}^c$ (right) plotted against log$P$, where 
both RT (reverse scale) and $\phi_{31}^c$ (reverse scale) increase with increasing log$P$.  
The bottom-right panel is the `{\it Kepler} non-Blazhko RRab only' version of the top-left panel of Fig.~7, 
and also is similar to Fig.~11.  The dashed line in the bottom-right panel
(from fig.15 of CCC5), the equation of which is $\phi_{\rm 31}^c = 3.124 + 5.128 {\rm log}P$,
represents the observed $V$-band relation for the RRab stars in M3 -- it is encouraging that the slope
of the line appears to agree with the slope of the {\it Kepler} metal-poor stars.
According to this calibration NR~Lyr is the most metal-poor star in the sample, followed by 
KIC~7021124 and FN~Lyr, and most of the RRab stars have metallicities similar to those in M3, which has
[Fe/H]$ = -1.50 \pm 0.05$ dex (Carretta {\it et al.} 2009 scale; W.~Harris' on-line catalog).   
Metallicities for the other stars, including the four stars suspected of having high [Fe/H] values, 
are discussed below.

{\bf Figure 10} contains the RT {\it vs.} $A_{\rm tot}$ diagram (left panel) and
the $\phi_{31}^s$  (which is $\pi$ larger than $\phi_{31}^c$) {\it vs.} $A_{\rm tot}$ diagram (right panel) for the {\it Kepler} stars. 
In general, as the total amplitude increases, both RT and $\phi_{31}^s$ decrease, with linear fits described
by the equations: $RT = -0.118 A_{\rm tot} + 0.256$ and $\phi_{\rm 31}^s = -1.75 A_{\rm tot} + 6.75$;  however,
owing to the large scatter about these lines, and the suspected large range in metallicities of the {\it Kepler} stars,
we believe these equations to be unreliable (and for this reason they were not plotted).  Theoretical models
(see the middle-left panels of Figs.14 and 15 below) suggest that it is more probable that there are families of lines (or curves), 
one for each metallicity.  Unfortunately, the present observational data 
are insufficient to establish these.    
It is noteworthy that AW~Dra, and to a lesser extent V784~Cyg and V894~Cyg, 
have larger $\phi_{31}^s$ values and larger RTs than other stars of comparable $A_{\rm tot}$.
And, while it appears that all the low-amplitude stars have large RTs and large $\phi_{\rm 31}^s$ values, 
this conclusion is based on only four low-amplitude candidate metal-rich stars and  
two low-amplitude long-period candidate metal-poor stars (KIC~7030715, AW~Dra), one of
which (AW~Dra) is a possible outlier in both graphs and which may be a long-period evolved star of 
intermediate-metallicity (if our preliminary CFHT spectra result is correct -- see $\S6.1.5$ below)
rather than a near-ZAHB star of low-metallicity.
In any case, both KIC~7030715 and AW~Dra have considerably longer periods
than the four high-metallicity stars. 


In {\bf Figure~11} the {\it Kepler} data are compared
with the galactic field star data analyzed by S04 (see his fig.~3).  
The field RR~Lyr stars are the same as those studied by Simon \& Lee (1981), 
Simon \& Teays (1982) and Simon (1988), the details of which are given in table~1 of S04.
Fig.~11 is comparable to  the bottom-right panel of our Fig.~9, which  shows only the
{\it Kepler} non-Blazhko stars. 
Because the Simon {\it et al.} stars
have known [Fe/H]$_{\rm ZW}$ values from Layden (1994) it is 
possible to sort them by metallicity.  Following Sandage we plotted 
two subsets of the data, which have average metallicities $\langle$[Fe/H]$\rangle = -0.35$ and 
$-1.63$ dex (ZW scale);
the diagonal lines were obtained by substituting these values into eq.~3 of S04. 
This diagram is also the galactic field star version of the upper-left panel of Figure 7, which included
RR~Lyr stars in galactic and Magellanic Cloud globular clusters.
The chief difference is that the calibration has now been extended to more metal-rich stars.

Fig.~11 shows that the stars KIC~6100702, V2470~Cyg, V782~Cyg and V785~Cyg, 
all of which have long risetimes, are solidly in the metal-rich region of the diagram,
with the highest $\phi_{31}^c$ values and all with periods shorter than
0.55~d.  The other {\it Kepler} stars are a better match to the
field stars that have  [Fe/H]$_{\rm ZW}$=$-1.63$ dex.  AW~Dra and KIC~7030715, with long periods,
intermediate risetimes, and high $\phi_{31}$ values, are seen to have metallicities $\sim -1.64$ dex (ZW scale).
Three of the four metal-rich stars (V782~Cyg, KIC~6100702 
and V2470~Cyg) have [Fe/H]$_{\rm ZW} = -0.4$ dex, and V784~Cyg is  the most metal-rich star
in our sample with [Fe/H]$_{\rm ZW} = -0.2$ dex. The most metal-poor stars are seen to be
more metal-poor than [Fe/H]$_{\rm ZW} = -1.6$ dex.

Equation 1 of J98, which is identical to equation~3 of Jurcsik \& Kov\'acs 1996, 
describes lines of constant [Fe/H]$_{\rm CG}$ in  
the period-$\phi_{31}^s$($V$) diagram.  
Using the offset formula for 
$\phi_{31}^s$ given in Eq.~2 the {\it Kp} version of the J98 equation becomes
\begin{equation}
\renewcommand{\arraystretch}{1.2}\begin{array}{l}
  {\rm [Fe/H]_{CG}}  = -5.241 - 5.394 \thinspace P + 1.345 \thinspace \phi_{31}^s({\it Kp}).  
\label{eq}
\end{array}
\end{equation}
Metallicities calculated with this formula are listed in column~2 of Table~4 under 
`J98eq1'.  For comparison purposes the corresponding values on the ZW and C9 scales 
also are listed.

Our second estimate of [Fe/H] is derived from the  
[Fe/H]$_{\rm ZW}$ - log$P$ - $\phi_{31}^c$ relation of S04 (his eq.~3 - note that the
cosine version of the Fourier phase parameter is employed here) -- see lower-right panel of Fig.~9.
Applying the  $\phi_{31}^c$ offset given in Eq.~2, Sandage's relation becomes
\begin{equation}
{\rm [Fe/H]_{\rm ZW}}  = -6.217 -7.012 \thinspace {\rm log}P +  1.411 \thinspace  \phi_{31}^c({\it Kp}), 
\end{equation}
where the intercept term  has a total uncertainty of
$\pm$0.023, and the 
$\phi_{31}^c$ and log$P$  coefficients have respective uncertainties 
of $\pm$0.014 and $\pm$0.071 (from S04).  Metallicities derived from Eq.~4  are listed in Table~4 under
`S4eq3'.

\begin{figure}
\centering
\includegraphics[width=8cm] {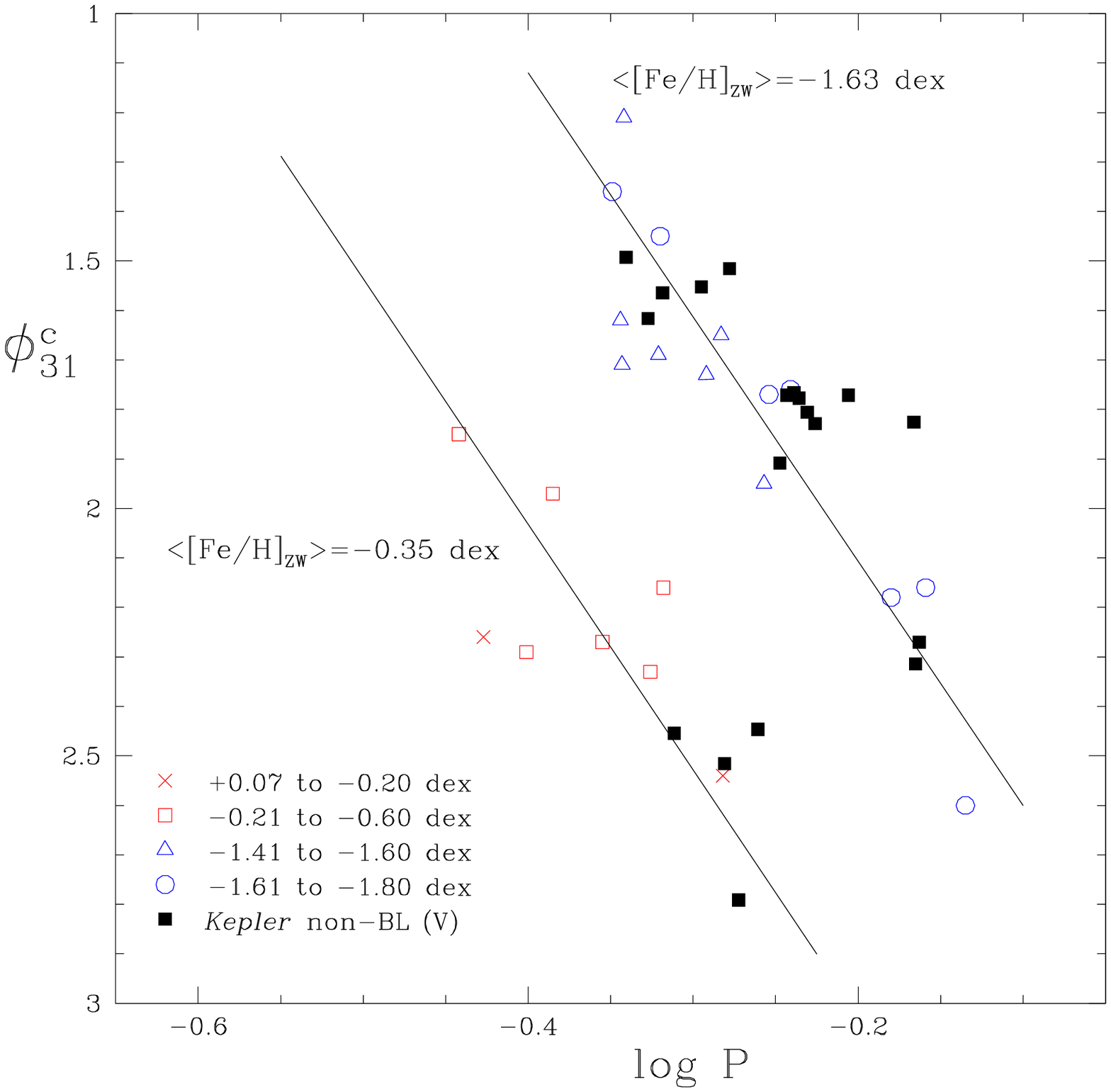} 
\caption{Period-$\phi_{31}^c$ diagram, comparing the {\it Kepler} 
non-Blazhko RR~Lyr stars and the Lub-Simon galactic field stars (from S04). 
The latter have metallicity estimates (on the ZW scale) from Layden (1994).   The symbols
are as indicated on the graph: {\it Kepler} stars (large black squares); 
two very metal-rich RR~Lyr stars (red crosses);
six metal-rich RR~Lyr stars (red open squares);
seven intermediate-metallicity RR~Lyr stars (blue open triangles); and 
nine intermediate-to-low [Fe/H] RR~Lyr stars (blue open circles). 
The {\it Kepler} stars have been shifted upwards by 0.151 (in accordance
with Eq.~2) and thus are $V$ values (as are the values for the Lub-Simon stars).
 }
\label{PHI21vsA1and PHI31}
\end{figure}


Implementation of the $P$-$A$ diagram (see upper-left panel of Fig.~9) for deriving metallicities
was via eq.~6 of S04, combined with the  {\it Kp}-$V$ offset for  $A_{\rm tot}$
(Eq.~2):   
\begin{equation}
\renewcommand{\arraystretch}{1.2}\begin{array}{l}
  {\rm [Fe/H]_{\rm ZW}}  = -1.453 \thinspace A_{\rm tot}({\it Kp}) - 7.990 \thinspace {\rm log}P - 2.348,  
\label{eq}
\end{array}
\end{equation}
where the errors in the coefficients (from S04) 
are $\pm0.027$, $\pm 0.091$ and $\pm 0.043$, respectively.   
The derived [Fe/H]$_{\rm ZW}$ values are listed in Table~4 under `S4eq6'.
According to this formula the most metal-poor star is AW~Dra, 0.2 dex more metal-poor than NR~Lyr,
and $0.6$ dex more metal-poor than the estimates derived form the $\phi_{31}$ equations!
Given its location in the $\phi_{\rm 31}^c$ vs log$P$ diagram, and our recently derived 
spectroscopic metallicity of $-1.33\pm0.08$~dex, it is our opinion that this is an
evolved RR~Lyrae star;  curiously, it follows the pattern of evolved RR~Lyr stars seen in
M3 by CCC5 (their fig.16).  
The metal-rich stars also are predicted to be more-metal poor than the values given by the
$\phi_{\rm 31}$ estimates.


Our fourth set of [Fe/H] estimates was based on the RT {\it vs.} log$P$ diagram (lower-left panel of Fig.~9).
The metallicities were calculated using Sandage's relation between RT, $P$ and [Fe/H] (his eq.~7),
combined with our $V$-{\it Kp} offset (Eq.2):
\begin{equation}
{\rm [Fe/H]_{\rm ZW}}  = 6.33 \thinspace {\rm RT}({\it Kp}) - 9.11 \thinspace {\rm log}P - 4.59,   
\end{equation}
where the S04 intercept has been increased by $0.01$ (representing the 
RT conversion from $V$ to {\it Kp}).   
Application of this equation gave the metallicities listed in Table~4 under `S4eq7'.

\begin{table}
\caption{Metal abundances based on the {\it Kp} photometry and the 
equations of Jurcsik (1998) and Sandage (2004).}
\label{cfreq}
\begin{flushleft}
\begin{tabular}{lcccc}
\hline
\multicolumn{1}{c}{Star}  &  \multicolumn{4}{c}{ [Fe/H] }      \\   
\cline{2-5}
             & $\phi_{31}^s$&  $\phi_{31}^c$& $A_{\rm tot}$& RT     \\  
            &   J98eq1&  S4eq3& S4eq6& S4eq7            \\
             &CG, ZW, C9 &  ZW  &    ZW  &    ZW     \\
 \multicolumn{1}{c}{(1)} & (2) & (3) & (4) & (5)  \\
\hline
NR~Lyr       &$ -2.04, -2.34, -2.32$ & $ -2.28$& $-2.13$& $-2.16$  \\
V715~Cyg     &$ -1.18, -1.44, -1.34$ & $ -1.45$& $-1.17$& $-0.74$  \\
V782~Cyg     &$ -0.25, -0.47, -0.29$ & $ -0.51$& $-0.86$& $-0.76$  \\
V784~Cyg     &$ +0.06, -0.14, +0.07$ & $ -0.18$& $-1.09$& $-0.87$  \\
KIC~6100702  &$ -0.14, -0.35, -0.17$ & $ -0.38$& $-0.69$& $-0.48$  \\
NQ~Lyr       &$ -1.55, -1.83, -1.77$ & $ -1.86$& $-1.68$& $-1.57$  \\
FN~Lyr       &$ -1.62, -1.90, -1.84$ & $ -1.94$& $-1.70$& $-1.31$  \\
KIC~7021124  &$ -1.79, -2.08, -2.04$ & $ -2.08$& $-1.91$& $-1.83$  \\
KIC~7030715  &$ -1.39, -1.66, -1.58$ & $ -1.60$& $-1.97$& $-1.96$  \\
V349~Lyr     &$ -1.46, -1.73, -1.66$ & $ -1.77$& $-1.47$& $-1.08$  \\
V368~Lyr     &$ -1.27, -1.53, -1.44$ & $ -1.53$& $-1.27$& $-0.64$  \\
V1510~Cyg    &$ -1.56, -1.83, -1.77$ & $ -1.86$& $-1.89$& $-1.57$  \\
V346~Lyr     &$ -1.55, -1.83, -1.76$ & $ -1.86$& $-1.84$& $-1.52$  \\
V350~Lyr     &$ -1.56, -1.84, -1.78$ & $ -1.86$& $-1.96$& $-1.62$  \\
V894~Cyg     &$ -1.51, -1.79, -1.72$ & $ -1.82$& $-2.01$& $-1.46$  \\
V2470~Cyg    &$ -0.48, -0.71, -0.55$ & $ -0.74$& $-1.13$& $-0.99$  \\
V1107~Cyg    &$ -1.30, -1.56, -1.48$ & $ -1.60$& $-1.56$& $-1.38$  \\
V838~Cyg     &$ -1.30, -1.56, -1.48$ & $ -1.58$& $-1.40$& $-0.84$  \\
AW~Dra       &$ -1.47, -1.74, -1.67$ & $ -1.68$& $-2.34$& $-2.06$  \\
\hline
\end{tabular}
\end{flushleft}
\end{table}

\subsubsection{Discussion of the derived metallicities}

In the absence of spectroscopic metal abundances\footnote[8]{High-resolution echelle spectra of the 
brightest 18 {\it Kepler} RR~Lyr stars (non-Blazhko and
Blazhko, RRab and RRc) recently have been acquired with the ESPaDOnS spectrograph on the
Canada-France-Hawaii (CFHT) 3.6-m telescope (Nemec, Ripepi, Chadid {\it et al.} 2011, in preparation).  
These spectra currently are being measured and {\it preliminary} informative results are available for two of the 
non-Blazhko RRab stars:  (1) For KIC~6100702 a spectroscopic metal abundance of $-0.18 \pm 0.06$ dex has been derived -- this compares very favourably
with the two $\phi_{\rm 31}$-based metallicities, [Fe/H]$_{\rm C9} = -0.17$ dex and
[Fe/H]$_{\rm ZW} = -0.38$ dex (see columns 2 and 3 of Table~4), both of which are more metal-rich than the $A_{\rm tot}$ and RT values.  
Thus for one of the four suspected metal-rich non-Blazhko RRab stars
there is spectroscopic confirmation of high metallicity;   and (2) for AW~Dra a spectroscopic [Fe/H]$ = -1.33 \pm 0.08$ dex 
has been derived.  If correct this is more metal-rich than any of the derived photometric estimates.  Our suspicion is that AW~Dra, which 
exhibits a large period-shift in the $A_{\rm tot}$ {\it vs.} log$P$ diagram  and in the  $A_1$ {\it vs.} log$P$ diagram (see upper two panels of Fig.~9), and
is an outlier in the $\phi_{\rm 21}^s$ {\it vs.} $A_1$ diagram and in the $\phi_{\rm 21}^s$ {\it vs.} $\phi_{\rm 31}^s$ diagram (see bottom two panels of Fig.~6), and 
an outlier in both Fig.10 panels,  may be in an advanced post-ZAHB evolutionary state (see $\S7$).  
The spectroscopic metallicities for these two stars currently
are being firmed up and data reductions are continuing on the other 16 stars for which spectra are in hand.} 
which of the [Fe/H] values in Table~4 are the most reliable?  One area of potential bias comes 
from the fact that all the metallicity correlations used in $\S6.1.4$ assume that the stars are unevolved.  This probably is true for most of 
the {\it Kepler} non-Blazhko stars;  however, there may a few evolved stars in the sample, the most likely candidate being AW~Dra.  In such a
case the period will have been increased by luminosity evolution and the derived [Fe/H] will be systematically smaller than the true value.
Another potential bias occurs for the most metal poor stars where the Fourier-based metallicity calibrations may give values
systematically too high (see JK96 and Nemec 2004). The implication here is that the most
metal-poor stars in our sample are even more metal deficient than the values given in Table~4.
  
In general the metallicities that are based 
on the Fourier $\phi_{31}$ phase parameter ({\it i.e.}, those 
derived using equation~1 of J98, and equation~3 of S04) are very similar.  This can be seen by comparing the
$\phi_{\rm 31}^s$ and  $\phi_{\rm 31}^c$ values (ZW scale) given in columns~2 and 3 of Table~4. 
When the  $\phi_{\rm 31}$ metallicities are compared with those calculated using $A_{\rm tot}$ (column~4) and RT (column~5) 
a number of systematic differences are present.   
For all the low-amplitude stars (the four suspected metal-rich stars, and AW~Dra and KIC~7030715)
the RT and $A_{\rm tot}$ estimates tend to be systematically more metal-poor than the 
$\phi_{\rm 31}$ estimates.  On the other hand, for most of the suspected metal-poor stars the $A_{\rm tot}$ and RT equations give  
more metal-rich estimates of [Fe/H] than those derived with the $\phi_{\rm 31}$ equations.  And for several of the stars 
({\it e.g.}, V1510~Cyg, V346~Lyr, V350~Lyr, V894~Cyg and V1107~Cyg) the RT gives [Fe/H] values richer than the $\phi_{\rm 31}$ values, while the
$A_{\rm tot}$ formula gives [Fe/H] values more metal poor. 
The largest difference occurs for V784~Cyg, where $\phi_{\rm 31}$ suggests [Fe/H]$_{\rm ZW} = -0.16$ while $A_{\rm tot}$ and RT imply [Fe/H]$_{\rm ZW}$$=-1$ dex. 
Our suspicion is that the slope in the RT term of Equation~6 is too steep (see M3 line in lower-left panel of Fig.9).

AW~Dra is the only star for which the estimated [Fe/H] can be compared with a previously published
metallicity estimate, and, unfortunately, that estimate is quite uncertain.  Based on its location in the $P$-$A_V$ diagram,
Castellani {\it et al.} (1998) found the metallicity of AW~Dra to be in the 
range $-1.9$ to $-1.4$ dex.    Our estimate, 
[Fe/H]$_{\rm C9}$=$-1.67$ dex, is consistent with theirs, but should be more reliable because it 
is based on much more extensive and higher precision photometry (see Table~3, and footnote~8).

\begin{table*}
\caption{Galactic coordinates, reddenings, extinctions, dereddened colours and distances }
\label{cfreq}
\begin{flushleft}
\begin{tabular}{lccllcccc}
\hline
\multicolumn{1}{c}{Star} &$\langle Kp \rangle$&  $\langle V \rangle$     &\multicolumn{1}{c}{ $l$(II) }  &\multicolumn{1}{c}{ $b$(II) } &   $E(B-V)$   & A$_V$   & ($B-V$)$_0$          &  $d$ \\   
               & [mag]    & [mag]   &\multicolumn{1}{c}{  [deg] }   &\multicolumn{1}{c}{  [deg]  } &     [mag]    & [mag]   & J98, KW01 & [kpc]   \\ 
\multicolumn{1}{c}{(1)} & (2)      &  (3)  &\multicolumn{1}{c}{   (4)  }   &\multicolumn{1}{c}{   (5)   } &     (6)        &  (7)  & (8)        &   (9)      \\
\hline
NR~Lyr         & 12.683   & 12.44 & 69.86686  & 13.55258  & 0.175(0.009) & 0.532 &0.369, 0.372 & 3.0 \\
V715~Cyg       & 16.265   &       & 72.92658  & 07.72890  & 0.170(0.009) & 0.517 &0.321, 0.327 & 14.2 \\
V782~Cyg       & 15.392   &       & 75.14435  & 06.94725  & 0.509(0.024) & 1.606 &0.358, 0.363 & 15.1 \\
V784~Cyg       & 15.370   &       & 76.41126  & 06.56689  & 0.343(0.010) & 1.088 &0.351, 0.353 & 11.8 \\
KIC~6100702    & 13.458   & 13.64 & 71.0171   & 17.7007   & 0.103(0.006) & 0.324 &0.348, 0.353 & 3.9 \\
NQ~Lyr         & 13.075   & 13.36 & 73.11219  & 15.06421  & 0.068(0.003) & 0.207 &0.352, 0.358 & 3.7 \\
FN~Lyr         & 12.876   & 12.79 & 73.45862  & 14.69286  & 0.104(0.007) & 0.320 &0.323, 0.328 & 2.9 \\
KIC~7021124    & 13.550   &       & 73.5605   & 14.7208   & 0.113(0.006) & 0.366 &0.357, 0.361 & 4.1     \\
KIC~7030715    & 13.452   & 13.24 & 74.5488   & 12.5345   & 0.108(0.006) & 0.334 &0.376, 0.377 & 3.8 \\
V349~Lyr       & 17.433   &       & 72.23472  & 18.37838  & 0.102(0.013) & 0.301 &0.324, 0.330 & 22.7  \\
V368~Lyr       & 16.002   &       & 74.40557  & 14.98618  & 0.061(0.003) & 0.189 &0.306, 0.308 & 11.0 \\
V1510~Cyg      & 14.494   &       & 78.7859   & 07.2303   & 0.484(0.030) & 1.480 &0.340, 0.345 & 10.4 \\
V346~Lyr       & 16.421   &       & 73.65539  & 19.49070  & 0.052(0.001) & 0.160 &0.340, 0.345 & 13.8 \\
V350~Lyr       & 15.696   &       & 75.65435  & 19.58972  & 0.051(0.001) & 0.158 &0.341, 0.345 & 9.9 \\
V894~Cyg       & 13.393   & 12.92 & 78.73702  & 12.60755  & 0.112(0.005) & 0.346 &0.329, 0.331 & 3.2 \\
V2470~Cyg      & 13.300   & 13.54 & 78.3633   & 14.9140   & 0.068(0.001) & 0.211 &0.357, 0.360 & 3.7 \\
V1107~Cyg      & 15.648   &       & 78.54873  & 15.03295  & 0.070(0.002) & 0.219 &0.348, 0.355 & 9.7 \\
V838~Cyg       & 13.770   & 14.24 & 79.20399  & 16.34690  & 0.067(0.002) & 0.210 &0.314, 0.318 & 5.3 \\
AW~Dra         & 13.053   & 12.85 & 80.23054  & 19.05392  & 0.047(0.001) & 0.146 &0.363, 0.365 & 3.0  \\
\hline
\end{tabular}
\end{flushleft}
\end{table*}

\subsection{Reddenings and Mean Colours}

\subsubsection{Dereddened colours}

J98 showed that the dereddened mean $B-V$ colours correlate well with 
pulsation period $P$ and the Fourier $A_1$ coefficient, and derived the following 
relationship (her eq.~3): 
\begin{equation}
(B-V)_0 = 0.308 + 0.163 P - 0.187 A_1.   
\end{equation}
Replacing  $A_1$($V$) with $A_1$(Kp) from Eq.~2 
simply reduces the intercept from 0.308 to 0.298.  Dereddened $B-V$ colours calculated with 
our modified equation for the {\it Kp}-photometry, and with 
the J98 equation for the $V$-photometry of FN~Lyr and AW~Dra,  are given in 
column~8 of {\bf Table~5} (under`J98').  The colour range is from 0.306 (for V368~Lyr) to 0.376 
(for KIC~7030715), with a mean colour of 0.343.

Another relationship for ($B-V$)$_0$ was derived by KW01.  Their eq.~6 includes an
additive  $A_3$ term, and the  $A_1$ and $A_3$ are derived from $V$-photometry.  Using our
offsets to convert to the {\it Kp} scale gives:  
\begin{equation}
\renewcommand{\arraystretch}{1.2}\begin{array}{l}
  (B-V)_0  = 0.448 + 0.189 \thinspace {\rm log}P  -0.313 \thinspace A_1 + 0.293 \thinspace A_3   
\label{eq}
\end{array}
\end{equation} 
where $A_1$ and $A_3$ now are derived from {\it Kp} photometry and the KW01 intercept has been modified
accordingly.
Dereddened colours derived using this equation  (and the KW01 equation for the 
$V$ data of AW~Dra and FN~Lyr)  are also given in column~8 of Table~5 (under `KW01').  These estimates 
are practically identical to the J98 colours and have an  overall mean of 0.347.  
In subsequent analyses we adopt for each star the average of the two colour estimates.

\begin{figure*}
\begin{center}$
\begin{array}{cc}
\includegraphics[width=8cm]{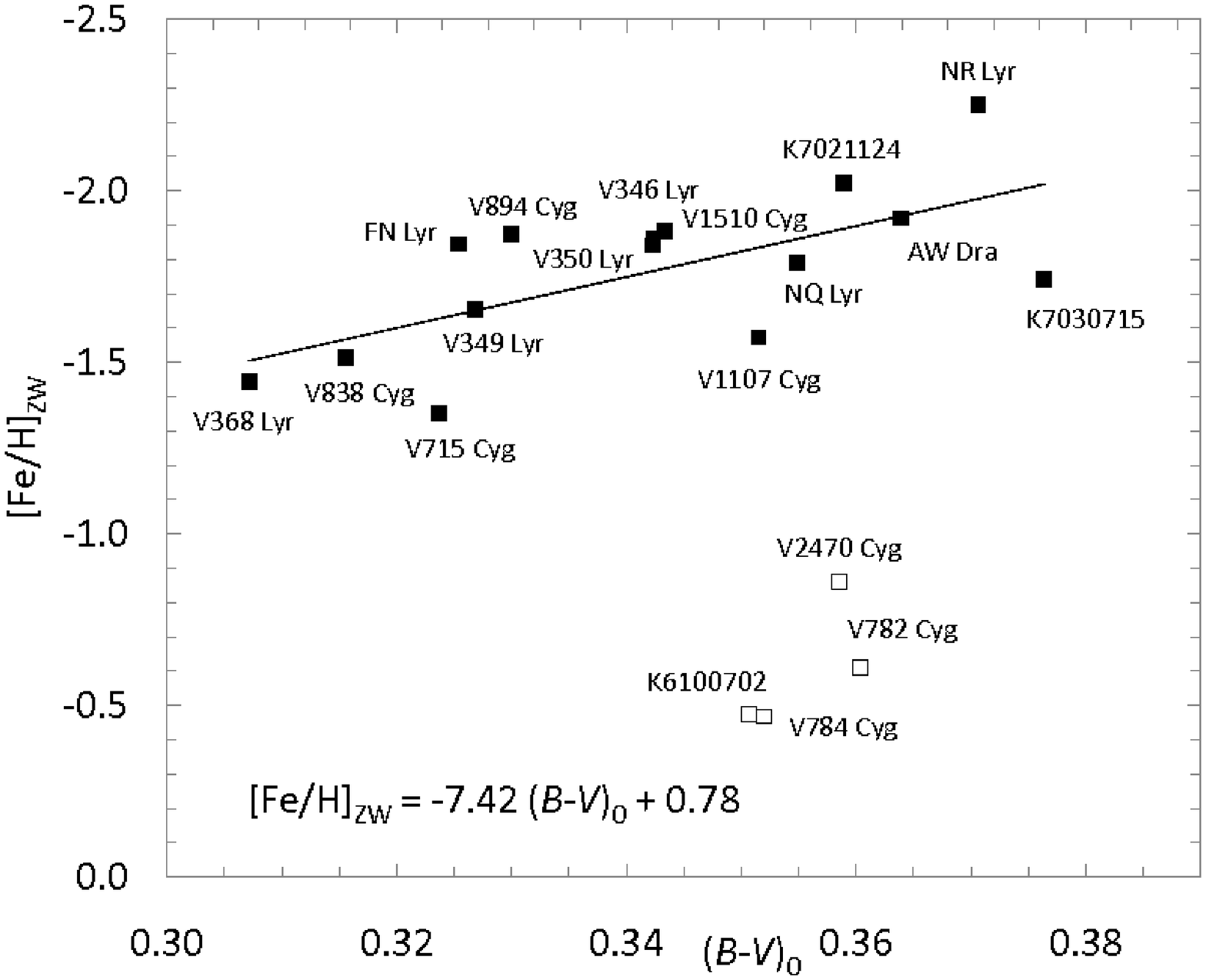}& 
\includegraphics[width=8cm]{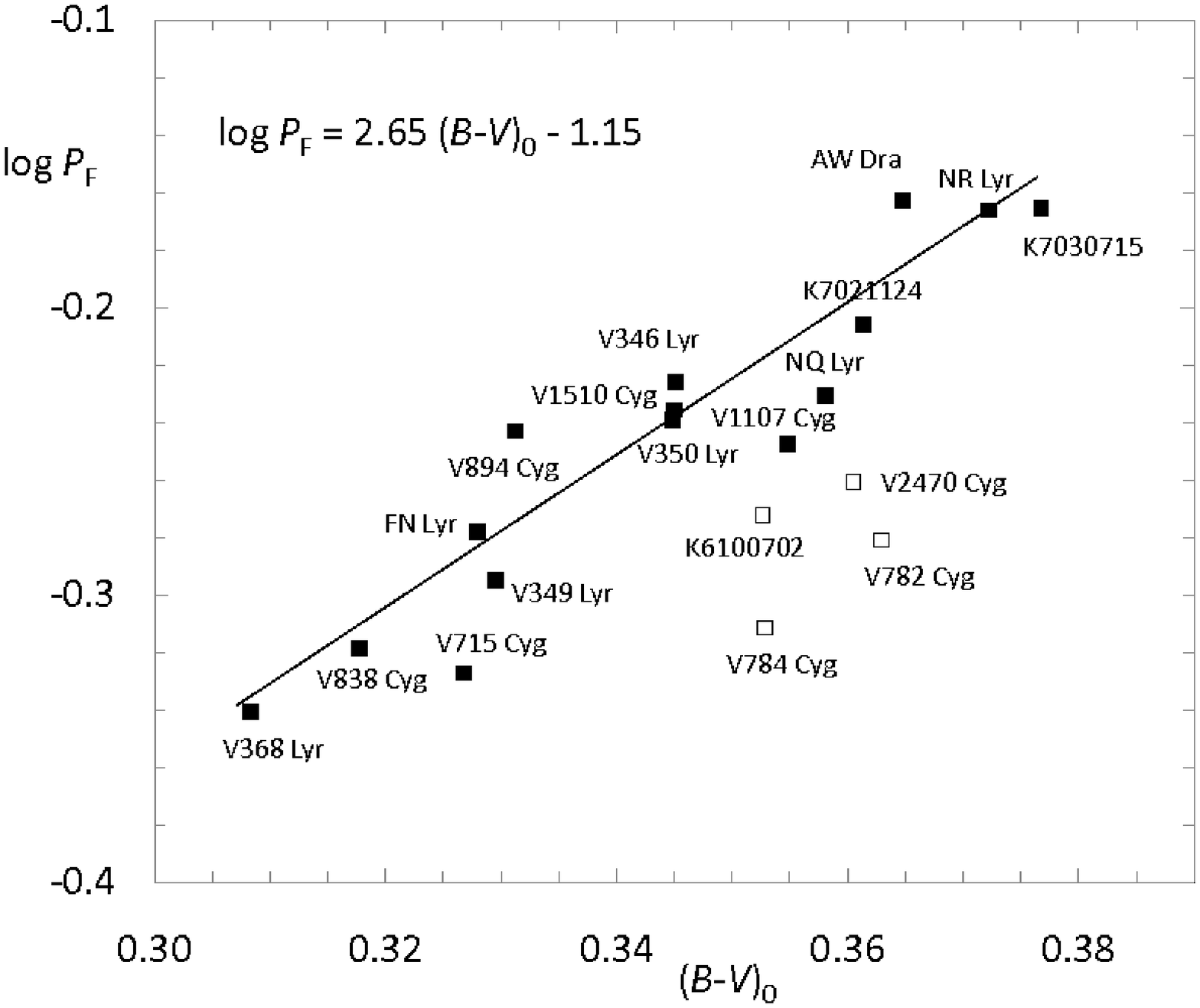} \\
\includegraphics[width=8cm]{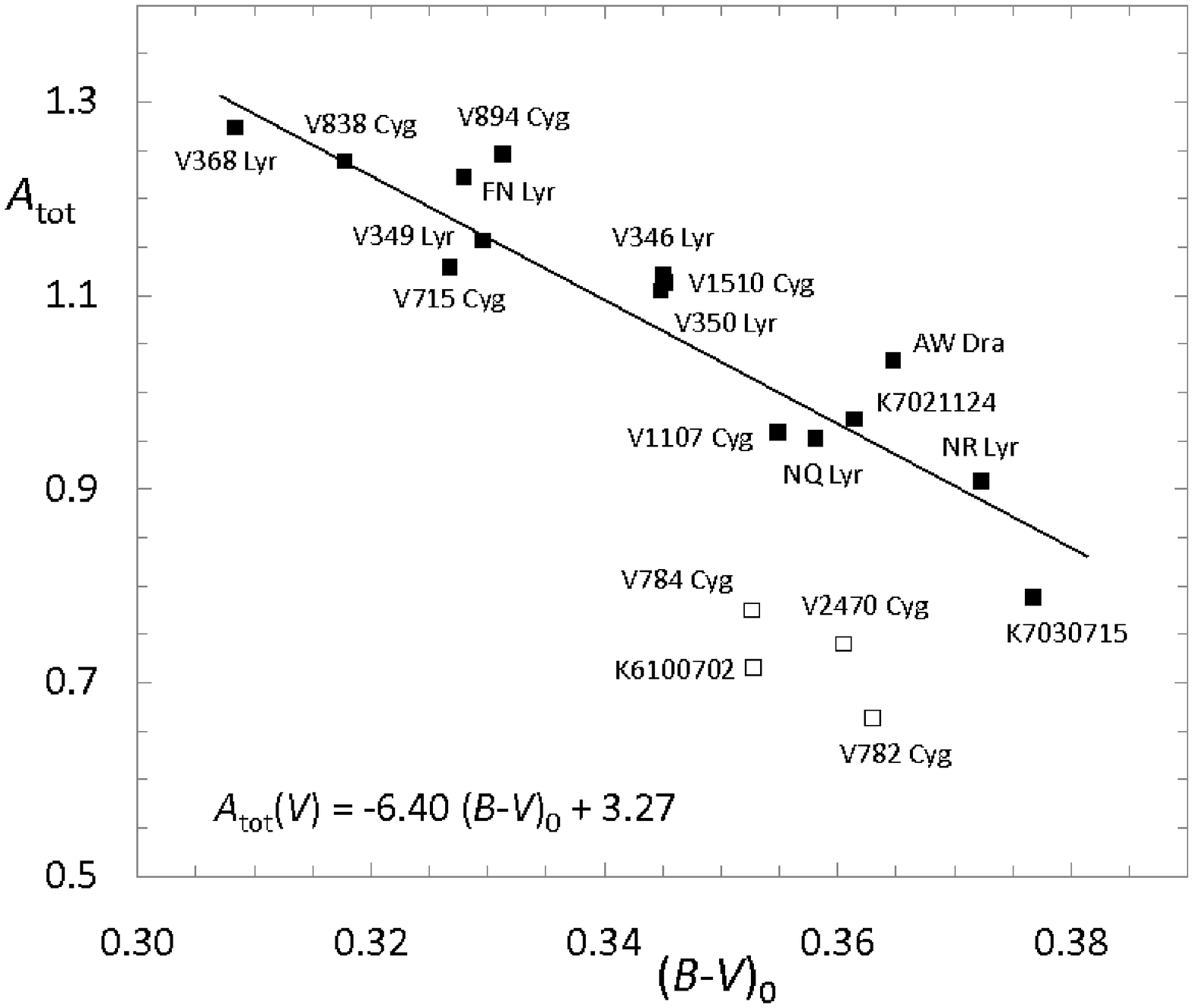}& 
\includegraphics[width=8cm]{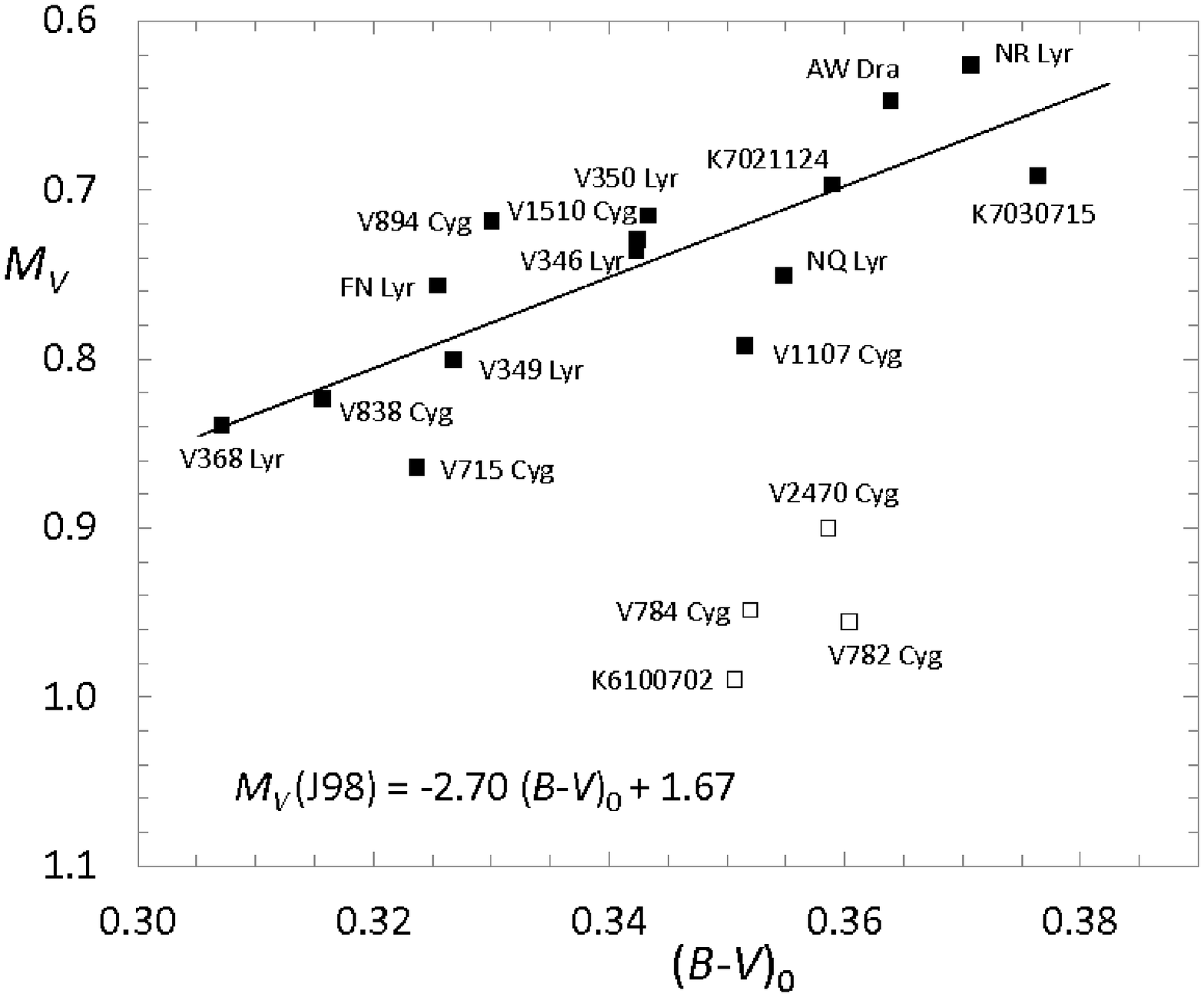} \\
\end{array}$
\end{center}
\caption{Four diagrams for the non-Blazhko RR~Lyr stars, all with dereddened mean colour  ($B-V$)$_0$
(average of the two values given in column~8 of Table~5)  along the abscissa:  
metallicity {\it vs.} colour (Top left), period {\it vs.} colour (Top right), 
total amplitude {\it vs.} colour (Bottom left), and absolute magnitude {\it vs.} colour (Bottom right).  
In each panel the line (and its equation) is from a least squares fit to the points for the 
15 metal-poor stars ({\it i.e.}, those with [Fe/H]$<-1.0$ dex).
The linear correlations show that the reddest metal-poor stars (those nearest the red edge
of the Instability Strip) are more metal poor, have longer periods and smaller total amplitudes, 
and are more luminous than the bluest metal-poor stars.
The diagrams also show that at a given colour the four metal-rich stars (plotted with open squares)  
have shorter periods, smaller amplitudes and are less luminous than the metal-poor stars.
  }
\label{ColourRelationships}
\end{figure*}

{\bf Figure~12} shows four graphs for the non-Blazhko RR~Lyr stars, all with dereddened mean colour, $(B-V)_0$, 
plotted along the x-axis.  In every graph the four high metallicity stars (V782 Cyg, V784~Cyg, KIC~6100702 and
V2470~Cyg), plotted as open squares, stand apart from the other RR~Lyr stars.  One sees that they also tend
to be on the red side of the instability strip.
The three stars with the reddest colours (and coolest temperatures), NR~Lyr, AW~Dra and KIC~7030715, 
are all of very low metallicity and have the longest periods ($P\sim0.68$~d) in our sample,
and the three shortest period stars (V368~Lyr, V838~Cyg, V715~Cyg) have the bluest colours and 
are metal poor but of intermediate metallicity ([Fe/H]$_{\rm ZW} \sim -1.5$ dex).
Linear regressions were made to the points representing the 
15 metal-poor stars, and the equations of the lines are given on the graphs. 
The colour-metallicity diagram (top left) shows that the reddest
metal-poor stars tend to have the lowest metallicities;
the colour-period diagram (top right) shows that the metal-poor stars exhibit a correlation such that 
the reddest stars have the longest periods, and that at a given colour the metal-rich stars have 
shorter periods than the metal-poor stars; the colour-total amplitude diagram (bottom left) shows 
that the red metal-poor stars have lower amplitudes than the blue metal-poor stars, and that 
at a given colour the metal-rich stars have lower amplitudes than
the metal poor stars;  and the colour-absolute magnitude diagram (bottom right -- see discussion in $\S6.4$) 
shows that at a given colour the metal-rich stars have lower luminosity than the metal-poor stars, and 
that the reddest metal-poor stars are more luminous than the bluest metal-poor stars.

\subsubsection{Reddenings and extinctions}

Table~5 also contains $E(B-V)$ reddenings (column~6) derived from the
large-scale reddening maps of Schlegel, Finkbeiner \& Davis (1998).  
These were estimated using the NStEd on-line facility and the galactic ($l$,$b$) coordinates given in 
Table~5 (columns 4-5).   According to the maps the most reddened stars in our sample are V782~Cyg, V784~Cyg and V1510~Cyg.
Total extinctions in the $V$-passband, A$_V$, were calculated
assuming an extinction-to-reddening ratio A$_V / E(B-V) = 3.1$ and are given in column~7 of Table~5.

For AW~Dra, which has calibrated $B,V$ photometry, the above reddenings can be compared 
with reddenings derived from the Fourier-based dereddened colours and the observed reddened mean $B-V$ colours.
Castellani {\it et al.} (1998) observed ($B-V$)$_0$=0.368, 
and from the $P$-$A$ diagram estimated [Fe/H] = $-1.4$ dex.  Using the Burstein \& Heiles
(1982) reddening maps they adopted the value $E(B-V)=0.04-0.06$ mag, which gives  $(B-V)_0$=0.31-0.33.  
This estimate of the reddening is consistent with 0.047$\pm$0.001 given in Table~5, but 
the dereddened colour is bluer than the average of the two Fourier based values, ($B-V$)$_0$= 0.364 (J98, KW01). 
This difference explains why the Castellani {\it et al.} estimates of $T_{\rm eff}$ (see their table~7, and below) 
are higher than our estimates (column~2 of Table~6 below).

\begin{table*}
\caption{Fundamental physical properties of the {\it Kepler} non-Blazhko RR~Lyr stars }
\label{cfreq}
\begin{flushleft}
\begin{tabular}{lccccccc}
\hline
\multicolumn{1}{c}{Star}&$T_{\rm eff}$& $M_V$ &log$\thinspace g$& $L({\rm puls})$ & $L({\rm evol})$ & ${\cal M}({\rm puls})$ &  ${\cal M}({\rm evol})$ \\    
             & KW01, S06   &  J98, F98, BCD7, CC8&J98, KW99        & J98, J98           & S06, S06, S06      & J98, J98     &   S06, BCD7   \\
             & eq11, eq18  & eq2, eq2, eq10, eq4a & eq15, eq12      & eq16, eq17         & eq8, eq10, eq12    & eq14, eq22  &   eq15, eq7  \\  
\multicolumn{1}{c}{(1)} &(2)&         (3)          &     (4)      &     (5)    &     (6)          &     (7)    &    (8)     \\
\hline
NR~Lyr       & 6242, 6124 & 0.63, 0.64, 0.67, 0.47 & 2.677, 2.689 & 46.9, 47.3 & 58.7, 54.1, 61.1 & 0.66, 0.62 & 0.73, 0.72 \\
V715~Cyg     & 6607, 6603 & 0.86, 0.81, 0.81, 0.67 & 2.874, 2.882 & 38.0, 36.7 & 47.6, 45.1, 48.6 & 0.60, 0.55 & 0.64, 0.64 \\
V782~Cyg     & 6493, 6512 & 0.96, 0.95, 1.03, 0.84 & 2.817, 2.826 & 32.0, 34.6 & 30.5, 34.4, 40.3 & 0.53, 0.50 & 0.57, 0.58 \\
V784~Cyg     & 6546, 6577 & 0.95, 0.98, 1.08, 0.88 & 2.807, 2.815 & 31.0, 33.3 & 27.2, 32.2, 38.9 & 0.49, 0.49 & 0.56, 0.56 \\
KIC~6100702  & 6564, 6605 & 0.99, 0.98, 1.08, 0.87 & 2.855, 2.863 & 31.0, 33.1 & 27.4, 32.3, 38.9 & 0.52, 0.49 & 0.56, 0.56 \\
NQ~Lyr       & 6382, 6358 & 0.75, 0.73, 0.72, 0.57 & 2.756, 2.767 & 42.1, 42.2 & 55.2, 50.3, 54.3 & 0.63, 0.58 & 0.68, 0.68 \\
FN~Lyr ({\it Kp})& 6531, 6482 & 0.76, 0.72, 0.71, 0.56 & 2.814, 2.822 & 42.7, 40.2 & 55.8, 50.8, 55.1 & 0.60, 0.59 & 0.69, 0.68 \\
FN~Lyr ($V$) & 6559, 6551 & 0.76, 0.72, 0.72, 0.56 & 2.814, 2.823 & 42.5, 39.4 & 55.6, 50.6, 54.8 & 0.60, 0.59 & 0.69, 0.68 \\
KIC~7021124  & 6332, 6240 & 0.70, 0.69, 0.69, 0.52 & 2.725, 2.737 & 44.5, 44.4 & 57.6, 52.4, 57.6 & 0.64, 0.60 & 0.71, 0.70 \\
KIC~7030715  & 6267, 6181 & 0.69, 0.74, 0.73, 0.58 & 2.676, 2.688 & 41.7, 43.7 & 54.5, 49.8, 53.7 & 0.58, 0.58 & 0.68, 0.67 \\
V349~Lyr     & 6550, 6517 & 0.80, 0.76, 0.74, 0.60 & 2.835, 2.843 & 40.8, 39.0 & 53.2, 48.8, 52.5 & 0.61, 0.57 & 0.67, 0.66 \\
V368~Lyr     & 6683, 6689 & 0.84, 0.80, 0.79, 0.65 & 2.891, 2.898 & 38.9, 36.2 & 49.4, 46.3, 49.8 & 0.57, 0.56 & 0.65, 0.65 \\
V1510~Cyg    & 6435, 6370 & 0.73, 0.72, 0.71, 0.56 & 2.762, 2.772 & 42.8, 41.8 & 56.0, 51.0, 55.3 & 0.60, 0.59 & 0.69, 0.68 \\
V346~Lyr     & 6439, 6376 & 0.74, 0.72, 0.71, 0.56 & 2.766, 2.776 & 42.6, 41.6 & 55.8, 50.8, 55.0 & 0.60, 0.59 & 0.69, 0.68 \\
V350~Lyr     & 6426, 6359 & 0.72, 0.71, 0.71, 0.55 & 2.750, 2.760 & 43.1, 42.1 & 56.3, 51.2, 55.6 & 0.58, 0.59 & 0.69, 0.69 \\
V894~Cyg     & 6496, 6443 & 0.72, 0.71, 0.71, 0.55 & 2.771, 2.781 & 42.9, 41.0 & 56.2, 51.1, 55.5 & 0.56, 0.59 & 0.69, 0.68 \\
V2470~Cyg    & 6469, 6470 & 0.90, 0.91, 0.94, 0.79 & 2.793, 2.802 & 33.9, 36.1 & 36.4, 38.1, 42.9 & 0.53, 0.51 & 0.59, 0.60 \\
V1107~Cyg    & 6425, 6375 & 0.79, 0.77, 0.76, 0.62 & 2.776, 2.786 & 40.1, 40.3 & 51.8, 47.9, 51.4 & 0.61, 0.57 & 0.66, 0.66 \\
V838~Cyg     & 6627, 6617 & 0.82, 0.78, 0.77, 0.64 & 2.863, 2.871 & 39.5, 37.3 & 50.8, 47.2, 50.7 & 0.58, 0.56 & 0.66, 0.65 \\
AW~Dra ({\it Kp})& 6306, 6217 & 0.65, 0.71, 0.70, 0.54 & 2.673, 2.685 & 43.4, 44.1 & 56.7, 51.5, 56.1 & 0.55, 0.59 & 0.70, 0.69 \\
AW~Dra ($V$) & 6303, 6276 & 0.65, 0.70, 0.70, 0.54 & 2.673, 2.684 & 43.3, 43.3 & 56.6, 51.4, 56.0 & 0.51, 0.59 & 0.70, 0.69 \\
\\
\hline
\end{tabular}
\end{flushleft}
\end{table*}

\subsection{Effective Temperatures}

Mean $T_{\rm eff}$ values for the RR~Lyr stars were estimated using two prescriptions, one from KW01 
and the other from S06 (where the relationship between ($B-V$)$_0$ colour, $T_{\rm eff}$, and [Fe/H], 
including the effects of such other factors as surface gravity and turbulent velocity, has been reviewed). 
For the  colour interval 0.20$<$$(B-V)_0$$<$0.30 S06 adopts the 
formula given by Carney, Storm \& Jones (1992):
\begin{equation}
  (B-V)_0  = -2.632 \thinspace {\rm log}T_{\rm eff} + 0.038 \thinspace [{\rm Fe/H}]_{\rm ZW} + 10.423.   
\end{equation}
Inverting this equation (eq.~18 of S06) gives:   
\begin{equation}
  {\rm log}T_{\rm eff} = -0.380 \thinspace (B-V)_0 + 0.0144 \thinspace [{\rm Fe/H}]_{\rm ZW} + 3.960.    
\end{equation}
This equation, along with our colour and metallicity estimates, 
was used to calculate $T_{\rm eff}$ for the 19 non-Blazhko stars.  These  
are given in column~2 of {\bf Table~6} (under `S06').   
Since the {\it Kepler} RR~Lyr stars have redder colours than the recommended
range of applicability of the S06 equation $T_{\rm eff}$ values also were calculated using eq.~11 of KW01. 
For [$\alpha$/Fe]=0 the metallicity term [M/H]$_{\rm CG}$ is equal to [Fe/H]$_{\rm CG}$ and we can write
\begin{equation}
\renewcommand{\arraystretch}{1.2}\begin{array}{l}
   {\rm log} \thinspace T_{\rm eff} = 3.8840 - 0.3219 \thinspace (B-V)_0 + 0.0167 \thinspace {\rm log} \thinspace g \\ 
    + 0.0070 \thinspace [{\rm Fe/H}]_{\rm CG},
\label{eq}
\end{array}
\end{equation}
which is valid over the colour range of the ab-type RR~Lyr stars studied here.
Estimates of $T_{\rm eff}$ based on this formula also are given in Table~6 (column 2, under KW01).  
In general the temperatures agree, with an average difference of only 20~K, and one could simply take
the average for the best estimate;  however, the 
KW01 temperatures are higher (lower) than the S06 temperatures for the metal-poor (metal-rich) stars.
We shall see below that there is reason to believe that the KW01 formula is to be preferred (see $\S6.9$, 
and in particular the bottom left panel of Fig.~13).  

Castellani {\it et al.} (1998, Table~7) found $T_{\rm eff} \sim 6700$~K for AW~Dra, 
assuming fundamental mode pulsation, mass $\sim$0.65-0.75 $\cal M_{\sun}$, and luminosity $\sim$65-80 $L_{\sun}$. 
This temperature is $\sim$400~K hotter than the corresponding 
Fourier-based estimate of $T_{\rm eff} \sim 6300$~K.  The difference is not surprising given
that their assumed mass and luminosity are higher than the Fourier-based values.

\subsection{Absolute magnitudes}

Following on from the Jurcsik \& Kov\'acs (1996) and Kov\'acs \& Jurcsik (1996, 1997) papers, 
J98 gives an equation for 
absolute magnitude (based on $V$-photometry) that depends on $A_1$ and $\phi_{31}^s$ (her eq.~2).  
Application of our  $V$-{\it Kp} offsets (Eq.~2) gives   
\begin{equation}
  M_V  = 1.179 - 1.396 \thinspace P - 0.477 \thinspace A_1 + 0.103 \thinspace \phi_{31}^s,  
\end{equation}
where  $A_1$ and $\phi_{31}^s$ are Fourier parameters derived from the {\it Kp}-photometry.
The resulting $M_V$ values are given in column~3 of Table~6 (under `J98, eq2').
 
The Fernley {\it et al.} (1998) equation relating $M_V$ and metallicity,  
\begin{equation}
\renewcommand{\arraystretch}{1.2}\begin{array}{l}
  M_V  =  (0.20\pm0.04) \thinspace [{\rm Fe/H}]_{\rm CG}  + (1.03\pm0.14),   
\label{eq}
\end{array}
\end{equation}
assumes that RR~Lyr has  $M_V=0.78(\pm0.29)$ at [Fe/H]$_{\rm CG}=-1.39$ dex
and is consistent with the statistical parallax solution for 84 halo RR~Lyr stars
and with various Baade-Wesselink analyses.   $M_V$ values derived using
this equation are listed in column~3 of Table~6 (under `F98, eq.2').
In general the F98 and J98 values are similar, with a mean difference (J98 minus F98)
of only 0.01($\pm$0.01) mag.  On the downside there has been mounting evidence that the calibration
of $M_V$ with [Fe/H] is non-linear (see S06 and TSR8).

Equation 10 of Bono, Caputo \& Di Criscienzo (2007) provides a third estimate of $M_V$.
They show that over the entire metallicity range from [Fe/H]$_{\rm ZW}= -2.5$ to 0.0 dex the galactic field variables
with Layden (2007) metallicities are described by
\begin{equation}
M_V = 1.19 \thinspace (\pm0.10) + 0.50\thinspace {\rm [Fe/H]}_{\rm ZW} + 0.09\thinspace {\rm [Fe/H]}_{\rm ZW}^2.
\end{equation}
Absolute magnitudes calculated with this quadratic formula also are given in column~3 of Table~6 (under
'BCD7').   For the metal-poor stars the BCD7 $M_V$ values are very similar to the F98 values;
however, because of the non-linear metallicity dependence they are fainter
than the F98 values for the four metal-rich stars.  

Most recently, Catelan \& Cortes (2008) argue that revised values for the trigonometric 
parallax and reddening of RR~Lyr imply that   
the luminosity scale for RR~Lyr stars should be brighter and give the following 
equation: 
\begin{equation}
\renewcommand{\arraystretch}{1.2}\begin{array}{l}
  M_V  =  (0.23\pm0.04) \thinspace [{\rm Fe/H}]_{\rm ZW} + (0.984\pm0.127).  
\label{eq}
\end{array}
\end{equation}
Absolute magnitudes calculated with this equation (given in column~3 of Table~6 under `CC8') are, 
on average, $\sim$0.15 mag brighter than the values from F98 and BCD7, which in turn are
$\sim$0.05 brighter than the J98 values.  Note that these differences are smaller than the
uncertainties in the individual $M_V$ estimates (which are $\sim$0.20 mag). 
The colour-$M_V$ diagram plotted in Fig.12 (bottom right panel) was constructed using the 
$M_V$ values calculated with eq.~2 of J98.

\subsection{Distances}

Approximate distances for the non-Blazhko RR~Lyr stars are given in column~9 of Table~5.
These were computed assuming: (1) the
Fernley {\it et al.} (1998) $M_V$ values (which lie between the J98 and CC08 values);
(2) the ASAS $\langle V \rangle$ values (given in column~3 of Table~5) for the nine 
stars having ASAS $V,I$ photometry; and for the other
stars the $\langle Kp \rangle$-magnitudes brightened by 0.15 mag (the observed average
$V$-{\it Kp} offset for FN~Lyr and AW~Dra); 
and (3) the visual extinctions given in column~7 of Table~5.  
With uncertainties of $\sim$0.15 mag for $M_V$, $\sim$0.08 mag for the $V$-{\it Kp} offset, and 
$\sim$0.03 mag for $A_V$, these distances are quite uncertain.  Taken at face value 
the nearest of the stars have distances $\sim$3 kpc, and the most distant star is V349~Lyr at $\sim$22 kpc.
These estimates will be considerably improved when calibrated $BVRI$-photometry becomes available.

\begin{figure*}
\begin{center}$
\begin{array}{cc}
\includegraphics[width=8cm]{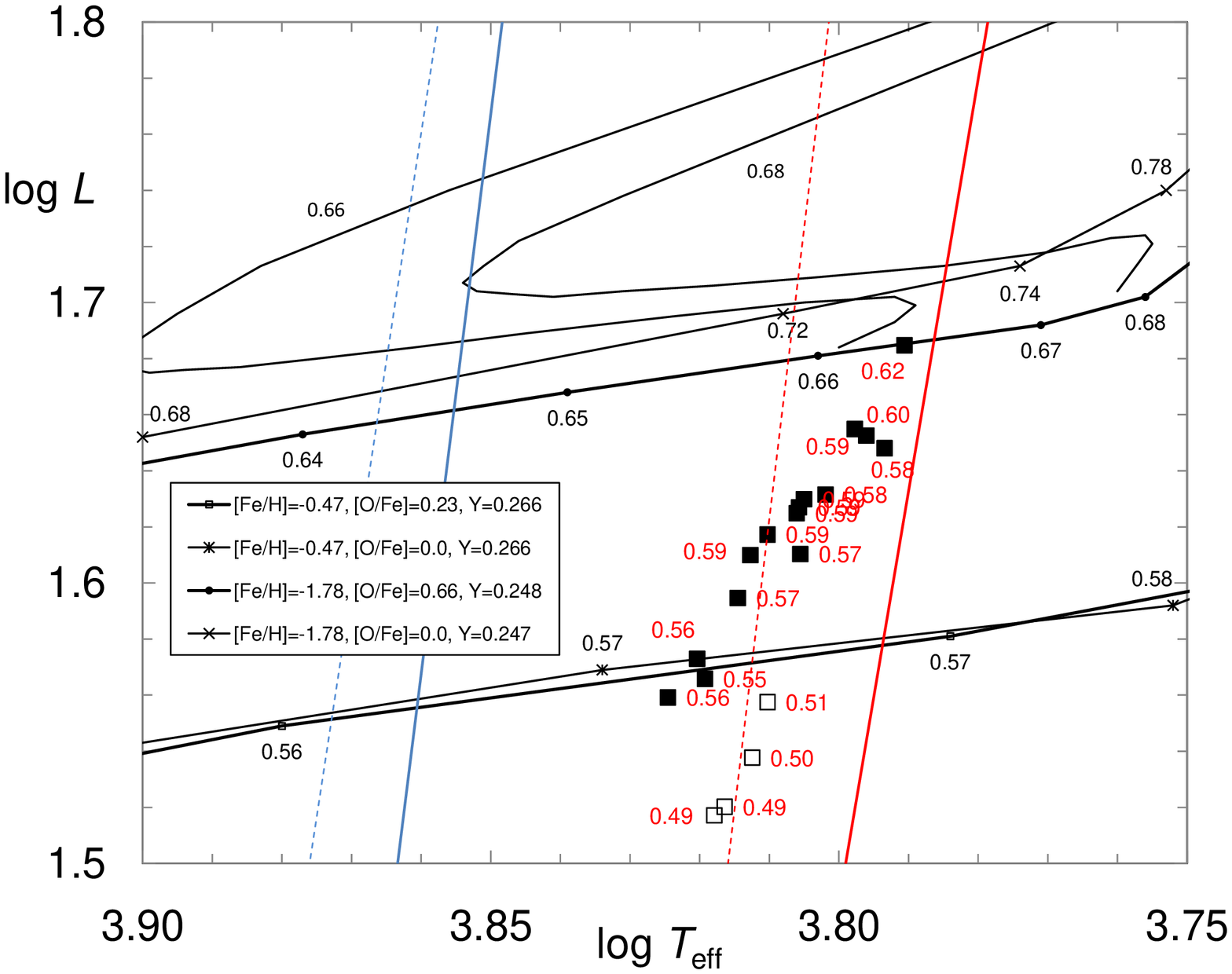} &
\includegraphics[width=8cm]{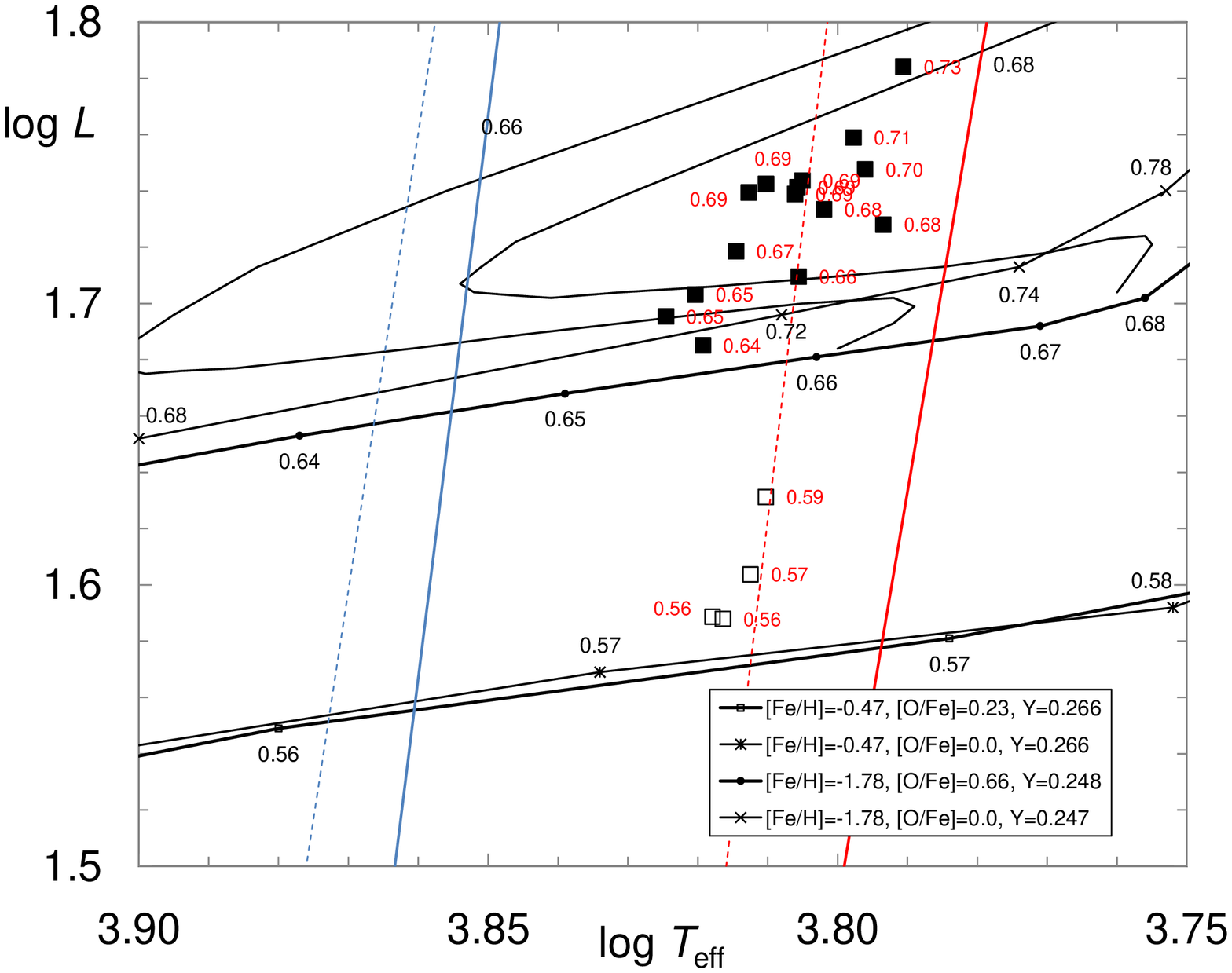} \\
\includegraphics[width=8cm]{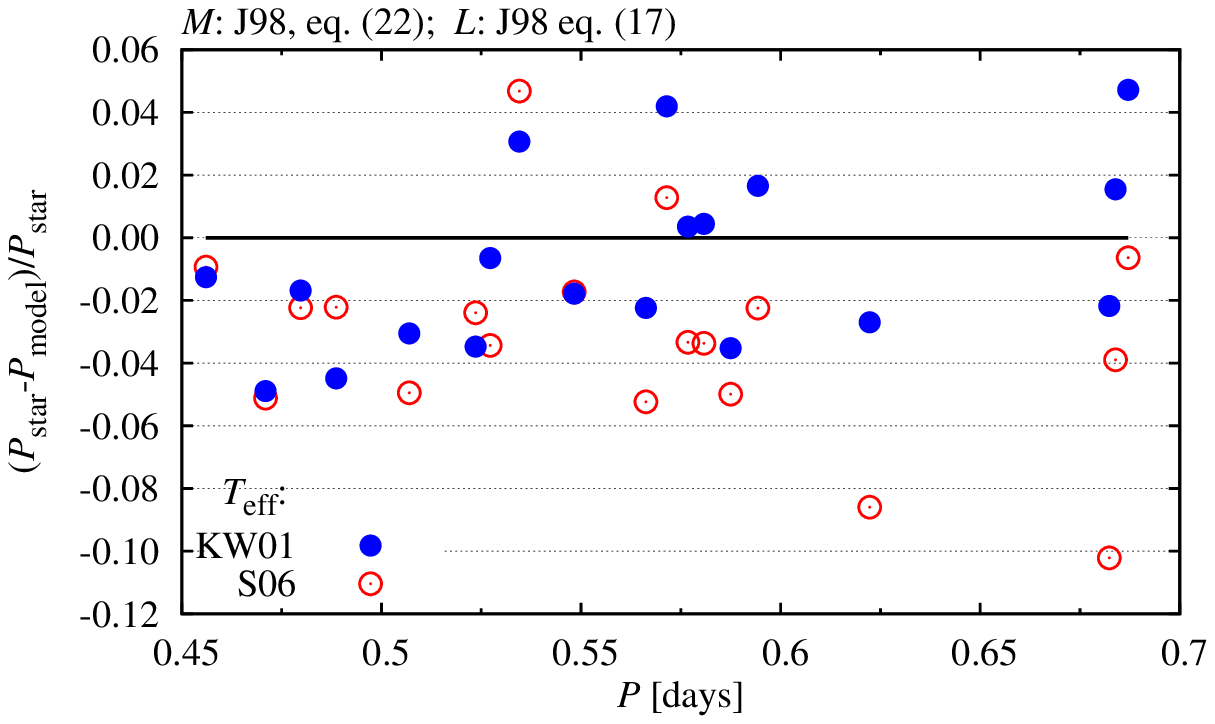} &
\includegraphics[width=8cm]{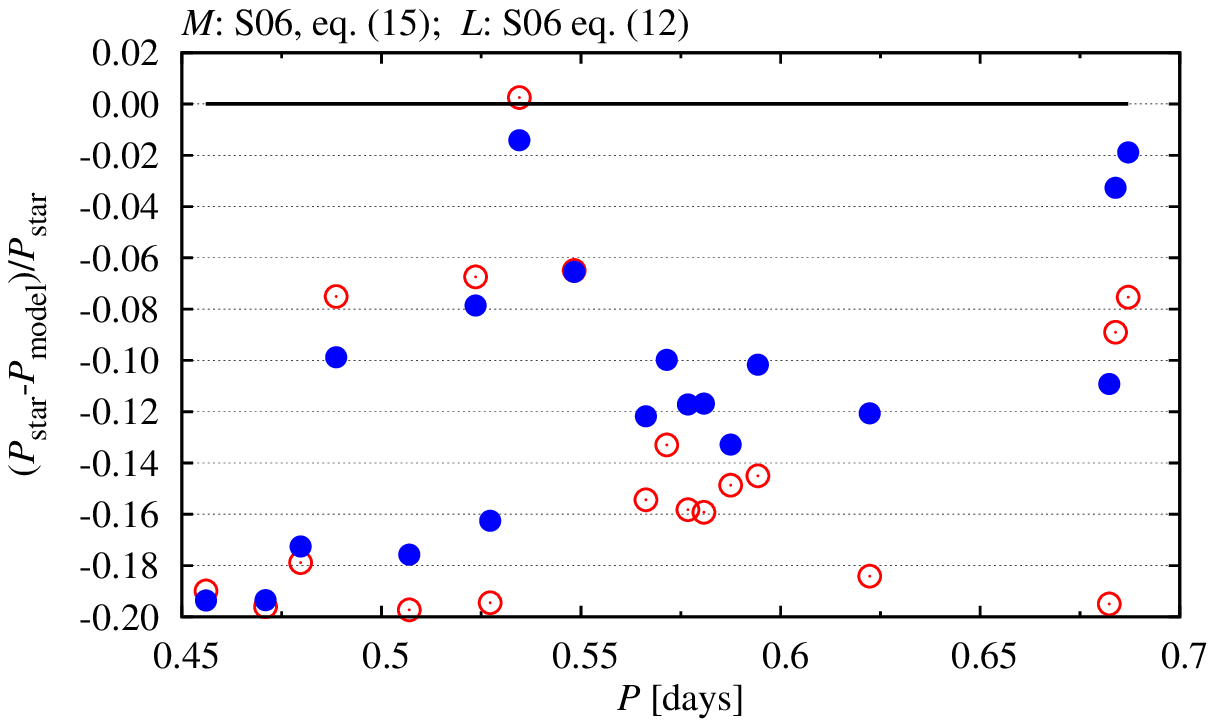} \\
\end{array}$
\end{center}
\caption{({\bf Top panels}) HR-diagrams showing the locations of the {\it Kepler} non-Blazhko RR~Lyr stars compared
with two sets of horizontal branch models (low- and high-metallicity) from Dorman (1992).  
In both graphs the four metal-rich RR~Lyr stars have been plotted with open squares and the metal-poor
stars with filled squares;  the black numbers next to the ZAHB symbols and the two evolutionary tracks are the 
assumed masses for the individual stellar evolution models.  
For the non-Blazhko RR~Lyr stars in the top left panel the $L({\rm puls})$ were calculated with 
eq.~17 of J98 and labelled (in red) with the ${\cal M}({\rm puls})$ calculated with
eq.~22 of J98;  in the top right panel the $L({\rm evol})$ were calculated with eq.~12 of S06 and labelled
(in red) with the ${\cal M}({\rm evol})$ calculated with eq.~15 of S06.
Both panels (left and right) show the blue and red edges for the instability strip
computed using the Warsaw pulsation code; the fundamental mode edges are plotted as solid lines, 
and first-overtone mode edges as dashed lines.
({\bf Bottom panels}) Graphs comparing the observed pulsation periods and the 
pulsation periods derived with the Warsaw pulsation code. The assumed masses and luminosities
are the values given in the top panels.  For each star two points are plotted,
computed assuming the $T_{\rm eff}$ from eq.~11 of KW01 (blue dots) and from eq.~18 of S06 (red circles).
The best agreement is seen in the left panel, with the KW01 temperatures favoured over the S06 values.
In both panels the largest differences are seen for the longest period stars, with no metallicity
dependence.
 }
\label{HR diagram}
\end{figure*}

\subsection{Surface gravities}

Mean surface gravities were calculated using eq.~15 from J98, and eq.~12 from KW99.
The J98 formula, which is accurate to $\pm$0.004, depends only on the period and is given by 
\begin{equation}
\renewcommand{\arraystretch}{1.2}\begin{array}{l}
{\rm log} \thinspace g = 2.473 - 1.226 \thinspace {\rm log} P.
\label{eq}
\end{array}
\end{equation}
The KW99 formula depends on period, mass and effective temperature:
\begin{equation}
\renewcommand{\arraystretch}{1.2}\begin{array}{l}
{\rm log}\thinspace g = 2.938 + 0.230 \thinspace {\rm log} {\cal M} 
- 0.110 \thinspace {\rm log} T_{\rm eff}  - 1.219 \thinspace {\rm log} P
\label{eq}
\end{array}
\end{equation}
Both estimates are given in column~4 of Table~6 and the results are in good agreement.

\subsection{Pulsational luminosities and masses}

Following on from the basic equation of stellar pulsation given by the Ritter (1879) relation,  
$P\sqrt\rho = Q$ (where $\rho$ is the mean density in cgs units and $Q$ is the pulsation constant),
van Albada \& Baker (1971, 1973) derived an equation relating the pulsation period to
mass, luminosity and $T_{\rm eff}$.  When this equation and similar more recent equations that 
include a dependence on metal abundance are used to derive the mass
and luminosity for an RR~Lyr star such quantities are referred to as pulsational mass 
${\cal M}$(puls) and pulsational luminosity $L$(puls).  Both are expressed here
in solar units. 

Pulsational luminosities were calculated with two different formulae 
given by J98  and are reported in column~5 of Table~6.
In the first case (her eq.~16) the luminosity depends only on the metallicity:  
\begin{equation}
 {\rm log} L = 1.464 - 0.106 \thinspace {\rm [Fe/H]_{CG}}.
\end{equation}
The second formula (her eq.~17) also takes into account $T_{\rm eff}$: 
\begin{equation}
{\rm log} L = 10.260 - 0.062 \thinspace {\rm [Fe/H]_{CG}}   - 2.294 \thinspace {\rm log}T_{\rm eff}. 
\end{equation}
In both cases the lower metallicity stars have the higher luminosities.  In the second equation the 
KW01 effective temperatures (Table~6, column~2) were used for the calculations.
With both equations one sees in the HR-diagram two approximately parallel lines, one for the metal-rich stars and one
for the metal-poor stars, each with the luminosity increasing as $T_{\rm eff}$ decreases (see
the top left panel of Fig.13).  At a given $T_{\rm eff}$ the luminosity difference between 
the metal-rich and metal-poor stars is greater with eq.~16 than it is with eq.~17.

Column~7 of Table~6 contains ${\cal M}$(puls) values computed  using eq.~14 and 22 of J98.
The former is given by  
\begin{equation}
\renewcommand{\arraystretch}{1.2}\begin{array}{l}
  {\rm log} {\cal M} =  1.477 \thinspace {\rm log}L - 1.754 \thinspace {\rm log}P  
  - 6.272 \thinspace {\rm log} T_{\rm eff}  \\ + \thinspace 0.037 \thinspace {\rm [Fe/H]_{CG}} + 20.884,
\label{eq}
\end{array}
\end{equation}
and the latter by  
\begin{equation}
  {\rm log} {\cal M}  =  -0.328 - 0.062 \thinspace {\rm [Fe/H]_{CG}}.   
\end{equation}
In both cases the adopted [Fe/H] values given in Table~4 (transformed to the 
CG system) were used, and in the first equation we used the ${\rm log}\thinspace L$ from J98 eq.~17 and 
the ${\rm log}\thinspace T_{\rm eff}$ from eq.~11 of KW01.
The average mass for the four metal-rich stars is $\sim$0.50
${\cal M}_{\sun}$ compared with the average mass for the metal-poor stars of $\sim$0.60 
${\cal M}_{\sun}$.

\subsection{$L$ and ${\cal M}$ from stellar evolution models}

A ZAHB stellar evolution model takes as its input the mass, ${\cal M}$(evol), and
chemical composition ($X,Y,Z$).  A subsequent evolutionary track for a given mass and composition 
gives the luminosity $L$(evol) and 
effective temperature as a function of time.  Examples of such 
models are those by Dorman (1992), Bono {\it et al.} (1997) and VandenBerg {\it et al.} (2000). 
Based on three different sets of stellar evolution models S6 gives formulae for $L$(evol) 
as a function of [Fe/H].  His eq.~8, which follows from the models of Caputo {\it et al.} (2000), is given by
\begin{equation}
{\rm log} L = 1.245 - 0.451 \thinspace {\rm [Fe/H]}   - 0.097 \thinspace {\rm [Fe/H]}^2.
\end{equation}
His eq.~10, which is based on the alpha-enhanced ZAHB models of Catelan, Pritzl \& Smith (2004),
is given by
\begin{equation}
{\rm log} L = 1.404 - 0.243 \thinspace {\rm [Fe/H]}   - 0.043 \thinspace {\rm [Fe/H]}^2.
\end{equation}
Both of these equations, like the BCD7 absolute magnitude formula discussed above,
assume a quadratic dependence on [Fe/H].  This is not the case for 
eq.~12 which is derived from the models of Clementini {\it et al.} (2003) and is given by
\begin{equation}
{\rm log} L = 1.538 - 1.110 \thinspace {\rm [Fe/H]}.
\end{equation}
Luminosities computed with these three formulae are given in column~6 of Table~6.
For the metal-poor stars all three $L$(evol) are systematically larger than the 
$L$(puls) values given in column~5.  For the four metal-rich stars the agreement is
better but there is a wide range of $L$(evol) owing to the linear or non-linear [Fe/H] dependencies.
Regardless of which formula was used the most 
luminous stars have the lowest metallicities and there is internal consistency (as was the case
for the J98 luminosities).  It is not clear whether the $L$(puls) or $L$(evol) are correct.  
Further comparison of the derived luminosities is given in $\S6.9$ (after masses are
discussed).  

A mass equation with ${\cal M}$ varying linearly with [Fe/H] is also given by S06 (his eq.~15):
\begin{equation}
  {\rm log} \thinspace {\cal M}  =  -0.283 - 0.066 \thinspace {\rm [Fe/H]_{\rm ZW}}.
\end{equation}
Since this equation was derived from the 
Bono {\it et al.} (1997) horizontal branch models it gives evolutionary masses.  
The ${\cal M}$(evol) derived with this equation
are given in column~8 of Table~6.  A more recent mass formula by 
Bono, Caputo \& Di~Criscienzo (2007), their eq.~7, which is based on the 
Pietrinferni {\it et al.} (2004, 2006) HB models, is given by
\begin{equation}
\langle log {\cal M} \rangle = -0.2675 - 0.063 \thinspace {\rm [Fe/H]}.
\end{equation}
This formula is very similar to the S06 equation. 
The masses computed with this formula are given in column~8 of Table~6 and are
seen to agree to within 0.01 ${\cal M}_{\odot}$ with the S06 masses.  
As was the case for the luminosities, the ${\cal M}$(evol) are all larger than the corresponding 
${\cal M}$(puls) values.

\subsection{Comparison with evolutionary models}

Having computed ${\cal M}$, $L$ and $T_{\rm eff}$ values for the non-Blazhko RR~Lyr stars 
from observed Fourier parameters using equations that derive from both pulsation and stellar evolution theory 
it is of interest to compare the results with the locations of model ZAHBs and evolutionary tracks. 

The top two panels of {\bf Figure~13} show HR diagrams with ${\rm log} \thinspace T_{\rm eff}$ as abscissa and ${\rm log} \thinspace L$ as ordinate, 
and two sets of ZAHB loci from Dorman (1992) computed for different masses along the ZAHB.  
The more luminous horizontal branch assumes [Fe/H]$=-1.78$ dex, and the less luminous branch [Fe/H]$=-0.47$ dex.
The numbers next to the symbols are the  masses for the individual models, which are seen to be higher for the
low-[Fe/H] tracks than for the high-[Fe/H] tracks.
For both assumed metallicities oxygen enhanced and non-enhanced ZAHBs have been plotted 
-- the effect of increasing the oxygen to iron ratio from [O/Fe]=0 to 0.66 for the low-metallicity ZAHBs 
is to lower the luminosity and reduce the mass at a given temperature.  
For the high-metallicity tracks an oxygen enhancement from [O/Fe]=0 to 0.23 has little
effect on the derived $L$ or ${\cal M}$.  Also plotted in the low-[Fe/H] case 
are the evolutionary paths away from the ZAHB for two masses, 0.66 and 0.68 ${\cal M}_{\odot}$.
In both panels the non-Blazhko RR~Lyr stars with low metallicities are represented by large black squares,
the four high-metallicity stars are plotted with open squares, 
and the $T_{\rm eff}$ are the average of the KW01 and S06 values (column~2 of Table~6).

In the top left panel of Fig.~13 the luminosities and masses (labelled in red) of the {\it Kepler} 
RR~Lyr stars were calculated  
with eqs.~17 and 22 of J98 and thus are based on pulsation theory.  The $L$(puls) and ${\cal M}$(puls)
are seen to be systematically smaller than values derived from the ZAHB tracks 
(for the appropriate metal abundance).  The reddest non-Blazhko RR~Lyr stars lie
close to the fundamental mode red-edge and have the smallest amplitudes.  
This graph also shows blue and red edges of the instability strip for the 
fundamental mode (red and blue solid lines) and first-overtone mode (red and blue dashed lines).  The edges were calculated
with the Warsaw pulsation code (see Section 7) assuming a mass of 0.65~${\cal M}_{\sun}$.  The {\it Kepler} non-Blazhko stars 
all lie in the fundamental mode region of the variability strip, and the smallest amplitude RR~Lyr stars 
(the four metal-rich stars, KIC~7030715 and NR~Lyr) have locations near the fundamental red edge (FRE) of the instability 
strip.  As expected, all the stars near the FRE have low $R_{31}$ values.

In the top right panel of Fig.~13  the luminosities were calculated with eq.~12 of S06, and the 
masses (labelled in red) with eq.~15 of S06; thus they are evolutionary $L$ and ${\cal M}$ values.  In this case there is 
very good agreement with the stellar evolution models, as one expects since they are based on stellar 
evolution models.  Enhancing the oxygen to iron ratio by the plotted amounts makes little difference. 

It is unclear which are correct, the $L$(puls) and ${\cal M}$(puls), or the $L$(evol) and ${\cal M}$(evol)?   
The mass and luminosity discrepencies go in the same direction as seen for Cepheids.  
Pietrzynski {\it et al.} (2010) recently derived a dynamical mass ${\cal M}$(dynam)  
for a classical Cepheid in a well detached, double-lined eclipsing binary in the LMC.  The mass they derive 
is very accurate and favours ${\cal M}$(puls).   The reason for the discrepancies
may be the same, as suggested by Pietrzynski {\it et al}. -- not enough mass loss has been taken 
into account in the evolution models.  For further guidance on these questions we turn to 
pulsation models.

\section{HYDRODYNAMIC MODELS}

Smolec \& Moskalik (2008) recently have developed the Warsaw convective pulsation programs
for studying stellar pulsation.
The codes, both linear and non-linear, are one dimensional and use a single equation to describe 
the generation of turbulent energy; this is done according to the model proposed by 
Kuhfu\ss{} (1986, see also Wuchterl \& Feuchtinger 1998).  
Even though a simple diffusion approximation is used  
to describe the radiation field the models are able to reproduce quite well the dynamics of 
RR~Lyrae pulsations.  

The most recent application of these programs (Smolec {\it et al.} 2011) has been to construct 
hydrodynamic models for the purpose of testing Stothers' (2006) proposed explanation of 
the Blazhko phenomenon that is observed in about half of all RR~Lyr stars.  In these models the 
strength of the turbulent convection was modulated and the resulting models were compared in detail 
with the Fourier descriptions of the {\it Kepler} observations of the RR~Lyr stars 
(Kolenberg {\it et al.} 2011).

The initial unmodulated models of the Smolec {\it et al.} (2011) study are also applicable for 
the non-Blazhko stars considered here.  The predicted  bolometric light curves were transformed 
to $V$-band light curves by applying to each pulsation phase a bolometric correction derived 
from Kurucz (2005).

\subsection{Pulsational or Evolutionary $L$, ${\cal M}$?}

As a check on the luminosities and masses derived from stellar pulsation theory and
from the horizontal branch models the Warsaw pulsation code was used to calculate pulsation periods.
Such periods are very robust and depend very weakly on the assumed convective parameters.
The calculations assumed the metallicities given in Table~4, the
$L$(puls) and ${\cal M}$(puls) values (calculated with the J98 equations) given in the top left panel of Fig.~13, 
and the $L$(evol) and ${\cal M}$(evol) values (from the S06 equations) given in the top right panel of Fig.~13.
The results, summarized in `percent-difference' diagrams comparing the observed pulsation periods and the 
model pulsation periods, are shown in the bottom panels of Fig.~13. 
Because the assumed effective temperatures are important, in both panels 
two points are plotted for each star, 
one computed assuming the $T_{\rm eff}$ from eq.~11 of KW01 (blue dots), and the other
assuming the $T_{\rm eff}$ from eq.~18 of S06 (red circles).  
For the pulsational values (left panel) the best formulae seem to be those in J98, and the
KW01 temperatures (blue dots) seem to be preferred over the S06 temperatures (red circles) -- in this case the observed and
calculated periods always agree to within $5\%$ and $\sim$half the stars are within $2\%$.  
On the other hand, the $L$ and ${\cal M}$ values from the stellar evolution models (right panel) result in 
model periods systematically too large by as much as $20\%$, with almost no agreement.
In both panels the largest differences between the KW01 and S06 temperatures 
are seen for the longest period stars.  The four metal-rich
stars all have periods shorter than 0.55~d and there is no systematic metallicity
dependence.

\subsection{Predicted Fourier parameters}

\begin{figure*}
\centering
\includegraphics[width=16cm]{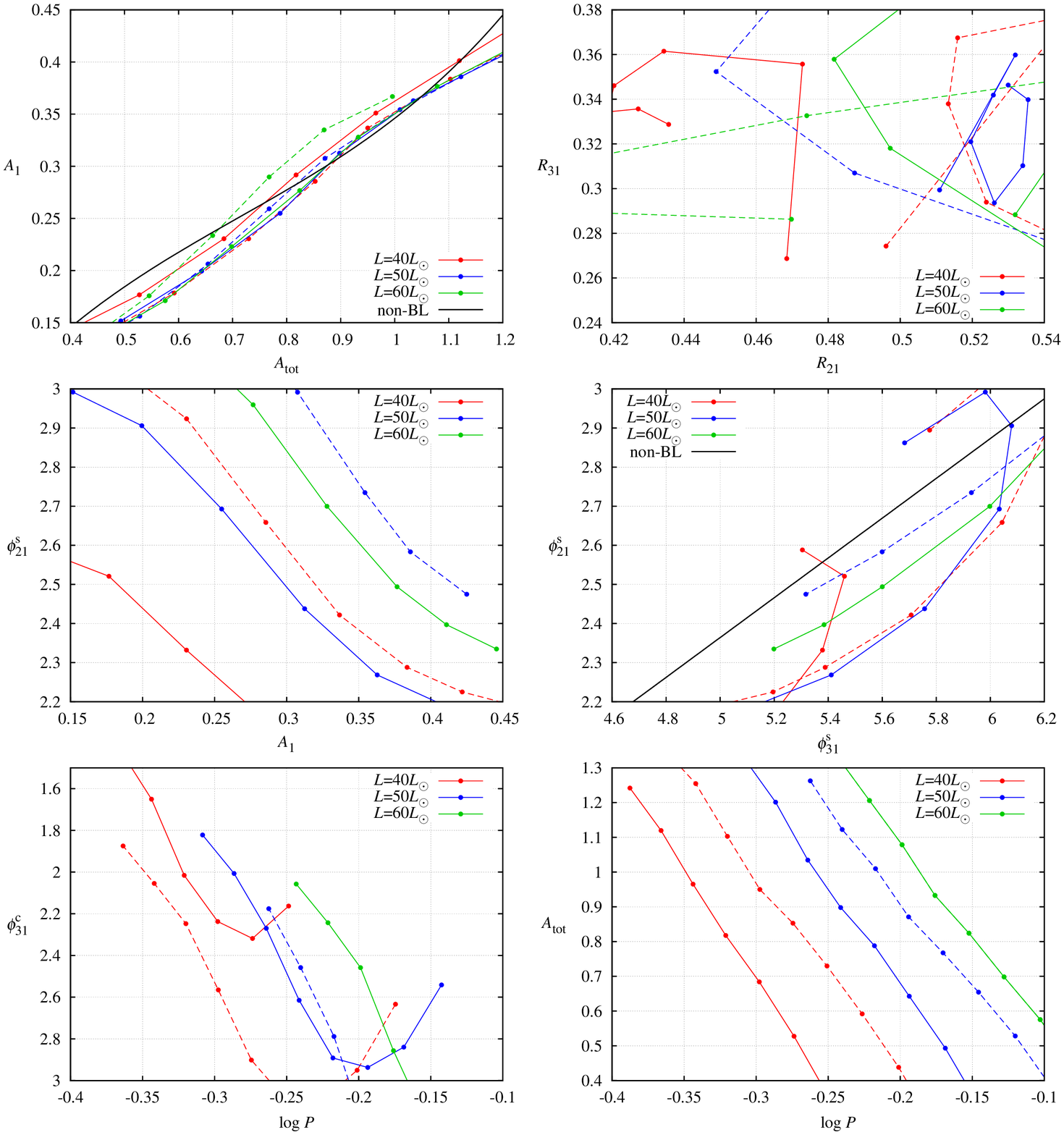}
\caption{Graphs showing the effect on the pulsation-model Fourier parameters of varying the 
$L$, and varying ${\cal M}$, while keeping the composition constant at ($X,Z$)= (0.76, 0.001),
{\it i.e.}, [Fe/H]=$-1.2$ dex. The different colours represent luminosities 40, 50 and 60 $L_{\odot}$.  The 
solid lines are for ${\cal M} = 0.65 {\cal M}_{\odot}$ and the dashed lines are
for ${\cal M} = 0.55 {\cal M}_{\odot}$.  All ordinates are for the $V$-passband.  }
\label{Models}
\end{figure*}

\begin{figure*}
\centering
\includegraphics[width=16cm]{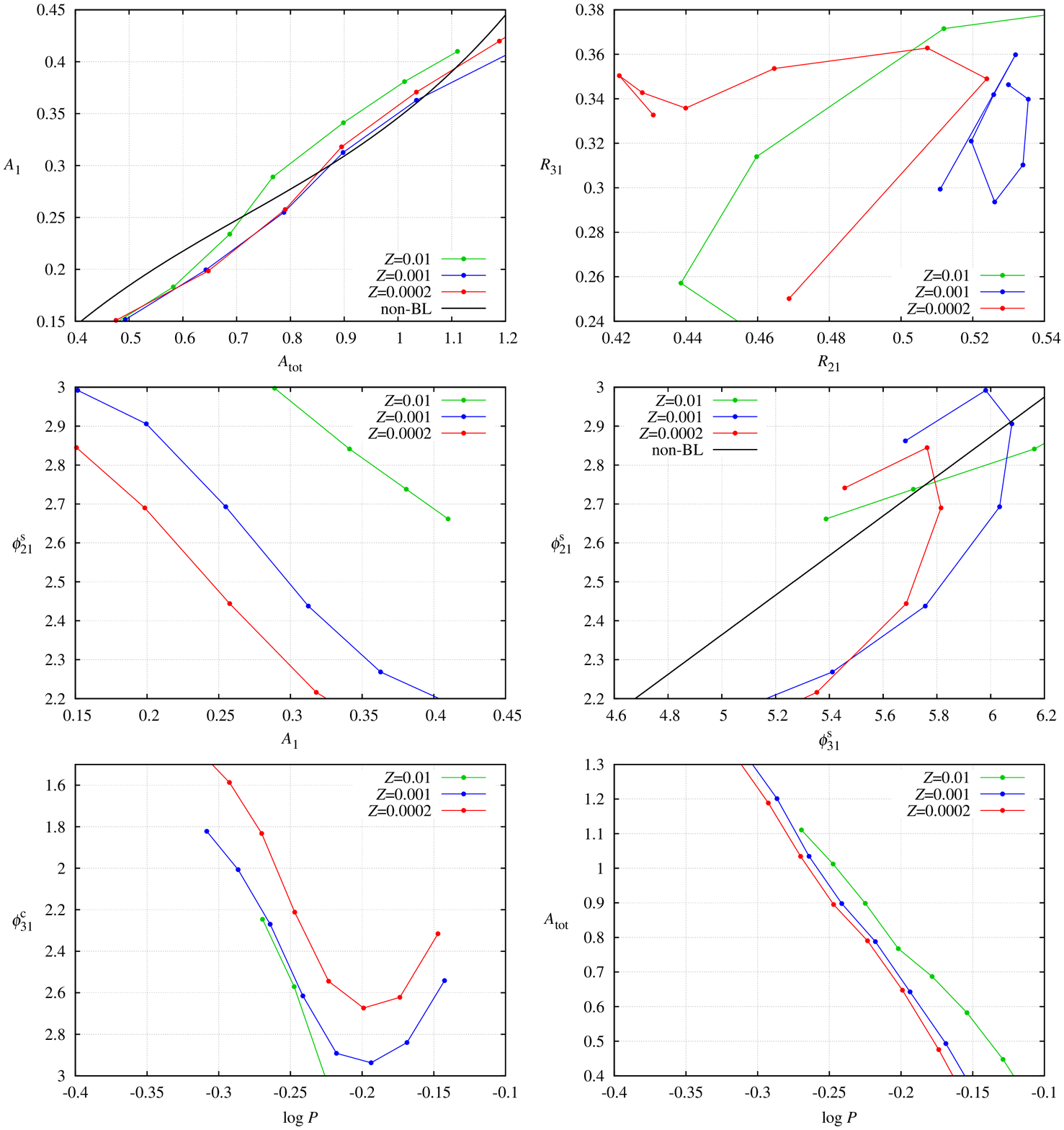}
\caption{Graphs showing the effect on the Fourier parameters derived with the Warsaw convective pulsation code 
of varying [Fe/H] from $-0.24$ dex ($Z=0.01$, green), to $-1.24$ dex ($Z=0.001$, blue), to $-1.93$ dex ($Z=0.0002$, red),  
while keeping the mass, luminosity and hydrogen content constant at 0.65${\cal M}_{\odot}$, 
50$L_{\odot}$ and $X$=0.76. All ordinates are for the $V$-passband.  }
\label{Models}
\end{figure*}

Three sets of models were computed with the Warsaw code, varying $L$ while holding the other 
variables constant, then ${\cal M}$, then [Fe/H].  The resulting Fourier parameters were plotted in diagrams such as 
those found in Figs.~2 and 6-8.   In {\bf Figure~14} the luminosity was varied ($L/L_{\sun} = 40, 50, 60$) while the metallicity was 
held constant at [Fe/H]$=-1.30$~dex ($Z=0.001$) with the hydrogen fraction fixed at $X$=0.76.  
Two masses were considered, ${\cal M}/{\cal M}_{\sun} = 0.65$ (solid lines) and 0.55 
(dashed lines).   
In {\bf Figure~15} the metallicity was varied from [Fe/H]$=-0.24$~dex ($Z=0.01$), to $-1.24$~dex ($Z=0.001$), to $-1.93$~dex ($Z=0.0002$)\footnote[9]{When
transforming between $Z$ and [Fe/H] we used [Fe/H]$= {\rm log}Z -  {\rm log}Z_{\odot} = {\rm log}Z + 1.765$, where the solar metallicity was assumed
to be $Z_{\odot} = 0.01716$ (Sweigart \& Catelan 1998).}  while keeping the other variables constant at 
$L/L_{\sun} = 50$,  ${\cal M}/{\cal M}_{\sun} = 0.65$, and $X$=0.76. 

The $A_1$ vs. $A_{\rm tot}$ diagrams (top left panels of Figs.~14-15) show very good agreement with 
the estimates for the non-Blazhko stars (right panel of Fig.~2).  
The approximately linear trend seems to be independent of luminosity or mass or metallicity variations. 
However, the fit is not exact and the observed non-linearities (which are small) appear to go in 
opposite directions to the model predictions.  

The $\phi_{21}^s$ vs $A_1$ graphs (middle left panels of Figs.~14-15) appear 
to be quite sensitive to $L$, ${\cal M}$ and metallicity effects (see the lower left panels of Figs.6-7).  
A star of given mass and composition will move up and to the right as its luminosity increases.
Alternatively, a star of given $L$ and composition will move up and to the right if the mass decreases.
And, if $L$, ${\cal M}$ and $X$ are kept constant metal-rich stars ought to occupy the high $A_1$, high $\phi_{21}$ 
region of the diagram. 

To attempt an application consider the stars AW~Dra and V784~Cyg, both of which are
seen in Fig.6 with higher than average ${\phi_{21}^s}$ values at a given $A_1$. 
If the models are indicative, then the two stars with the largest
luminosities might be expected to be AW~Dra and V784~Cyg.  According to column 5 of Table~6 AW~Dra is one of the
most luminous stars in the sample, but V784~Cyg is not -- it has one of the lowest luminosities.
On the other hand, V784~Cyg is one of the four metal-rich stars, suspected of having low $L$ and low mass. 
The high location in Fig.~6 of V784~Cyg (and the other high metallicity stars) 
might be explained by its low mass, whereas the location of AW~Dra would seem
to be due primarily to its high luminosity (either as a low-[Fe/H] star near the ZAHB or as a more metal-rich
star in an advanced evolutionary state).   The highest $L$ (and highest ${\cal M}$) star in the sample, 
NR~Lyr (see Table~6), also has the lowest metallicity (see Table~4), even lower than that of AW~Dra; 
for it the lower [Fe/H] would seem to cancel out the higher $L$, thus explaining its smaller amplitude and 
smaller $\phi_{21}$ than AW~Dra.  

The top right and middle right diagrams of Figs.14-15 are somewhat noisy.  
The top-right panel of Fig.~14 suggests 
that the lowest $L$ stars have the smallest $R_{21}$ values (note that the scale matches that
shown in the upper right panel of Fig.~6, but is much reduced from
that shown in Fig.~7 for globular cluster stars). Basically, $R_{21}$ and $R_{31}$ have to be
small very close to the red edge of the instability strip.  At the red edge the amplitude
goes to zero, as do $R_{21}$ and $R_{31}$.
The middle right panels of Fig.~14 suggest that the lowest-$L$ stars are not
expected to have high $\phi_{21}$ values. This is to be compared with 
the {\it Kepler} data shown in the lower right panel of Fig.~6 that suggest that there is a 
nearly linear relationship between $\phi_{21}$ and $\phi_{31}$ -- this is
not seen in the models.

The bottom left panels of Figs.~14-15 show that at a given $\phi_{31}$
a shift to the right can be caused by a luminosity increase or a higher mass.  The bottom left panel
of Fig.~15 shows the traditional result that at a given $\phi_{31}$ metallicity decreases as period
increases; but it also seems to be suggesting that at a given period $\phi_{31}$ increases with
decreasing metallicity.  

Finally, the bottom right panel in Fig.~14 shows that at a given $A_{\rm tot}$
and metal abundance stars of higher $L$ for a given mass, or of lower mass for a given $L$, have longer periods; 
this is as expected, and similar results are seen in fig. 6 of Dall'Ora et al. (2003).
The bottom right panel of Fig.~15 shows that for a given $L$ and ${\cal M}$ 
varying the metallicity has a relatively small effect on the period.   Taken together these two panels suggest that the main factor 
shifting the periods to longer values in the Bailey diagram (or any of its surrogate diagrams - see Fig.9)
seems to be higher luminosities (which can be caused either by higher ZAHB mass or
by advanced HB evolution) and not lower metallicities. 

Obviously more work needs to be
done to optimize the estimation of the physical variables in these observational planes, 
but the potential seems high.

\section{SUMMARY}

The main results of this paper are as follows:

(1) Fourier decomposition has been performed on the 19 least modulated ab-type RR~Lyr stars observed with
the {\it Kepler} space telescope at long cadence (every 30 min) and short cadence (every 1 minute) 
during the first 417 days of its operation (Q0-Q5);

(2) While none of the RRab stars shows the recently discovered `period-doubling'
effect seen in Blazhko variables, the star KIC~7021124 was discovered to 
pulsate in the fundamental and second overtone modes with a period ratio $P_2/P_0 = 0.59305$ 
and to have properties similar to those of V350~Lyr;  

(3) Period change rates and improved periods have been derived from O$-$C diagrams for several of the stars
that have historical data.  For AW~Dra, data from the last 12 years suggest that its period is increasing at the rate
d$P$/d$t$=3.79 d/Myr, while data spanning the last 100 years suggest a slower rate, d$P$/d$t$=0.32 d/Myr.  
FN~Lyr appears to have a very slowly increasing period, and the periods of NR~Lyr and NQ~Lyr appear to be constant.  

(4) Because the differences between the {\it Kp} and $V$ Fourier parameters for three stars (AW~Dra, FN~Lyr an NR~Lyr) are found to be 
small and systematically different we were able to use extant $V$-band correlations (with small modifications) 
to derive underlying physical characteristics for the {\it Kepler} stars.
This procedure seems to be validated through comparisons of the {\it Kepler} stars 
with other galactic and LMC field RR~Lyr stars 
and with RR~Lyr stars in galactic and LMC globular clusters.   

(5) Preliminary metal abundances have been derived for all the non-Blazhko RR~Lyr stars.  Thirteen of the 
stars appear to be similar to those found in intermediate metallicity globular clusters ({\it i.e.}, [Fe/H]$_{\rm C9}$$\sim -1.6$ dex);  
the most metal deficient star appears to be NR~Lyr with [Fe/H]$_{\rm C9} = -2.3$ dex,
and the four lowest amplitude stars (KIC~6100702, KIC~9947026, V782~Cyg and V784~Cyg) appear to be metal-rich with 
[Fe/H]$_{\rm C9}$ between $-0.55$ and $+0.07$ dex. 

(6) In general the luminosities of the metal-rich RR~Lyr stars are found to be lower than those of the metal-poor stars;
however, the luminosities derived from stellar evolution models are systematically higher than those derived
from stellar pulsation models. It is not clear which are correct.    

(7) The three stars with the longest periods, AW~Dra, NR~Lyr and KIC~7030715 (all with periods $\sim$0.68~d)
also are the reddest and coolest stars.  We suspect that AW~Dra may be in an evolved state.  In general, the stars with the
lowest amplitudes are found to be located nearest the red edge of the instability strip.

(8) The mass range for the entire sample (approximate) is from 0.50 to 0.65 ${\cal M}_{\sun}$ if based on 
pulsation theory, or from 0.56 to 0.72 if based on HB evolution models.  

(9) Finally, the Fourier parameters of the stars have been compared with values newly computed with
the Warsaw convective pulsation codes. We find that in the $P$-$A$ diagram varying the metallicity 
for a given $L$ and ${\cal M}$ has a relatively small effect on the period-shift at a given 
amplitude (in fact, the small shift is toward shorter periods for more metal poor stars),
and that the main factors causing the period shifts must be $L$ and ${\cal M}$.



\section*{Acknowledgments}
Funding for this Discovery mission is provided by NASA's Science Mission Directorate. The authors 
acknowledge the entire {\it Kepler} team, whose outstanding efforts have made these results possible. 
JMN gratefully acknowledges Dr. Amanda F. Linnell Nemec (International Statistics \& Research Corporation) 
for  a critical reading of the paper, and the Camosun College Faculty Association for funding his attendance
at the 3rd Kepler Asteroseismic Science Consortium Workshop held in Aarhus, Denmark in June 2010; he also
thanks Dr. Geza Kov\'acs for his software and discussions on Fourier methods, and Dr. Dorota Szczygiel 
for discussions on the ASAS data. 
Also, we thank Dr. K. Mighell for sending us his Guest Observer LC:Q2-Q4 data for FN~Lyr, and the
anonymous referee for a very thoughtful, knowledgeable and helpful report. 
KK, EG and RSm acknowledge support from the Austrian Fonds zur F\"orderung der wissenschaftlichen Forschung, project numbers AP~21205-N16 (RSm),
T359-N16 and P19962 (KK and EG). 
RSz and JB are supported by the Lend\"ulet program of the Hungarian Academy of Sciences (HAS) and OTKA Grants K76816 and K83790 and MB08C 81013. 
RSz acknowledges support of the J\'anos Bolyai Research Scholarship of the HAS.  The research leading to these
results has received funding from the European Community's Seventh Framework Programme (FP7/2007-2013) under
grant agreement no. 269194.  Finally, as this paper was nearing completion we learned of the
death of Dr. Allan Sandage -- this paper is dedicated to his memory.

\section*{Supporting Information}
To illustrate the constancy of the pulsations of the stars studied here `animated gifs'
have been prepared for all the stars.  These may be found in the online version of this
article, along with the data files.

\appendix

\section{KIC~7021124}

During the course of our analyses the star KIC~7021124 was discovered to be a doubly-periodic
RR~Lyr star, similar to V350~Lyr (see B10).   
From an analysis of the Q1 data (1626 long cadence
observations made over a 30-day period) its period was found to be $P_0 = 0.6224925(7)$~d, corresponding
to the frequency $f_0 = 1.606445(2)$~d$^{-1}$.
A light curve phased with this period is plotted in the upper panel
of {\bf Figure A1}.
The bottom panel shows the Fourier transform after prewhitening with $f_0$ and its harmonics.  
In addition to residual power seen at the location of the removed primary frequency and its
harmonics, we also see a family of at least six additional peaks (labelled with arrows). 
Adopting the feature at 2.7088~d$^{-1}$ as the independent frequency $f_2$, the other frequencies
correspond to $f_2\pm k\thinspace f_0$, where $k$ is an integer.  
With this assumption the period ratio is $P_2/P_0 = 0.59305$,
which is almost identical to that derived by B10 for V350~Lyr ($P_2/P_0 = 0.592$). 
The two stars also are similar in their fundamental periods, in their Fourier characteristics
(see Figs. 6,9,10), and in their masses and luminosities (see Table~5).

\begin{figure}
\begin{center}$
\begin{array}{c}
\includegraphics[width=8cm]{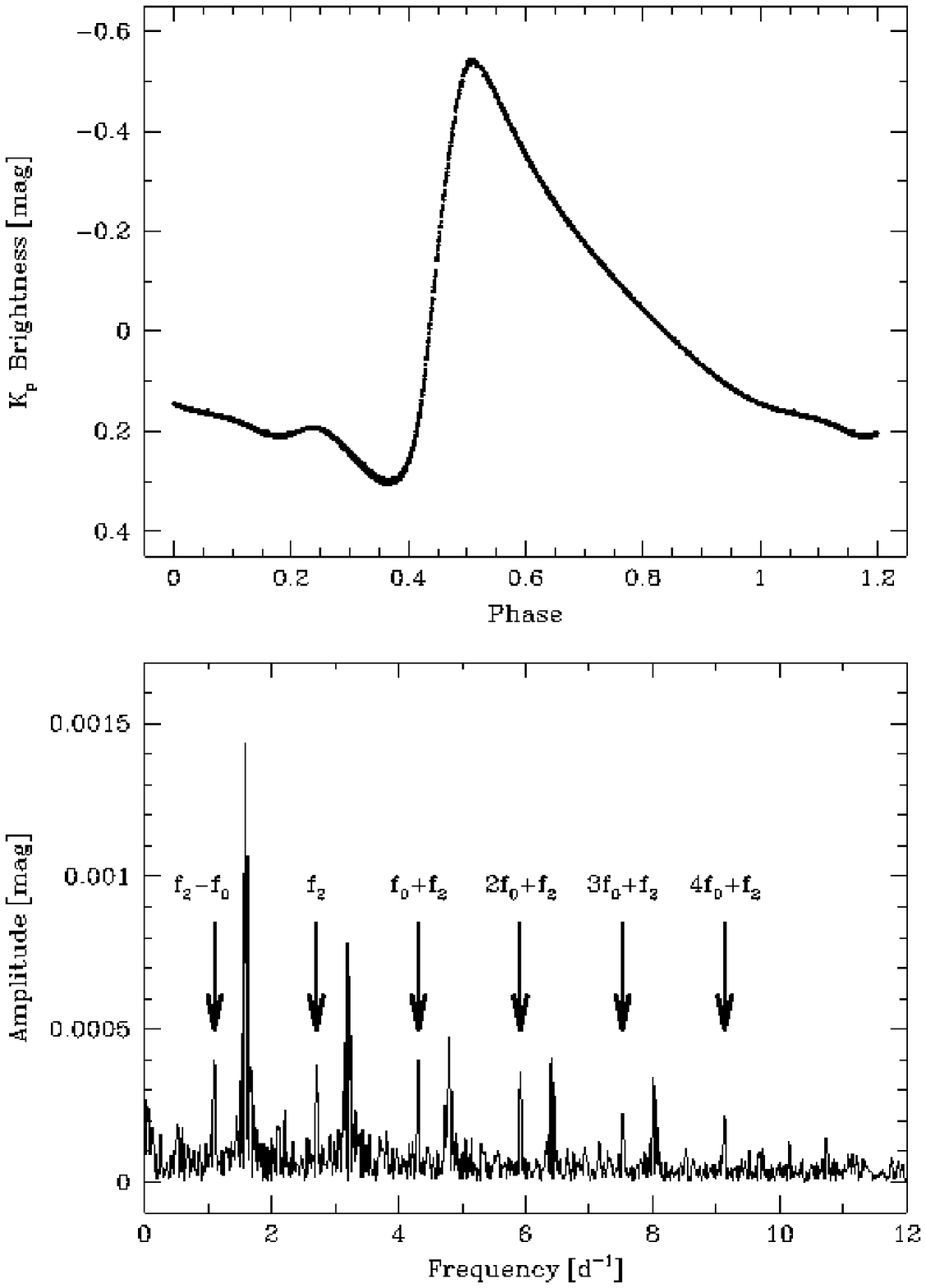} 
\end{array}$
\end{center}
\caption{(Top) Light curve for KIC~7021124, phased with the period 0.6224925(7)~d and $t0=54953.0$ (the
mean {\it Kp} magnitude, 13.550, has not been added to the ordinate). 
(Bottom) Fourier amplitude spectrum of the detrended Q1 long-cadence data, after prewhitening with the above period.  
The pattern seen here is similar to that seen for V350~Lyr in fig.7 of Benk\H o {\it et al.} (2010).  }
\label{V350}
\end{figure}

\begin{figure}
\begin{center}$
\begin{array}{c}
\includegraphics[width=8cm]{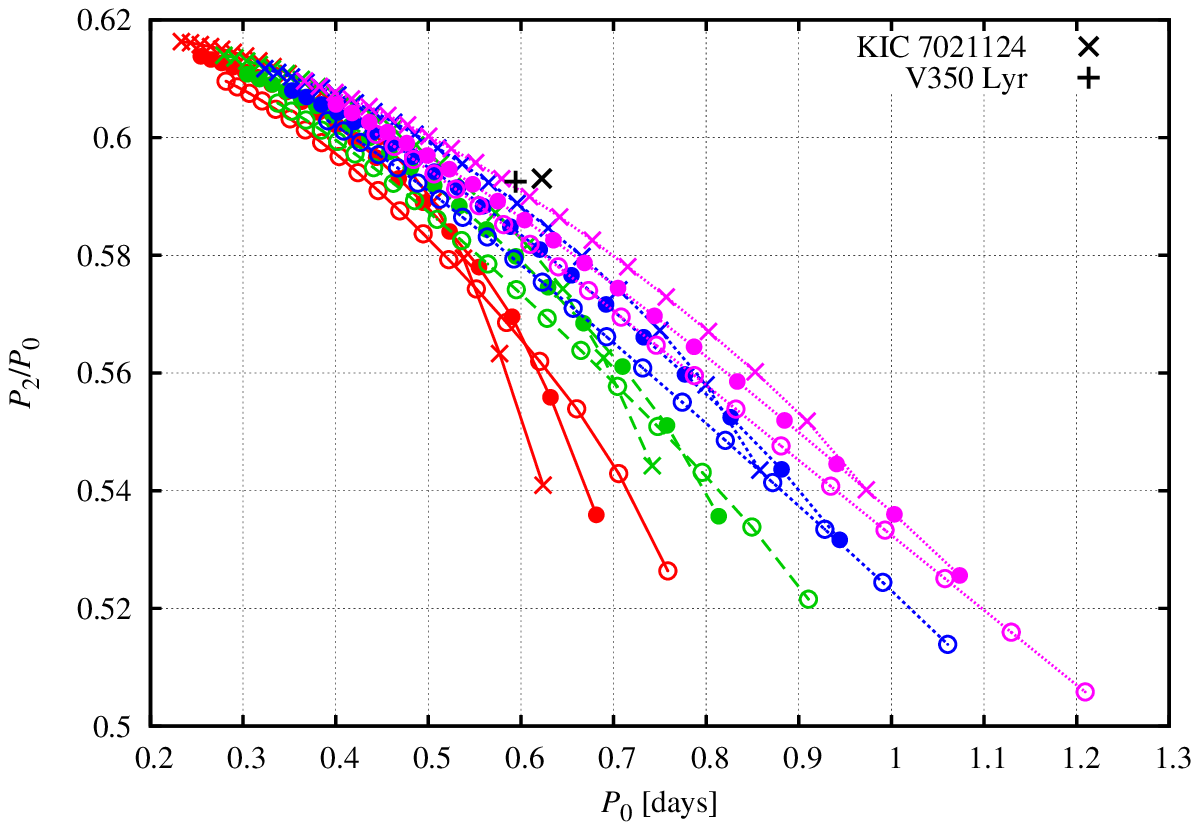}
\end{array}$
\end{center}
\caption{
KIC~7021124 (and V350~Lyr) in the Petersen ($P_2/P_0$ {\it vs.} $P_0$) diagram. The mass inferred
from this diagram is $\sim$0.75 ${\cal M}_{\odot}$, higher than both ${\cal M}$(puls) and
${\cal M}$(evol).}
\label{V350_Petersen}
\end{figure}

{\bf Figure~A2} shows KIC~7021124 (and V350~Lyr) plotted in the $P_2/P_0$ {\it vs.} $P_0$ `Petersen' diagram. 
The curves were computed with the Warsaw pulsation hydrocode including turbulent convection.  
The periods and period ratios do not depend strongly on
the convective parameters entering the model (which in this case were set C, adopted in
Baranowski {\it et al.} 2009). Other model parameters are:  $Z$=0.0001 (or [Fe/H]=-2.2);
$X$=0.76 (latest opacities, with new solar mixture); 
three masses -- 0.55${\cal M}_{\odot}$ (open circles), 
0.65${\cal M}_{\odot}$ (filled circles),  0.75${\cal M}_{\odot}$ (crosses); 
and four luminosities:  40${L}_{\odot}$ (red), 
50${L}_{\odot}$ (green),
60${L}_{\odot}$ (blue),
70${L}_{\odot}$ (purple).
The best agreement for KIC~7021124 (and V350~Lyr) is obtained for a high luminosity and a high mass:
$L/L_{\odot}=70$ and  ${\cal M}/{\cal M}_{\odot} = 0.75$.
In Table~6 both stars are among the highest $L$ and ${\cal M}$ stars in our sample.
The high values inferred from the Petersen diagram are more in accord with the evolutionary
values than the pulsation values.

\section{ASAS $V,I$-Photometry}

Very little colour information is available for the non-Blazhko RR~Lyr stars.  
Fortunately the ASAS-North survey (see Pigulski {\it et al.} 2009)
includes calibrated $V$ and $I$ photometry for nine of the brightest 
{\it Kepler} non-Blazhko stars (see $\S 4.5$).
{\bf Figure~B1} shows the cyclic magnitude and colour behaviour of the nine stars 
in the ($V-I$,$V$)-diagram.  In every case one observes the well known trend of
bluest colour when the star is brightest (i.e., when phase equals 0.0).  The largest
colour range is for NQ~Lyr and V894~Cyg (the faintest ASAS star in the sample), and 
the smallest colour range is for 
the brightest star in the sample, NR~Lyr.  The apparent red mean colour for NR~Lyr probably is due to
its relatively large reddening (see column~5 of Table 5).  The `wiggles' are artificial, a result of noise 
and imperfect fitting of the mean light curves; nevertheless, the mean trends, magnitudes and colours should
be significant.

\begin{figure*}
\begin{center}$
\begin{array}{c}
\includegraphics[width=17cm]{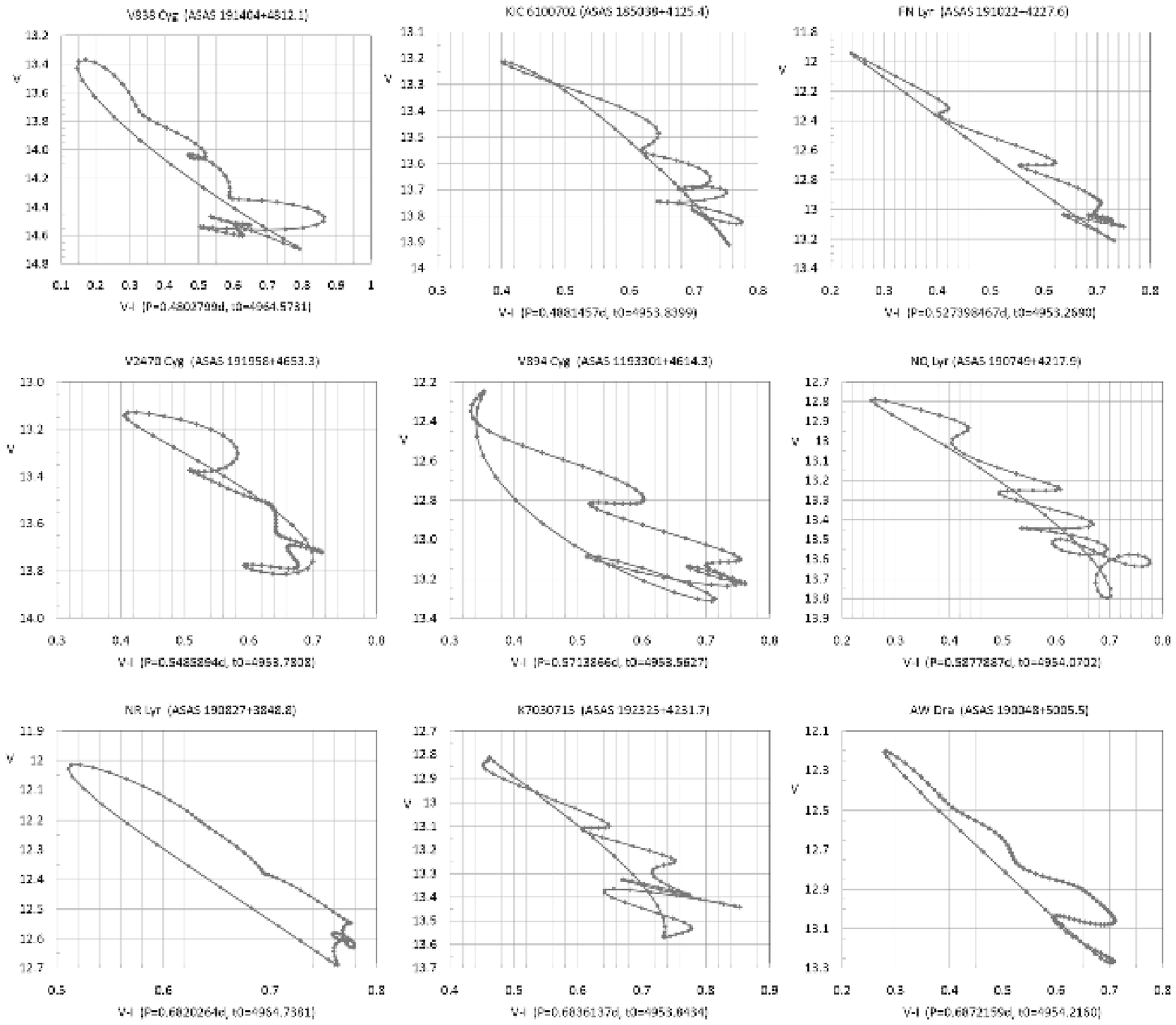}
\end{array}$
\end{center}
\caption{
Looping behaviour in the H-R diagram for the nine {\it Kepler} non-Blazhko RR~Lyr stars with $V,I$ photometric
data in the ASAS-North catalog.  The stars are ordered according to increasing period, and the points
along the mean light curves occur at every $1/100$th of the phase.   Total risetimes, $V$ amplitudes, $\langle V \rangle$ magnitudes 
and $\langle V-I \rangle$ colours for each star are given in the last four columns of Table~2. }
\label{ASAS_mosaic}
\end{figure*}

\end{document}